\documentclass[aps,prd,nofootinbib,twocolumn,superscriptaddress,preprintnumbers,balancelastpage,longbibliography]{revtex4-1}
\usepackage{tabularx}
\usepackage[utf8]{inputenc}
\usepackage[english]{babel}
\usepackage[nolist]{acronym}
\usepackage{xfrac}
\usepackage{xspace}
\usepackage{tabularx}
\usepackage{array}
\usepackage{makecell}
\usepackage{booktabs}
\newcolumntype{C}[1]{>{\centering\let\newline\\\arraybackslash\hspace{0pt}}m{#1}}
\newcolumntype{P}[1]{>{\centering\arraybackslash}p{#1}}
\newcommand{\ra}[1]{\renewcommand{\arraystretch}{#1}}

\usepackage{placeins}
\usepackage{amsmath,amssymb,mathtools,bm}
\usepackage{graphicx, color, hepunits}
\usepackage[dvipsnames]{xcolor}
\usepackage{cancel}
\usepackage[normalem]{ulem}
\usepackage{float}
\usepackage{multirow}
 \usepackage{hyperref} 
\hypersetup{
    colorlinks=true,       
    linkcolor=blue,        
    citecolor=blue,        
    filecolor=magenta,     
    urlcolor=blue          
}
\usepackage[utf8]{inputenc}
\usepackage[english]{babel}

\newcommand{\es}[2] {\begin{equation} \label{#1} \begin{split} #2 \end{split} \end{equation}}

\makeatletter
\newcounter{savesection}
\newcounter{apdxsection}
\renewcommand\appendix{\par
  \setcounter{savesection}{\value{section}}
  \setcounter{section}{\value{apdxsection}}
  \setcounter{subsection}{0}
  \gdef\thesection{\@Alph\c@section}}
\newcommand\unappendix{\par
  \setcounter{apdxsection}{\value{section}}
  \setcounter{section}{\value{savesection}}
  \setcounter{subsection}{0}
  \gdef\thesection{\@arabic\c@section}}
\makeatother

\begin{document}

\title{Axion mass prediction from adaptive mesh refinement cosmological lattice simulations}

\author{Joshua N. Benabou}
\affiliation{Berkeley Center for Theoretical Physics, University of California, Berkeley, CA 94720, U.S.A.}
\affiliation{Theoretical Physics Group, Lawrence Berkeley National Laboratory, Berkeley, CA 94720, U.S.A.}

\author{Malte Buschmann}
\affiliation{GRAPPA Institute, Institute for Theoretical Physics Amsterdam, University of Amsterdam, Science Park 904, 1098 XH Amsterdam, The Netherlands}

\author{Joshua W. Foster}
\affiliation{Astrophysics Theory Department, Theory Division, Fermilab, Batavia, IL 60510, USA}
\affiliation{Kavli Institute for Cosmological Physics, University of Chicago, Chicago, IL 60637}

\author{Benjamin R. Safdi}
\affiliation{Berkeley Center for Theoretical Physics, University of California, Berkeley, CA 94720, U.S.A.}
\affiliation{Theoretical Physics Group, Lawrence Berkeley National Laboratory, Berkeley, CA 94720, U.S.A.}

\date{\today}

\begin{abstract}
The quantum chromodynamics (QCD) axion arises as the pseudo-Goldstone mode of a spontaneously broken abelian Peccei-Quinn (PQ) symmetry. If the scale of PQ symmetry breaking occurs below the inflationary reheat temperature and the domain wall number is unity, then there is a unique axion mass that gives the observed dark matter (DM) abundance. Computing this mass has been the subject of intensive numerical simulations for decades since the mass prediction informs laboratory experiments. Axion strings develop below the PQ symmetry-breaking temperature, and as the string network evolves it emits axions that go on to become the DM. A key ingredient in the axion mass prediction is the spectral index of axion radiation emitted by the axion strings. We compute this index in this work using the most precise and accurate large-scale simulations to date of the axion-string network leveraging adaptive mesh refinement to achieve the precision that would otherwise require a static lattice with 262,144$^3$ lattice sites. We find a scale-invariant axion radiation spectrum to within 1\% precision.  Accounting for axion production from strings prior to the QCD phase transition leads us to predict that the axion mass should be approximately $m_a\in(45,65)$  $\mu \mathrm{eV}$.  However, we provide preliminary evidence that axions are produced in greater quantities from the string-domain-wall network collapse during the QCD phase transition, potentially increasing the mass prediction to as much as 300 $\mu$eV.  
\end{abstract}
\maketitle

\preprint{FERMILAB-PUB-24-0912-T}

The quantum chromodynamics (QCD) axion is currently the subject of a rapidly growing worldwide experimental program~\cite{Adams:2022pbo}.  The axion is well motivated because it can explain the dark matter (DM) of our Universe, resolve the strong-{\it CP} problem of the neutron electric dipole moment~\cite{Peccei:1977hh,Peccei:1977ur,Weinberg:1977ma,Wilczek:1977pj,Preskill:1982cy,Abbott:1982af,Dine:1982ah}, and appear in string theory compactifications~\cite{Svrcek:2006yi, Arvanitaki:2009fg}.  On the other hand, laboratory experiments for axion DM are notoriously difficult because the mass of the QCD axion particle is currently unknown to within roughly 10 orders of magnitude; many axion laboratory experiments and astrophysical probes, in contrast, need to know the axion mass to roughly one part in a million. This work aims to accelerate the search for axion DM by predicting the QCD axion DM mass in the cosmological scenario where the axion is generated as a pseudo-Goldstone boson from a spontaneously broken $\mathrm{U}(1)$ symmetry -- the Peccei-Quinn (PQ) symmetry -- at temperatures below the inflationary reheating temperature (see~\cite{Safdi:2022xkm,OHare:2024nmr} for reviews). 

\begin{figure*}[!htb]
    \centering
    \includegraphics[width=.9\textwidth]{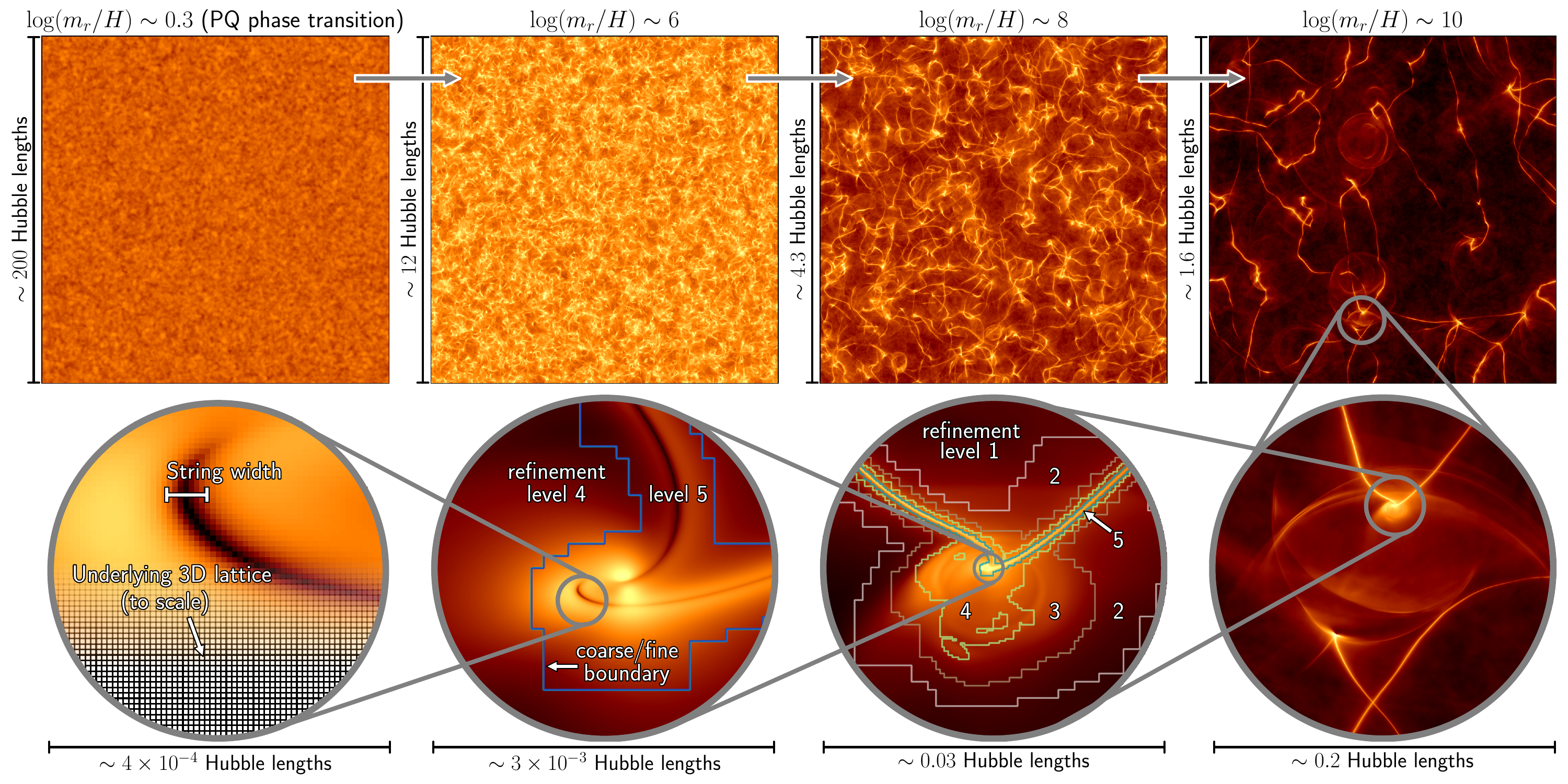}
    \caption{ 3D$\rightarrow$2D projections of the axion energy density on a logarithmic color scale. Each top row panel covers the entire simulation volume at different times, starting at the PQ phase transition \textsl{(left)} to the final state \textsl{(right)}. The bottom row is a set of nested zoom-ins of the final state at $\log(m_r/H)\sim 10$ illustrating the large scale separation between the size and width of the strings that make the AMR technique necessary. Overlaid in the three bottom left panels are the refinement boundaries along the line of sight. Animations are available \href{https://tinyurl.com/AxionStringsAMR}{here}.}
    \label{fig:MainSim}
\end{figure*}

The axion field $a$ would be massless but for its interactions with QCD of the form \mbox{${\mathcal L} \supset \frac{g^2}{ 32 \pi^2 f_a} a G^a_{\mu \nu} \tilde G^{a \, \mu \nu}$}, with $g$ the strong coupling constant and $G$ the QCD field strength.  This operator generates an axion mass \mbox{$m_a \approx 5.7 \, (10^{12} \, \, {\rm GeV} / f_a) $ $\mu$eV}~\cite{diCortona:2015ldu} at temperatures below the QCD phase transition.  The parameter $f_a$ is the axion decay constant, which is related to the vacuum expectation value (VEV) of the complex scalar that undergoes $\mathrm{U}(1)_{\rm PQ}$ symmetry breaking. Let us denote that complex scalar field as $\Phi = (r + \frac{v_a}{\sqrt{2}}) e^{i a / v_a}$, where $r$ is a real scalar degree of freedom referred to as the radial mode, and $v_a$ is the VEV of $\Phi$ in the broken phase. The VEV $v_a$ and $f_a$ are related by $f_a = v_a / N_{\rm dw}$, with $N_{\rm dw}$ the domain wall number.  In this work, we restrict to $N_{\rm dw} = 1$ to avoid long-lived domain walls. In the thermal universe, the PQ symmetry is restored at high temperatures, such that for $T \gg f_a$ the VEV is $\langle |\Phi|^2 \rangle = 0$. The universe undergoes a phase transition at $T \sim f_a$, where $|\Phi|^2$ acquires a VEV. By the Kibble-Zurek mechanism~\cite{Kibble:1976sj,Zurek:1985qw}, a defect network consisting of axion strings develops for temperatures below that of the PQ phase transition. The string network evolves to a scaling solution~\cite{Davis:1986xc}, with roughly a constant number of strings per Hubble patch -- up to logarithmic corrections~\cite{Gorghetto:2018myk} -- for temperatures $T < f_a$. At low temperatures, the radial mode is everywhere frozen at its VEV except in the vicinity of the string cores.  In Fig.~\ref{fig:MainSim} we illustrate the axion-string network in a snapshot of one of our simulations.

The axion-string network persists in the scaling regime until the QCD phase transition at $T \sim$ GeV.  The axion-string network maintains the scaling solution by radiating energy mostly in the form of axions (but see~\cite{Hindmarsh:1994re,Saikawa:2017hiv,Chang:2021afa,Gelmini:2021yzu,Gorghetto:2021fsn,Benabou:2023ghl}).  The axions are relativistic until the QCD phase transition.  At this point, the rapidly rising axion mass renders them nonrelativistic, and thereafter they act as cold DM. Also at the QCD phase transition domain walls develop between the strings, but with $N_{\rm dw} = 1$ the string-wall network collapses promptly. (See~\cite{Kawasaki:2014sqa,Hiramatsu:2012sc,Ringwald:2015dsf,Armengaud:2019uso,DiLuzio:2020wdo,Beyer:2022ywc} for $N_{\rm dw} > 1$.)

Significant effort has been dedicated to date to simulating the axion-string network in order to compute the axion DM abundance and, in turn, the axion mass that leads to the observed amount of relic DM~\cite{Vilenkin:1982ks,Sikivie:1982qv,Davis:1986xc,Harari:1987ht,Shellard:1987bv,Davis:1989nj,Hagmann:1990mj,Battye:1993jv,Battye:1994au,Yamaguchi:1998gx,Klaer:2017ond,Gorghetto:2018myk,Vaquero:2018tib,Drew:2019mzc,Drew:2022iqz,Drew:2023ptp,Gorghetto:2020qws,Dine:2020pds,Buschmann:2021sdq,Kim:2024wku,Saikawa:2024bta,Kim:2024dtq}.  On the other hand, recent simulations by different groups disagree dramatically on the predicted axion mass.  In particular, Refs.~\cite{Gorghetto:2020qws,Kim:2024wku,Kim:2024dtq} claim that the spectrum of axion radiation emitted by strings becomes infrared (IR) dominated at late times, which leads to a large DM abundance. For example, Ref.~\cite{Gorghetto:2020qws} finds that the correct DM abundance is achieved for $m_a > 500$ $\mu$eV. On the other hand, Ref.~\cite{Buschmann:2021sdq} found a nearly conformal spectrum of axion radiation, leading to the prediction $m_a \in (40,180)$ $\mu$eV.  Ref.~\cite{Gorghetto:2020qws} performed a suite of static cosmological lattice simulation of the classical equations of motion for the PQ field in co-moving coordinates with lattices up to 4,500$^3$ sites, evolving to temperatures as low as $\log(m_r / H) \sim 8$, where $m_r$ is the radial-mode mass and $H$ is the Hubble parameter. (As explained precisely below, $\log(m_r / H)$ is a proxy for time in logarithmic units between the PQ and QCD phase transitions.) More recently, Ref.~\cite{Saikawa:2024bta} performed larger static lattice simulations and considered a number of possible models for the time-evolution of the axion emission spectrum, finding them to be roughly equally compatible with the data but only some of them predicting IR-dominated emission at late times. In contrast, Ref.~\cite{Buschmann:2021sdq} used adaptive mesh refinement (AMR) simulations, with a dynamical lattice that tracks the string cores, to evolve out to $\log(m_r / H) \sim 9$.  The AMR setup in~\cite{Buschmann:2021sdq} would have required a static lattice consisting of 65,536$^3$ grid sites to achieve the same resolution. 

In this work, we extend the methodology developed in~\cite{Buschmann:2021sdq} by performing the largest AMR simulation of the axion-string network to date, evolving the network to $\log(m_r / H) \sim 10$; a static lattice would need 262,144$^3$ grid sites to achieve the same resolution as our AMR configuration.  Evolving a larger simulation box to lower temperatures both shrinks the statistical uncertainties on the measured quantities and, more importantly, allows us to reach deeper into the scaling regime. This minimizes the systematic uncertainties related to the finite IR and UV cut-offs as well as transients from the PQ phase transition. We additionally perform a suite of systematic simulations to account for axion production during the network collapse and to study the dependence of our results on the initial conditions. We conclude that the axion radiation spectrum shows no sign of becoming IR-dominated at late times and that the spectrum is nearly conformal, with a spectral index consistent with unity to 1\% precision.  On the other hand, our simulation of the string-wall network collapse during the QCD phase transition suggests that axion production from domain walls, even in the $N_{\rm dw} = 1$ scenario, could dominate over axion production by strings by a factor of a few, which would have dramatic implications for axion direct detection.

{\bf PQ Epoch Simulations.}---
We solve the equations of motion of the complex scalar field $\Phi$ on a three-dimensional grid with AMR. (See App.~\ref{sec:app_eom} and the Supplementary Materials (SM) for the equations of motion and simulation details.) We use the publicly available code \texttt{sledgehamr}~\cite{Buschmann:2024bfj} to perform one large, primary simulation and then a suite of systematic simulations. We begin by describing the setup of our primary simulation.

We start our primary simulation well before the PQ phase transition with an initial state based on a thermal distribution.  More precisely, we define a reference time $t_1$ such that $H(t_1) = f_a$, where $H$ is the Hubble parameter. Our simulations are evolved in terms of a re-scaled conformal time $\eta = R(t) / R(t_1)$, with $R$ the scale factor; in these units, we start our simulation at $\eta = \eta_i = 0.1$. We design our simulations such that the PQ phase transition occurs at $\eta \approx 0.75$.

We explicitly simulate the PQ symmetry breaking of $\Phi$ such that axion strings emerge dynamically within our simulation. Note that this approach differs from simulations performed by some other groups (see {\it e.g.}~\cite{Gorghetto:2018myk, Gorghetto:2020qws, Saikawa:2024bta}), where their simulations start after the PQ symmetry is already broken and include a pre-evolution stage to approach the attractor solution. 
In the SM we show simulation results using this alternate initial condition procedure. At late times, we find no difference between the two approaches. To avoid biases from transient effects at the PQ phase transition, we only analyze the emission spectrum at scale separations larger than  $\log(m_r/H)\sim 7.6$, with $m_r = \sqrt{2} f_a$, at which time our axion string network has approached the attractor solution. 

All simulations are performed in co-moving coordinates. The simulation volume of our fiducial simulation is chosen large enough to avoid finite volume effects until $\log(m_r/H)\sim 10$. The coarse level resolution at the final state is 8,192$^3$ cells with five levels of 2$\times$ refinement.  At $\eta_i = 0.1$ the simulation box contains 2,000$^3$ Hubble volumes of co-moving length $R(\eta_i) H(\eta_i)$, while this number drops to $200^3$ at $\eta = 1$. Our simulation ends at $\eta_f \sim 110$ with approximately 6 Hubble volumes. Our AMR setup ensures that all parts of the simulation volume are resolved, including the string emission wherever required. In particular, string cores at rest are resolved at all times by at least four grid cells. 

\begin{figure}
    \includegraphics[width=0.48\textwidth]{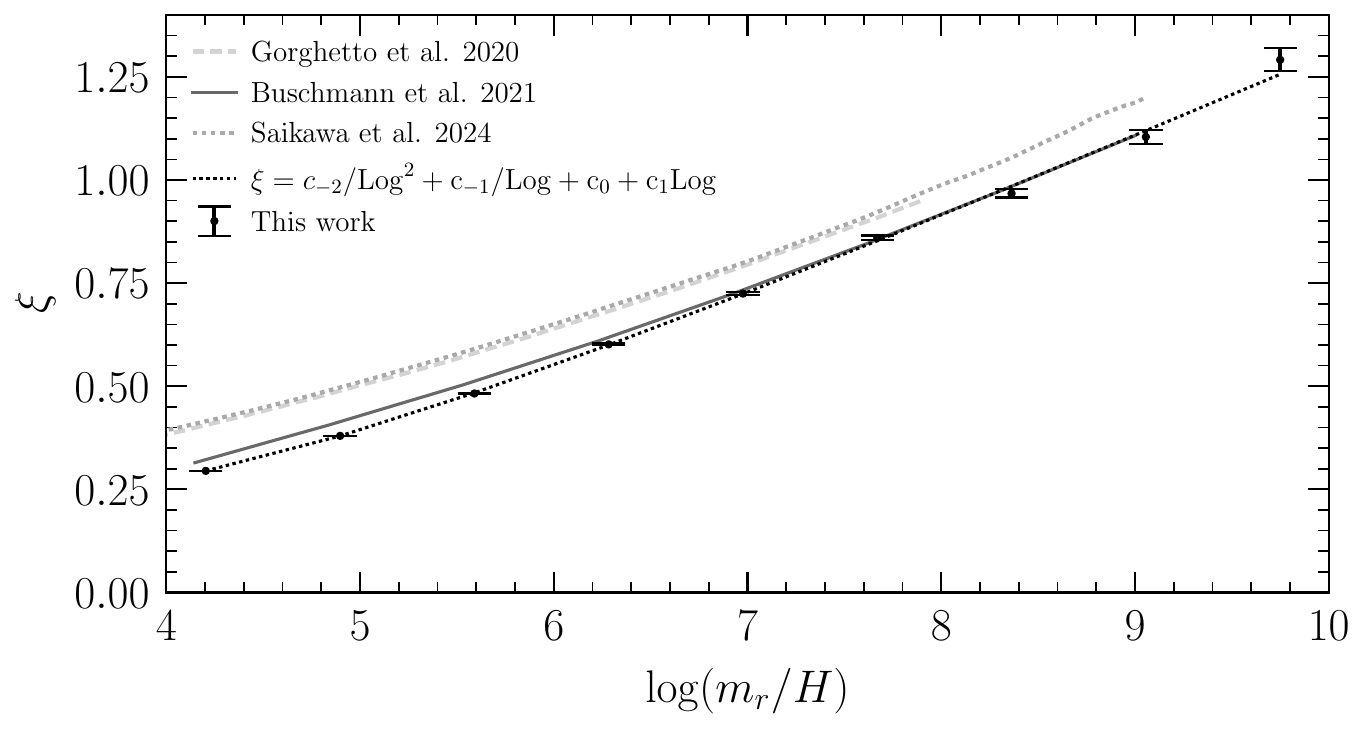}
    \caption{String length per Hubble volume $\xi$ as a function of the scale separation $\log(m_r/H)$. We compare our result with that of Buschmann et al.~\cite{Buschmann:2021sdq}, Gorghetto et al.~\cite{Gorghetto:2020qws}, and Saikawa et al.~\cite{Saikawa:2024bta}. While the simulations may differ at small values of $\log(m_r/H)$ due to differences in the initial state, they all approach the same attractor solution at large $\log(m_r/H)$. We also show a polynomial fit to our data (dotted black).}
    \label{fig:xi}
\end{figure}

Our axion mass prediction requires the extraction of two quantities from our primary simulation: (1) the string length per Hubble volume $\xi$; and (2), the instantaneous emission spectrum $F(k/H)$. We define the total string length per Hubble volume at time $t$ as $\xi=\ell t^2/\mathcal{V}$, where $\ell$ is the total string length within the simulation volume $\mathcal{V}$. We extract $\ell$ from our simulation numerically through the algorithm described in~\cite{Fleury:2015aca}. The result is shown in Fig.~\ref{fig:xi}, where data points are separated by a Hubble time ($\Delta\log (m_r/H)=\log 2$) to avoid correlations. (See~\cite{Buschmann:2021sdq} for a description of the data-driven procedure for estimating uncertainties.)

We overlay our result with that of previous simulations~\cite{Gorghetto:2020qws, Buschmann:2021sdq, Saikawa:2024bta}, to illustrate that they all approach the same attractor solution~\cite{Gorghetto:2018myk, Gorghetto:2020qws} where $\xi$ increases linearly with the scale separation $\log(m_r/H)$ at late times. The differences at small $\log(m_r/H)$ are due to the differences in the initial state but this matters very little at large $\log(m_r/H)$, as verified in the SM using alternate initial states. We verify the linear increase of $\xi$ by fitting a model of the form $\xi=c_{-2}/\mathrm{Log}^2 + c_{-1}/\mathrm{Log} + c_0 + c_1 \mathrm{Log}$ for $\mathrm{Log}\geq4$, where $\mathrm{Log}\equiv \log(m_r/H)$. We find $c_1=0.21 \pm 0.02$, which is similar to the $c_1=0.24 \pm 0.02$ of Gorghetto et al.~\cite{Gorghetto:2020qws}, $c_1=0.254\pm 0.002$ of Buschmann et al.~\cite{Buschmann:2021sdq}, and $c_1=0.23\pm 0.06$ of Saikawa et al.~\cite{Saikawa:2024bta}.  Note that our linear growth with $\log(m_r/H)$ may be in  some tension with the analytic velocity one-scale models~\cite{Martins:2018dqg, Hindmarsh:2019csc, Chang:2021afa, Correia:2024cpk}, which predict $\xi$ saturating at approximately $1.2$, whereas we observe $\xi= 1.29$ by $\log =9.75$.

We extrapolate our result to the beginning of the QCD phase transition at $\mathrm{Log}_*\in(60,70)$ and obtain $\xi_*\in(11,15)$.  More precisely, we define quantities such as $\xi_*$ to be the values at the time $t_*$ defined by $3 H(t_*)  = m_a(t_*)$, with $m_a(t)$ the time-dependent QCD axion mass. 

We now turn to the computation of the instantaneous emission spectrum $F(k/H)$. The function $F$ is defined such that $F \propto (1/R^3) \partial_t (R^3 \partial_k \rho_a)$, with $\rho_a$ the axion energy density in momentum space (see, {\it e.g.},~\cite{Gorghetto:2018myk,Buschmann:2021sdq}).  The function $F$ is important because while $\rho_a$ itself may be determined simply by energy conservation arguments, given the scaling solution for $\xi$, the spectrum $F$ is needed to compute the number density of axions $n_a$.  In particular, ignoring the effects of the axion mass for a moment, the number density of axions created by the string network at the time $t_*$ may be estimated by (see {\it e.g.}~\cite{Gorghetto:2018myk,Buschmann:2021sdq}) $n_a^* \approx 8 \pi f_a^2 H_* {\rm Log}_* \xi_* \langle H / k \rangle$, with the expectation value taken with respect to the instantaneous spectrum $F$ at $t_*$, in the limit of large ${\rm Log}_*$. We numerically determine $F$ from our simulation following the procedure described in App.~\ref{App:EmissionSpectrum}, similar to that of \cite{Buschmann:2021sdq}.

\begin{figure}
    \centering
    \includegraphics[width=\linewidth]{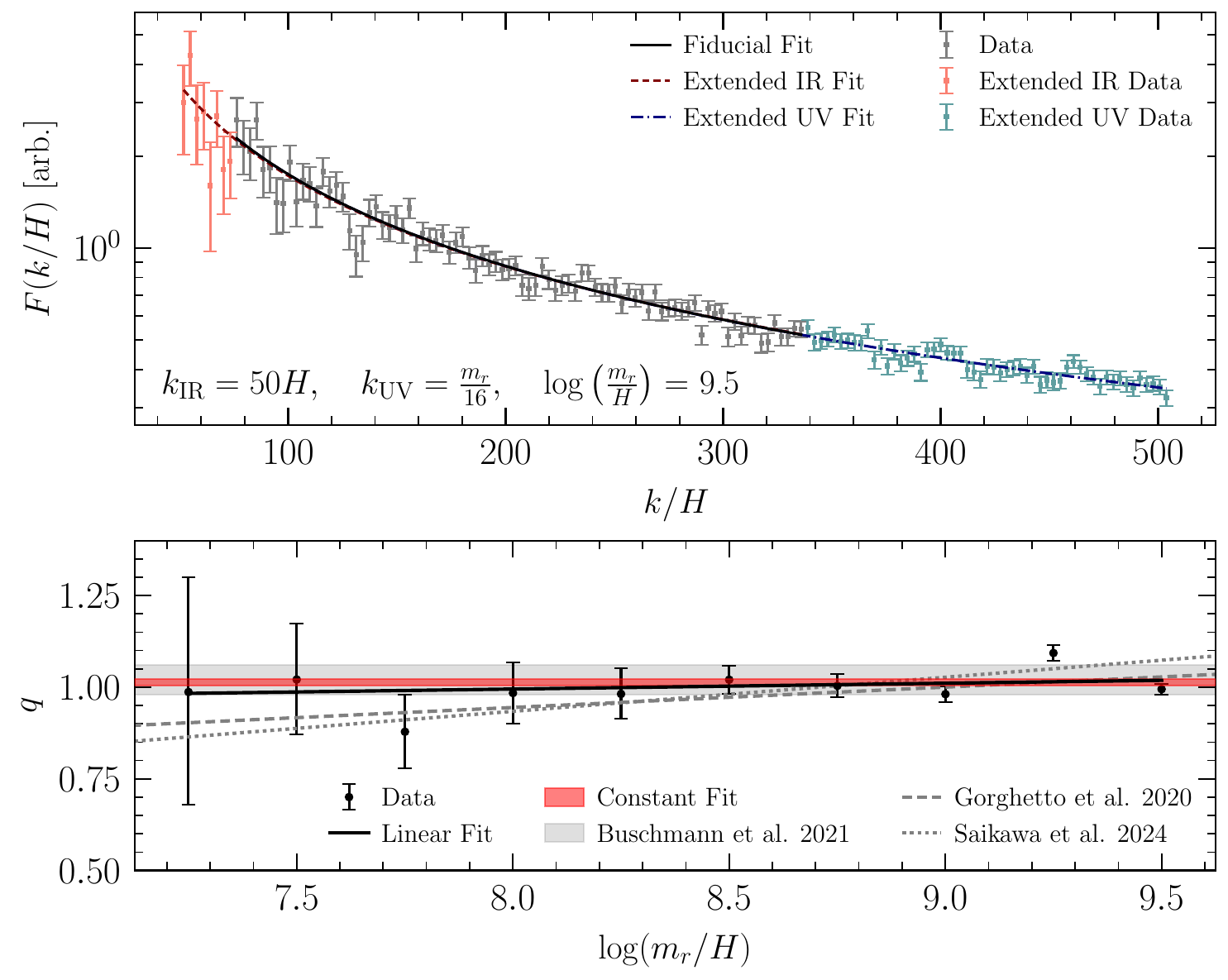}
    \caption{(\textit{Top}) Example of the emission spectrum calculated at $\log(m_r/H) = 9.5$ with associated power-law fit. The best-fit emission spectrum associated with our fiducial analysis choices with $x_\mathrm{IR} = 50$ and $x_\mathrm{UV} = 1/16$ is shown in black with the associated data points and inferred error bars shown in gray. Extending the fitting range by taking $x_\mathrm{IR} = 30$ ($x_\mathrm{UV} = 1/12$) results in the fit and error bars in red (blue). (\textit{Bottom}) The best-fit emission spectrum index using our fiducial fitting region over an interval between  $\log(m_r/H) = 7.25$ and $\log(m_r/H) = 9.5$. The red band illustrates the $1\sigma$ containment interval for the constant $q$ model from this work, and may be compared with the analogous containment interval from \cite{Buschmann:2021sdq} in gray. We also show the best fit to the linear growth in the emission spectrum with $\log(m_r/H)$ in black, along with the best-fit linear growth models of \cite{Gorghetto:2020qws} and \cite{Saikawa:2024bta}.}
    \label{fig:FiducialSpectrum}
\end{figure}

In Fig.~\ref{fig:FiducialSpectrum} (top panel) we illustrate the instantaneous spectrum at $\log (m_r / H) = 9.5$, with the time value referring to the earlier time used in the finite difference. We infer the statistical uncertainties on the spectral data points, which are illustrated in Fig.~\ref{fig:FiducialSpectrum}, through a data-driven procedure described in App.~\ref{App:EmissionSpectrum} and~\cite{Buschmann:2021sdq}. 

For $1 \ll k/H \ll m_r / H$ we expect $F \propto (k/H)^{-q}$ for some index $q$, since there is no other mass scale to influence the spectrum.  We fit this model to the data for $x_{\rm IR} H < k< x_{\rm UV} m_r$.  In our fiducial analysis, we chose $x_{\rm IR} = 50$ and $x_{\rm UV} = 1/16$, as in Ref.~\cite{Buschmann:2021sdq}, to ensure that our fit is performed sufficiently far from the UV and IR cut-offs.  The data outside of this interval are indicated by `Extended IR data' and `Extended UV data' in Fig.~\ref{fig:FiducialSpectrum}. In the bottom panel of Fig.~\ref{fig:FiducialSpectrum} we show our best-fit $q$ values at each of the time steps. 

We fit two models to the data for $q$ as a function of $\log (m_r / H)$: a linear model, where $q(t) = q_1 \log (m_r / H) + q_0 $, and a constant model where $q(t) = q_0$. In the linear fit, we find $q_1 = 0.015 \pm 0.023$, meaning that we find no evidence for a log-varying spectral index. The constant fit returns an index consistent with a conformal spectrum to 1\% precision: $q_0 = 1.01 \pm 0.01$.

It is useful to compare our results to those of previous works. Ref.~\cite{Buschmann:2021sdq} also found no evidence for a log-varying spectral index, with the constant fit returning $q_0 = 1.02 \pm 0.04$. Our results are consistent but with statistical uncertainties that are four times smaller. Our simulation contains $\sim$4.6 times more modes than that in~\cite{Buschmann:2021sdq} at fixed $\eta$, which partially accounts for our decreased statistical uncertainties. More importantly, however, is the fact that we can evolve to larger $\log$ values since the fit is dominated by the largest-$\log$ data points, which have the smallest statistical uncertainties since they have the largest number of $k$ modes. 

In contrast to our results, many works have found evidence for a logarithmically increasing spectral index $q$.  For example, using a suite of static lattice simulations Ref.~\cite{Saikawa:2024bta} finds $q_1 =  0.093 \pm 0.005$ when fitting a model that grows linear in the log; this disagrees with our result by more than 3 standard deviations. Ref.~\cite{Gorghetto:2020qws} also used a suite of static lattice simulations to find logarithmic growth in $q$, with $q_1 = 0.053 \pm 0.005$. Both Refs.~\cite{Gorghetto:2020qws} and~\cite{Saikawa:2024bta} use a similar methodology, and we believe that there are several possibilities for how their results could be biased to mimic logarithmic growth in $q$ (see also the discussion in~\cite{Buschmann:2021sdq}). While~\cite{Gorghetto:2020qws,Saikawa:2024bta} use an alternate procedure for constructing the initial conditions, we show in the SM that this is unlikely to be the source of the discrepancy. On the other hand, Refs.~\cite{Gorghetto:2020qws,Saikawa:2024bta} are more susceptible to finite-lattice-size effects, since the resolution to the string cores is quite low (at the level of around one lattice site per core width) towards the end of the simulations, given the finite lattice; Ref.~\cite{Buschmann:2021sdq} showed that finite lattice effects can lead to artificial logarithmic growth.  

Perhaps most importantly, however, we show in the SM that we can recreate the logarithmic growth seen in ~\cite{Gorghetto:2020qws,Saikawa:2024bta} by choosing less conservative values for the UV cut-off $x_{\rm UV}$. For example, taking $x_{\rm UV} = 1/4$ -- which is the fiducial value adopted in~\cite{Saikawa:2024bta} -- leads us to find strong evidence for logarithmic growth, with $q_1 \sim 0.18$. Our larger dynamical range allows us to implement stricter UV cut-offs than in~\cite{Gorghetto:2020qws,Saikawa:2024bta}. As we show in the SM, for our level of statistical uncertainties we find that for $x_{\rm UV} \gtrsim m_r / 12$ the end-point of the spectrum biases our fit to give artificial preference for logarithmic growth. In contrast, our results are consistent with no logarithmic growth for smaller $x_{\rm UV}$ (we test down to $x_{\rm UV} = m_r / 32$).   

{\bf Axion Mass Prediction.}---Previous works such as~\cite{Gorghetto:2020qws,Buschmann:2021sdq,Saikawa:2024bta} have calculated the axion relic DM abundance by accounting for axions produced prior to the QCD phase transition at some reference time defined by, for example, the condition $H(t) = m_a(t)$ in~\cite{Gorghetto:2020qws,Saikawa:2024bta}, or $3 H(t_*) = m_a(t_*)$ in~\cite{Buschmann:2021sdq}. Here, we adopt the latter definition of $t_*$, though we stress that both choices are somewhat arbitrary since the string network still exists for $t$ larger than these characteristic times. While it is true that the network begins to collapse for $t > t_*$ and is fully gone within a few Hubble times, as we argue below the axions produced during network collapse can still substantially affect the DM abundance.

If the axion field is in the linear regime, then we can simply estimate the DM abundance by computing the number density in axions at $t = t_*$ and redshifting this number density down to today, leading to the prediction~\cite{Buschmann:2021sdq} 
\es{eq:Omega_a_str}{
\Omega_a^{\rm str} \approx 0.12 \, h^{-2} \left(\frac{f_a}{1.4 \cdot 10^{11} \, \, {\rm GeV}} \right)^{1.17} \frac{318}{\delta} \sqrt{\frac{\xi_*}{13}} \frac{\rm Log_*}{70} \,.
}
Here we assume $m_a^2(T) \propto T^{-8.16}$~\cite{Borsanyi:2016ksw} along with the number of entropy degrees of freedom $g_*(T) \approx g_*^0 (T / {\rm MeV})^\gamma$ for $800 < T < 1800$ MeV and $g_*^0 \approx 50.8$, $\gamma \approx 0.053$~\cite{Lombardo:2020bvn}. The quantity $\delta$ is defined by~\cite{Buschmann:2021sdq} (for $q > 1$) $\langle H / k \rangle^{-1} \equiv \delta \sqrt{\xi}$, where the expectation value is taken with respect to the instantaneous spectrum $F$. The reason this quantity is expected to depend on $\sqrt{\xi}$ is because as $\xi$ increases logarithmically with time the strings necessarily become smaller, such that the IR cut-off of the spectrum moves towards the UV like $\sqrt{\xi}$.  As we show in the SM, this ansatz for the functional form for $\langle H / k \rangle^{-1}$ describes our spectral data well. We follow~\cite{Buschmann:2021sdq} and compute $\langle H / k \rangle$, at a given time step, by numerically integrating over $k$ with the simulation output for $F$ for $0 \leq k \leq x_{\rm IR} H$. For larger values of $k$ (assuming $q \geq 1$), we integrate to the physical UV cut-off assuming a fixed choice of $q$. 

Our maximum allowed choice of $q$ at 1$\sigma$ uncertainty is $q = 1.02$, while our minimal value is $q = 1.00$; lower values of $q$ lead to less DM. For an exactly conformal spectrum, one can show that we should instead of having a static $\delta$ value write $\delta = \delta_1 {\rm Log}_*$~\cite{Buschmann:2021sdq}. As we discuss in the SM, for $q = 1.02$ we measure $\delta \approx 318$, while for $q = 1.00$, we measure $\delta_1 \approx 8.6$. These values for $\delta$ are broadly consistent with the results found in~\cite{Buschmann:2021sdq}.  Varying the index $q$ between $1.00$ and $1.02$ while also adding in the abundance from misalignment ({\it e.g.},~\cite{OHare:2024nmr}) gives us the predicted QCD axion mass range $m_a \in (45,65)$ $\mu$eV to match the observed relic abundance~\cite{Planck:2018vyg}. Crucially, however, this estimate only accounts for axions produced before $t_*$. 

We must check that non-linear, number-changing processes do not violate the assumption of number-density conservation for $t > t_*$. Using the method discussed in~\cite{Gorghetto:2020qws,Buschmann:2021sdq} and explicit axion-only simulations with the predicted spectrum of axions, we verify (see the SM) that number-changing processes should change our DM abundance estimate by less than a few percent.

For $N_{\rm dw} = 1$, domain walls form shortly after $t_*$ and cause the string-domain-wall network to collapse by a time $t_{\rm coll}$. (In the SM we present an argument suggesting $t_{\rm coll} \sim 5 t_*$.) For $t > t_*$ it is important to account for axions produced both by strings and by the domain walls that stretch between them. To help compute the number of axions produced after $t_*$, we perform a simulation of the network collapse during the QCD phase transition with an AMR simulation based on an intermediate state of the PQ simulation performed in~\cite{Buschmann:2021sdq}. (See App.~\ref{app:QCD} for details.)  This simulation begins at $\eta = 0.1$, where the effects of the axion mass should be negligible. A non-zero axion mass is turned on at $\eta \approx 27$, with $t_*$ occurring at $\eta \approx 36$. The simulation ends at $\eta \approx 76$ after the string-wall network has disappeared.  In this simulation, we compute explicitly that the DM abundance increases by a factor $\sim$3 accounting for axions produced after $t_*$ relative to the prediction only accounting for axions produced prior to $t_*$. However, extrapolating from these results to the physical scenario requires us to understand analytically how the string and domain-wall contributions should scale with ${\rm Log}_*$. We estimate (see the SM) that accounting for string production post $t_*$ should increase the DM abundance such that $m_a \in (56, 89)$ $\mu$eV gives the observed DM abundance (where again the interval arises primarily from varying $q$ between $1.00$ and $1.02$), while accounting for domain wall production could increase the DM abundance by an additional factor $\sim$5, leading to a mass prediction up to 280 $\mu$eV.  We hypothesize that domain walls are more important than strings for DM production because they are extended field configurations with thickness $m_a^{-1}$ and thus deposit the majority of their energy into low-$k$, non- to semi-relativistic modes. 

{\bf Discussion.}---Our work suggests that the QCD axion should have a mass between roughly $40$ and $300$ $\mu$eV in order to explain the correct DM abundance if the axion is generated in a field theory UV completion after inflation with $N_{\rm dw} = 1$. Future experiments that could achieve sensitivity to QCD axions in this mass range include MADMAX~\cite{Caldwell:2016dcw,MADMAXinterestGroup:2017koy,MADMAX:2019pub,Garcia:2024xzc} and ALPHA~\cite{Lawson:2019brd,Wooten:2022vpj,ALPHA:2022rxj}. On the other hand, we note that our results are still subject to a large systematic uncertainty due to our only rough accounting of axions produced by domain walls, which we find could dominate the DM abundance.  Additionally, we note that the axion string cosmology could be more complicated than in the picture assumed in this work if, for example, $N_{\rm dw} >1$~\cite{Hiramatsu:2012sc,Beyer:2022ywc,Chang:2023rll}). The axion could also arise not from a field theory UV completion but from an extra dimension or string theory UV completion. As shown in {\it e.g.}~\cite{Benabou:2023npn} (see also~\cite{March-Russell:2021zfq,Reece:2024wrn}), string theory axions do not generically form axion strings. More work is needed to properly understand the dependence of our results on the UV completion of the axion theory.

\section*{Acknowledgements}

{\it
We thank Anson Hook and Tanmay Vachaspati for helpful conversations. We thank Javier Redondo, Kenichi Saikawa, Mark Hindmarsh, and Amelia Drew for their comments on the manuscript. M.B. acknowledges funding from the European Research Council (ERC) under the European Union’s Horizon 2020 research and innovation programme (Grant agreement No. 864035). J.W.F. is supported by Fermi Research Alliance, LLC under Contract DEAC02-07CH11359 with the U.S. Department of Energy. J.B. and B.R.S are supported in part by the DOE Early Career Grant DESC0019225 and in part by the DOE award DESC0025293. This research used resources of the National Energy Research Scientific Computing Center (NERSC), a U.S. Department of Energy Office of Science User Facility located at Lawrence Berkeley National Laboratory, operated under Contract No. DE-AC02-05CH11231 using NERSC award HEP-ERCAP0023978. 
}

\bibliography{main}
\clearpage
\appendix

\section{Model Lagrangian and Equations of Motion}
\label{sec:app_eom}
We model the dynamics of axions and associated topological defects with a complex scalar $\bm{\Phi}$ in a radiation-dominated background subject to the temperature-dependent potential
\begin{equation}
\begin{split}
V(\bm\Phi)  &=\bigg( | \bm\Phi|^2 - \frac{f_a^2}{2}  \bigg)^2 - \frac{\lambda T^2}{3}| \bm\Phi|^2  \\&+  m_a(T)^2 f_a^2 \left[1 - \frac{\sqrt{2}|\bm\Phi|}{f_a} \cos\mathrm{Arg}(\bm\Phi)]\right] \,.
\end{split}
\end{equation}
Here $f_a$ is the axion decay constant and $m_a(T)$ is the temperature-dependent axion mass.

We work in conformal time $\eta = R(t) / R(t_1)$ where $R(t)$ is the scale factor at time $t$ and $t_1$ is the time when the Hubble expansion rate $H(t_1) = f_a$. We also work in comoving spatial coordinates in units of $1/[R(t_1) H(t_1)]$. Then after decomposing $\Phi = f_a(\psi_1 + i \psi_2)/\sqrt{2}$, we obtain the dimensionless equations of motion
\begin{equation}
\begin{split}
        \psi_i'' + \frac{2}{\eta} \psi_i' - \bar\nabla^2 \psi_i
        &+ \psi_i \left[ \eta^2 (\psi_1^2 + \psi_2^2 - 1) +\frac{T_1^2}{3 f_a^2} \right] \\
        &- \frac{\eta^2 m_a(\eta)^2}{f_a^2}\delta_{i1} = 0\,,
\end{split}
\end{equation}
where primes indicate differentiation with respect to $\eta$, the $\bar\nabla$ is the Laplacian with respect to our dimensionless comoving coordinates, and the $\delta_{i1}$ indicates that this term is only applied to the equations of motion of $\psi_1$. We take $\lambda=1$ while the ratio $T_1^2/3 f_a^2$ represents a residual physical scale, which we take to have a value of $T_1^2 / 3 f_a^2 = 0.56233$. Though this value is unphysical, its effect is only to modify the hierarchy between the time that $\Phi$ begins to oscillate and when the PQ symmetry is broken; its contribution to subsequent dynamics is negligible.

\section{Details of the PQ Simulation}
In our PQ simulations, we may take the axion mass to be negligible by sending $m_a \rightarrow 0$. Our PQ simulation volume is a box with sidelength $\bar L = 200$, where an overbar indicates a quantity in units of the conformal horizon at time $t_1$. The simulation is resolved by 8,192$^{3}$ lattice sites at the base resolution and begins at an initial conformal time of $\eta_i = 0.1$, through the PQ symmetry breaking at $\eta \approx 0.75$, and until a final time of $\eta_f \approx 110$ when the simulation contains approximately $6$ Hubble volumes. At this time we reach a maximum scale separation of $\log(m_r / H) \approx 9.75$. Over the course of the simulation, as many as five levels of additional refinement are added. Performing an identically resolved uniform lattice simulation would then require a lattice of size 262,144$^3$ and exabytes of memory.

All simulations were performed at the facilities of the National Energy Research Scientific Computing Center (NERSC). The first part of the primary simulation ran for about 1.5 weeks on the now-retired Cori KNL cluster on up to 2,048 nodes with a total of 139,264 CPU cores, 786,432 CPU threads, and 129 TB of aggregate memory. The second part of the simulation ran for another 1.5 weeks on the new Perlmutter cluster on 256 nodes with a total of 32,768 CPU cores, 65,536 CPU threads, and 131 TB of aggregate memory. The simulation is memory-bound, \textsl{i.e.} the number of required nodes is primarily set by the amount of memory they can provide. 

\section{Details of the QCD Simulation}
\label{app:QCD}
In simulations of the QCD epoch, which we describe at greater length in the SM, we can no longer neglect the axion mass. We parametrize it by
\begin{equation}
    m_a(\eta) = \frac{f_a}{\mathcal{N}} \left(\frac{\eta}{\eta_\star} \right)^{n/2}
    \label{eq:axion_mass_parameterization} \,,
\end{equation}
where $\eta_\star$ is the conformal time when the domain wall network collapses (typically a factor of two in conformal time after axion oscillation) and $\mathcal{N}$ is a ratio parameterizing the hierarchy between the axion decay constant and axion mass at this time. We adopt $n = 6.68$, as previously implemented in \cite{Buschmann:2019icd}, from \cite{Wantz:2009it}, and we choose $\mathcal{N} \approx 19$.\footnote{See \cite{OHare:2021zrq} for the effect of varying $n$ on the axion DM abundance.} We use an intermediate state generated during the primary simulation of \cite{Buschmann:2021sdq} as the initial state of our QCD simulation. In particular, this provides an initial state at $\eta \approx 27$ in a simulation volume with $\bar L = 120$, which we resolve on a 4,096$^3$ base resolution lattice. Even though $m_a(\eta)$ is small at the beginning we multiply ~\eqref{eq:axion_mass_parameterization} by the logistic function $1/\sqrt{1+\textrm{exp}[-3(\eta-30)]}$ to ensure a smooth transition from the PQ epoch to the QCD epoch. Domain walls form in these simulations at $\eta \approx 32$ when the axion begins to oscillate, and the simulation concludes after the network has fully collapsed at $\eta \approx 76$. We implement identical refinement criteria in this simulation as in our primary string-only simulation, leading to as many as 5 refinement levels.

The simulation ran on the NERSC Perlmutter cluster using initially 256 nodes for 18 hours (32,768 CPU cores, 65,536 CPU threads, and 131 TB of aggregate memory). Once the string-domain wall network started collapsing we could reduce the number of required nodes to 128 for another $\sim3$ days of runtime.

\section{Fitting the Emission Spectrum}
\label{App:EmissionSpectrum}
We extract the function $F(k/H)$ numerically from the simulation output following the procedure described in detail in~\cite{Buschmann:2021sdq}.  
We compute the screened time-derivative of the axion field
\begin{equation}
\dot{a}_\textrm{screened}(x)=\dot{a}(x)\left(1+ r(x) / f_a\right)^2,    
\end{equation}
with $r(x)$ being the radial mode. The screening is necessary to remove string cores from $\dot{a}(x)$ as these are singularities in the axion field that otherwise would pollute the result. We perform a Fourier-transform of $\dot{a}_\textrm{screened}(x)$ to obtain $|\dot {\tilde a}_\textrm{screened}(k)|^2$ using the data at the lowest resolution. Those data are binned in $k$ with a resolution of $\Delta k = 2 \pi / \bar{L}$. We then compute $F$ by taking the appropriate finite difference over a time spacing of $\Delta \log(m_r / H) = 0.25$. We include all data with $\log (m_r / H) > 7.25$; this choice requires that we have at least $10$ $k$-points in the fit for $q$. 

We denote the binned instantaneous emission spectrum at time $\mathrm{Log}_i$ and physical wavenumber $k_j/H_i$ by $F_{ij}$. At each $\mathrm{Log}_i$, we assume the emission spectrum is modeled by a power-law mean and standard deviation of the form
\begin{gather}
\mu_{ij}(\mu, q) = \mu \left(\frac{k_j}{H_i} \right)^{q}, \quad  
\sigma_{ij}(\sigma, p) = \sigma \left(\frac{k_j}{H_i} \right)^{p} \,,
\end{gather}
so that the likelihood of the data at $\mathrm{Log}_i$  given the emission spectrum index $q$ is given by 
\begin{equation}
\log\mathcal{L}_i(q) = \max_{\{\mu, \sigma, p\}} \sum_j \Phi[F_{ij}, \mu_{ij}(\mu, q), \sigma_{ij}(\sigma, p)]
\end{equation}
after profiling over the nuisance parameters $\{\mu, \sigma, p\}$ where $\Phi(x, \mu,  \sigma)$ is the log of the gaussian probability density function with mean $\mu$ and standard deviation $\sigma$. For an individual spectrum, we can find the best-fit $\hat q$ at $\mathrm{Log}_i$ by $\hat{q} = \mathrm{argmax}_{q} \mathcal{L}_i(q)$.

To test the constant $q$ hypothesis, we then compute the likelihood of the data as a function of $q_0^\mathrm{const.}$ by 
\begin{equation}
\log\mathcal{L}(q_0^\mathrm{const.}) = \sum_i \log\mathcal{L}_i(q_0^\mathrm{const.}) \,.
\end{equation}
When we consider the case of a logarithmically growing emission spectrum index, we take
\begin{equation}
    q = q_1 \mathrm{Log} + q_0 
\end{equation}
and profile over the nuisance parameter $q_0$ so that we have 
\begin{equation}
\log\mathcal{L}(q_1) = \max_{q_0}\sum_i \log\mathcal{L}_i(q_1 \mathrm{Log}_i + q_0) \,.
\end{equation}
Following standard frequentist treatments~\cite{Safdi:2022xkm}, we then estimate the best fit constant index $\hat{q}_0^\mathrm{const.}$ by maximizing $\log\mathcal{L}(q_0^\mathrm{const.})$ with $1\sigma$ confidence intervals on our estimate of $\hat{q}_0^\mathrm{const.}$ given by $q_0^\mathrm{const.}$ such that $\log\mathcal{L}(\hat q_0^\mathrm{const.})- \log\mathcal{L}(q_0^\mathrm{const.}) = 1/2$. Estimating a best-fit and containment interval for $q_1$ and/or $q_0$ follows an identical procedure.

\section{Axion mass prediction}
\label{sec:axion_mass_prediction}
To extract the axion mass, we follow the procedure in Ref. \cite{Buschmann:2021sdq}. The number density of axions emitted by the string network at the time of the QCD phase transition, specified by $\log(m_r/H)=\textrm{Log}_*$, is, for $q>1$,  $n_a^{\text {string }} \approx\left(8 \pi f_a^2 H / \delta\right) \sqrt{\xi_*} \textrm{Log}_*$, where $\delta$ parametrizes $\langle H/k\rangle^{-1}=\delta \sqrt{\xi}$. Recall that, physically, $\delta$ encodes the dependence of the effective IR cutoff on the typical inter-string separation: $k_\mathrm{IR}/H \propto \sqrt{\xi}$. Redshifting $n_a^\mathrm{string}$ down to the present day, the abundance of QCD axion DM receives a contribution from string emission $\Omega_a^{\operatorname{str}}$ given by \eqref{eq:Omega_a_str} \cite{Buschmann:2021sdq}.
To obtain the total relic abundance, we multiply $\Omega_a^{\operatorname{str}}$  by a fudge factor $\mathcal{F}$ to account for the axions produced during the QCD phase transition at times $t>t_*$, with $t_*$ defined such that $3H(t_*)=m_a(t_*)$. In Ref. \cite{Buschmann:2021sdq} it was assumed that no axions are produced after the time $t_*$ such that $\mathcal{F}=1$. Following the arguments in the SM, we take $\mathcal{F}=1.7$  to account for axion production after $t_*$ from strings and $\mathcal{F}=8.2$ to account for production after $t_*$ from both strings and domain walls. Note that when matching the axion relic abundance to the observed DM abundance we also account for the contribution from misalignment, which we take as \cite{Borsanyi:2016ksw}
\begin{equation}
\Omega_a=0.12h^{-2}\left(\frac{28 \, \mu  \mathrm{eV}}{m_a}\right)^{1.17} \,.
\end{equation}

The inverse expectation value $\langle H/k\rangle^{-1}$ is shown for our fiducial spectrum fit in SM Fig. \ref{fig:delta_fid_fiducial}. Recall that for our fiducial emission spectrum, we measure the spectral index to be $q=1.013 \pm 0.0095$. Assuming the spectral index is at its $1\sigma$ upper limit $q=1.02$, we obtain $\delta=318.2 \pm 2.1$. Assuming a conformal spectrum $q=1$, we find $\delta_1=8.6 \pm 0.1$. In this case we use \eqref{eq:Omega_a_str} to compute the relic abundance from string emission with the replacement $\delta \to \delta_1\rm Log_*$. Note that Ref. \cite{Buschmann:2021sdq} found $\delta = 113\pm 7$, for the fiducial spectrum fit to their AMR simulation, for which $x_\mathrm{IR}=50$ and $q=1.06$. Supposing a conformal spectrum, they found $\delta_1 = 6.2 \pm 0.4$.

We compute $\xi_*$ assuming that the functional form  $\xi= c_{-2}/\textrm{Log}^2 + c_{-1}/\textrm{Log} + c_0 + c_1\textrm{Log}$, which is fit to the data for $\textrm{Log} \ge 4$ (see main text), remains valid until the QCD phase transition. We allow the fit coefficients $c_i$ to vary within their $1\sigma$ ranges. Varying $(\textrm{Log}_*,\xi_*, q)$ with $\textrm{Log}_* \in (60,70)$, and $q$ between $1$ and its upper limit at $1\sigma$, we obtain the range of axion masses $(150, 280) \,\mu \mathrm{eV}$.
Note that not including any emission from times $t>t_*$ would result in the mass range $(45, 65) \,\mu \mathrm{eV}$, consistent with Ref. \cite{Buschmann:2021sdq}, while ignoring the emission from domain walls corresponds to the range $(56, 89 ) \,\mu \mathrm{eV}$.

\clearpage
\unappendix
\clearpage
\onecolumngrid

\begin{center}
  \textbf{\large Supplementary Material for Axion Mass Prediction from adaptive mesh refinement cosmological lattice simulations}\\[.2cm]
  \vspace{0.05in}
  {Joshua N. Benabou, Malte Buschmann, Joshua W. Foster, and Benjamin R. Safdi}
\end{center}

\twocolumngrid

\setcounter{equation}{0}
\setcounter{figure}{0}
\setcounter{table}{0}
\setcounter{section}{0}
\setcounter{page}{1}
\setcounter{secnumdepth}{2}
\makeatletter
\renewcommand{\theequation}{S\arabic{equation}}
\renewcommand{\thefigure}{S\arabic{figure}}
\renewcommand{\thetable}{S\arabic{table}}
\onecolumngrid

This Supplementary Material provides additional details and results for the analyses discussed in the main Letter.

\section{Supplementary Figures}

\begin{figure*}[!h]
    \centering
    \includegraphics[width=0.48\textwidth]{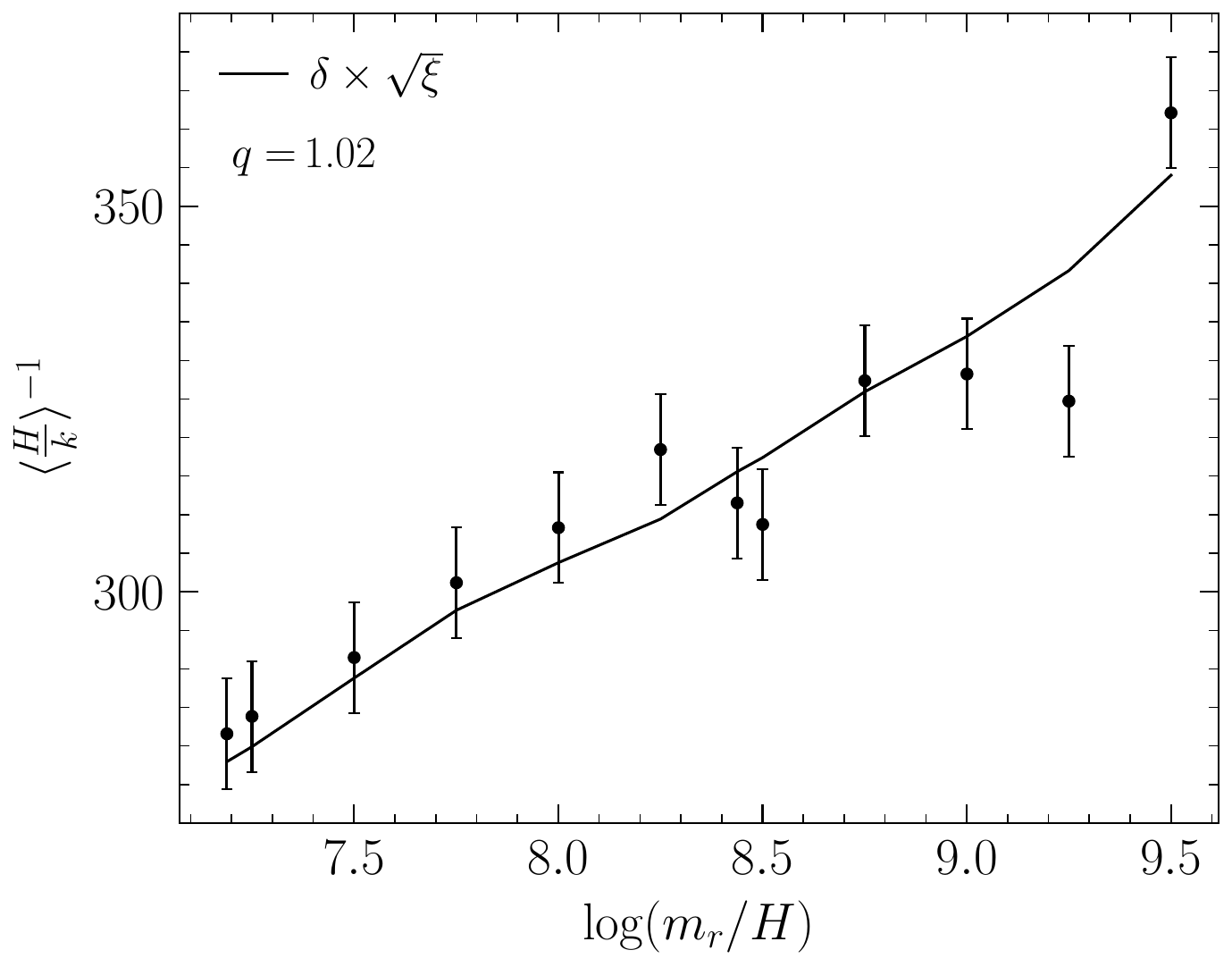}     \includegraphics[width=0.48\textwidth]{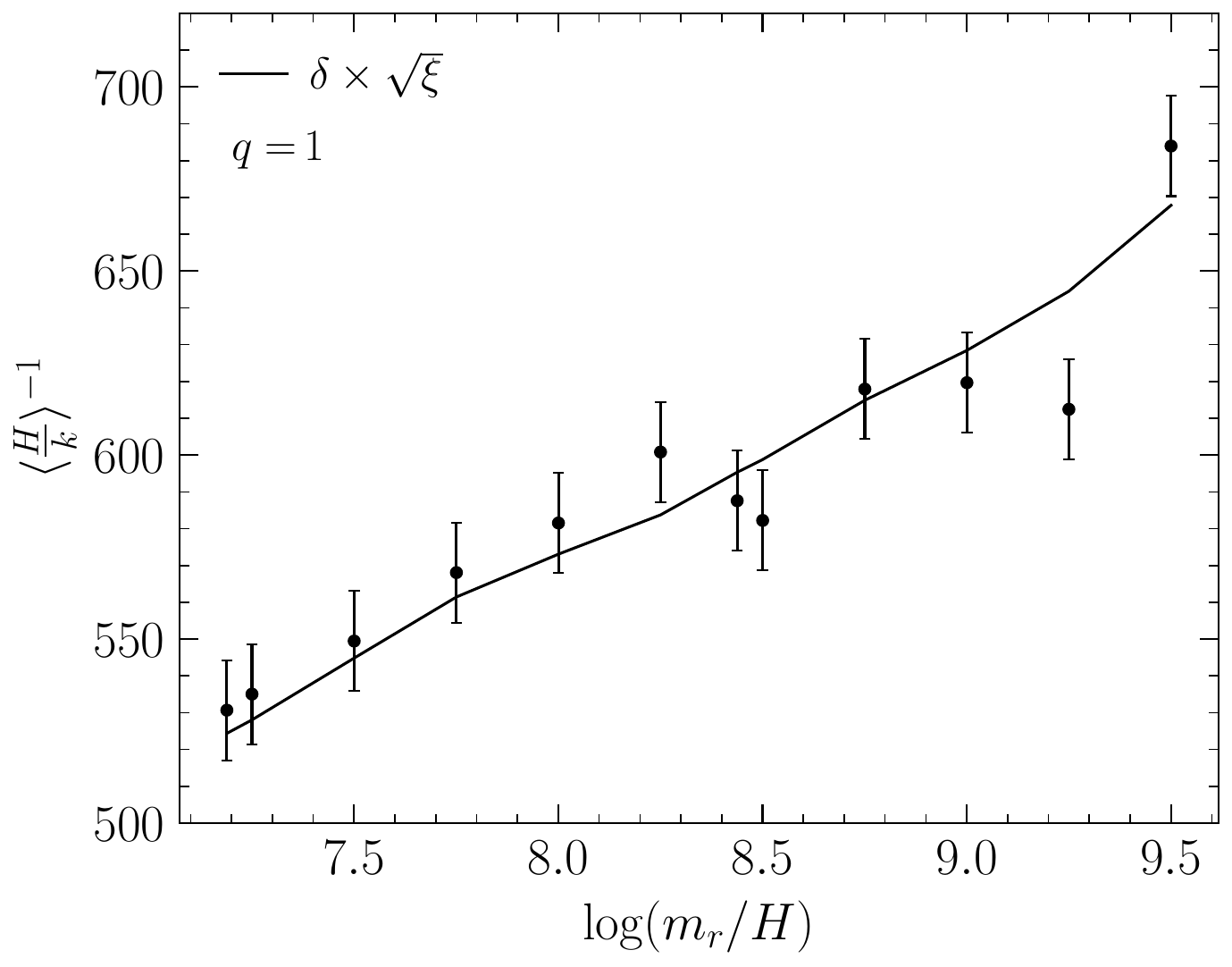}
    \caption{(\textit{Left})
    The inverse expectation value $\langle H/k\rangle^{-1}$ (data points) of the axion emission spectrum $F$ for our fiducial spectrum with $k_\mathrm{IR}/H = 50$ (plotted in Fig. \ref{fig:FiducialSpectrum}). Here we assume the spectral index is at its upper limit at $1\sigma$, $q=1.02$. The error bars are data-driven, obtained by fitting $\langle H/k\rangle^{-1}=\delta \sqrt{\xi}$, for a constant $\delta$, across the indicated log data points (best fit shown in solid). We assume a Gaussian likelihood and treat the standard deviation $\sigma$ as a nuisance parameter. We find $\delta=318.2 \pm 2.1$. (\textit{Right}) As in the left panel but assuming a conformal spectrum $q=1$. In this case, as discussed in the main text, we expect $\delta = \delta_1 {\rm Log}$; we measure $\delta_1 = 8.6 \pm 0.1$.  In both panels, the UV cutoff is set by  $\log(m_r/H_\mathrm{QCD})=70$.
    }
    \label{fig:delta_fid_fiducial}
\end{figure*}

\section{Simulation Setup and Refinement Criteria}
Our simulations are performed with the public code \texttt{sledgehamr}~\cite{Buschmann:2024bfj}, which is based on the block-structured adaptive mesh refinement framework  \texttt{AMReX}~\cite{Zhang2019, zhang2020amrex}. Using  \texttt{sledgehamr}, we model the dynamics of the PQ field, cosmic strings, and domain walls by evolving a complex field in a comoving volume with periodic boundary conditions in a radiation-dominated cosmological background using the method of lines. Our simulations, which we detail further here, are similar to but improve upon in several ways those implemented in \cite{Buschmann:2021sdq}. 

In making our spatial discretization, we evaluate the Laplacian of fields using the fourth-order accurate 13-point finite-difference stencil. Our equations of motion are integrated in time using a third-order strong-stability preserving forward Runge-Kutta. The time-step used in our simulations is set by a Courant-Friedrichs-Lewy (CFL) condition of $\Delta \eta \approx \Delta x / 3$ to sufficiently resolve the characteristic wave speed, which is $1$ in our units. Note that because the grid spacing $\Delta x$ depends on the refinement level, our time-step size varies from level to level as well. To coordinate the different time-step sizes we use a sub-cycling-in-time algorithm~\cite{Buschmann:2024bfj}.

In our simulations, the adaptive meshing technique is determined via user-defined criteria for tagging lattice sites. Lattice sites that are tagged, as well as those in their vicinity, are upsampled by a nested sub-lattice with twice the spatial resolution as the tagged cells. Tagging and refinement proceed iteratively so that lattice sites tagged at the base level are upsampled to the first refinement level; then, if lattice sites on the first refinement level are tagged, they are in turn upsampled to a second refinement level, and so on. As this procedure is computationally expensive it is done only periodically every ten time-steps. This explains why not only tagged cells but also those in their vicinity are upsampled, as we need to ensure that no string will be able to leave the refined region before it is readjusted ten time-steps later. Note that this also implies that higher refinement levels are necessarily readjusted more often. For details, see, \textit{e.g.}, \cite{Buschmann:2024bfj, Zhang2019, zhang2020amrex}.

We implement two independent refinement criteria, with lattice sites tagged for up-sampling if either criterion is satisfied. The first and simplest criterion is based on the identification of strings, allowing us to manually enforce high resolution of regions that contain these intrinsically small-scale features. The second and more general criterion is based on a \textit{self-shadow} method which automatically identifies areas of the simulation volume that must be maintained at high spatial resolution based on an estimate of the local absolute truncation error.

\subsection{String Tagging}
We localize strings in our simulation volume following an algorithm detailed in \cite{Fleury:2015aca} to identify all plaquettes (square loops with vertices at the lattice sites of our simulation) that are pierced by a string. The lattice sites that form the vertices of pierced plaquettes at level $\ell$ are then tagged if strings would be resolved by less than $4$ lattice sites as measured by if $m_r \Delta x < 1/4$ at the refinement level of interest. If strings are resolved by more than four lattice sites at level $\ell$, then those sites are not tagged for continued adaptive meshing as the string resolution is considered sufficient. At a given refinement level $\ell$ with spatial resolution $\Delta x_\ell$, we may estimate the number of lattice sites that resolve a string core by $1/(\eta m_r \Delta x_\ell)$, where the radial mode has mass $m_r = \sqrt{2}$ in our simulation units. Note that a factor of $\eta$ appears in estimating the string resolution as the fixed physical size of our strings appears to shrink on our lattice defined in comoving coordinates. As a result, the number of levels of refinement required to maintain sufficient resolution of strings will grow over the course of simulations. Note also that because the strings shrink continuously while our resolution improves only in factors of $2$, our string resolution will vary between four and eight lattice sites. Additionally, the number and location of string-plaquette piercings at the finest tagged level are tracked to determine the total string length $\xi$.

\subsection{Self-Shadow Tagging}
\label{sec:TEE}
Our second, and more general, refinement criterion is based on the \textit{self-shadow} technique, and its implementation is described in detail in \cite{Buschmann:2024bfj}. In brief, the self-shadow technique assesses local absolute truncation error by comparing the simulation state at identical locations at refinement levels $\ell$ and $\ell + 1$ after both levels have been independently evolved for $\Delta\eta_\ell$ as part of the sub-cycling-in-time algorithm. If the difference from level $\ell$ and $\ell+1$ exceeds some sufficiently small threshold $\epsilon_\ell$, then this indicates that although data at level $\ell+1$ remains trustworthy, that truncation error may become appreciable, and so a new level $\ell+2$ is generated to keep the simulation over-resolved at the finest level. We are free to choose truncation error thresholds which depend on the fields being compared and on the levels at which they are compared. We choose
\begin{equation}
    \epsilon_\ell = 3\times 10^{-4}, \qquad \epsilon_\ell' = 2^\ell \times 3 \times 10^{-4},
\end{equation}
using $\epsilon_\ell$ to compare $\psi_1$ and $\psi_2$ across levels $\ell-1$ and $\ell$, and similarly, using $\epsilon_\ell'$ to compare $\psi_1'$ and $\psi_2'$. While phenomenologically motivated, we found these provided effective data-driven tagging criteria sufficient for identifying domain walls, oscillons, and other localized small-scale features that arise over the course of our simulations without feature-specific tagging algorithms like those implemented for strings. Additionally, these criteria were previously found satisfactory in \cite{Benabou:2023ghl}. 

\section{Reconstructing the String and Domain Wall Network}
In this section, we review the algorithmic procedures by which we identify and track strings and domain walls in our simulations. Beyond just measuring the total string length $\xi$, we aim to more fully reconstruct the string and domain wall network in the interest of extracting additional properties of the network evolution, such as the local string curvatures. These properties help to confirm that the network behaves as expected and that we have indeed approached the scaling solution by the time we extract the spectral index $q$.

\subsection{Identifying Individual Strings}
We start by identifying string-plaquette piercings at the finest refinement level of a single simulation snapshot using the algorithm described in~\cite{Fleury:2015aca}. We then estimate the exact location of the string piercing within that plaquette using the interpolation method outlined in~\cite{Klaer:2019fxc}. After identifying the precise location at which each plaquette is pierced, we link nearby piercings together using a nearest-neighbor algorithm to identify continuous string loops.

String loop identification is computationally expensive finding nearest-neighbor pierced plaquettes involves evaluating their pairwise distances, and a single simulation snapshot may contain over $10^7$ string-pierced plaquettes. To alleviate this issue we divide our simulation volume into $16^3=4096$ equally-size subvolumes to cut down the number of potential locations that need to be considered at any given point. In each subvolume, we start by picking a random location that has not yet been assigned to a string chain and use it to start a new chain. We compute the distance from this location to all other locations within that subvolume. We find the closest unassigned location and add it to the chain as long as its distance is less than $\sqrt{3}\Delta x$. If no such location can be found we stop. This happens whenever the current string segment leaves the subvolume or when the chain closes on itself. Afterward, if there are still unassigned string locations, we start a new chain by repeating the above steps. We follow this procedure in all subvolumes in parallel. 

After identifying partial string chains, we link them by making pairwise connections of chain ends that are within a distance of $\sqrt{3}\Delta x$. Here, we must carefully take into account the periodicity of the simulation volume, and we manually confirm that after the linking procedure, all string chains form closed loops. Generally, the nearest-neighbor approach works well but can produce small errors in regions where strings are intersecting or where the string curvature is very large, \textit{i.e.}, comparable to the inverse lattice spacing at the refinement level where string tagging is performed, as nearest-neighbor identification can lead to isolated plaquette-piercings which are not assigned to any chain. However, this occurs in fewer than $10^{-5}$ of the plaquettes so we neglect them in the following.

\subsection{Extracting the String Curvature}
After identifying continuous string loops, we can parametrize each string as a curve by $\gamma(s)=\left(x(s), y(s), z(s)\right)$, where $s$ is the arc length along the string. From $\gamma(s)$, we compute the local string curvature $\kappa$ as
\begin{equation}
    \kappa = \frac{\left|\gamma'\times\gamma''\right|}{\left|\gamma'\right|^3}=\frac{\sqrt{(x''y'-y''x')^2+(x''z'-z''x')^2+(y''z'-z''y')^2}}{(x'^2 + y'^2 + z'^2)^{3/2}},
\end{equation}
where primes indicate a derivative with respect to $s$. We compute these derivatives numerically using a finite difference stencil. Since all string locations are unevenly spaced in $s$, we need to use a non-uniform stencil. For the first derivative, we use the fourth-order stencil, which we derive to be
\begin{equation}
    x'_i = \frac{A_0 + A_1 + A_2}{h_{-1}h_{+1}h_{-2}h_{+2}(h_{-1}+h_{+1})(h_{-2}+h_{+2})(h_{-1}h_{+1}-h_{-2}h_{+2})},
\end{equation}
with
\begin{equation}
    \begin{split}
        A_0 &= (h_{-1}+h_{+1})(h_{-2}+h_{+2})(h_{-1}^2h_{+1}^2(h_{+2}-h_{-2})+h_{-2}^2h_{+2}^2(h_{-1}-h_{+1}))x_i,\\
        A_1 &= h_{-2}^2 h_{+2}^2(h_{+2}+h_{-2})(h_{-1}^2x_{i+1} - h_{+1}^2x_{i-1}),\\
        A_2 &= h_{-1}^2 h_{+1}^2(h_{+1}+h_{-1})(h_{-2}^2x_{i+2} - h_{+2}^2x_{i-2}),
    \end{split}
\end{equation}
where $h_{\pm k}=|s_i - s_{i\pm k}|$. Not shown here are the trivial modifications to account for the periodic boundary conditions. For the second derivative, we use the second-order stencil which we derive to be
\begin{equation}
    x''_i = 2\frac{h_{-1}x_{i+1} + h_{+1} x_{i-1} - (h_{+1} + h_{-1}) x_i}{h_{-1}h_{+1}(h_{-1}+h_{+1})}.
\end{equation}
We also tested the third-order stencil\footnote{The non-uniform stencil involving $x_{i\pm2}$ is only of third-order in contrast to the corresponding uniform stencil which is of fourth-order. This is due to the lack of symmetry in $s$ which allows higher order terms to cancel in the uniform case.} for the second derivative but found it to be negatively affected by numerical noise at the grid level. We furthermore increase our stencil window to $s_{i\pm k}\rightarrow s_{i\pm 10k}$ as typically the curvature radius is much larger than the grid spacing, $R_c=1/\kappa \gg \Delta x$. This increases the numerical stability of the calculation. We go on to use these string curvatures as part of our algorithm which tracks the motion of identified string loops in time. 

\subsection{Tracking Strings Over Time}
We track individual strings over time to confirm that loops collapse as expected in our QCD epoch simulation. To do so, we need to identify which string, parameterized as $\gamma(s, \eta)$ at time $\eta$, corresponds to which other string $\gamma(s, \eta+\Delta \eta)$ at time $\eta+\Delta \eta$. This is a non-trivial task due to frequent string reconnections and break-ups. Moreover, it is ambiguous to identify a string with a pre-existing one or as a new one after a major merger or break-up event.

Here, we choose to identify a string at time $\eta + \Delta \eta$ with a pre-existing string at $\eta$ only if the difference in the string lengths between those times could have been realized by isolated string motion rather, \textit{i.e.} characterized by a loop radius that grows at the speed of light. If the string length changes faster than this causal bound, it must be due to a major merger or break-up, and we identify the string at time $\eta + \Delta \eta$ as a new string. Specifically, while a circular loop can change its radius $r$ by up to $r\rightarrow r \pm c\Delta\eta$ over a time interval of $\Delta\eta$, where $c$ is the speed of light, our strings are not circular. However, using the local curvature radius $R_c$ (see previous section) we can average over the relative radius change $\langle (R_c \pm c\Delta\eta)/R_c \rangle$ of each string segment. We use this average fractional change as an upper and lower bound on the allowed string length change. We find this criterion to work well in practice.

For a string at time $\eta$, we identify all string candidates at time $\eta+\Delta\eta$ within the allowed length range. To determine which candidate is the correct match, we pick up to 50 string segments at random and compute their minimum distance to each of the string candidates while ensuring proper treatment of periodic boundary conditions. If all minimum distances are closer than $c\Delta\eta$ it is considered a match. We only take a sample of 50 segments since it is sufficient to avoid false positives while not being too computationally demanding. If no match can be found we conclude the string must have been broken up significantly or merged with another string. Any string at time $\eta+\Delta\eta$ that has not been matched with another string is considered \textsl{new}. In our analysis, we only included strings with at least 20 string-plaquette piercings. Fig.~\ref{fig:individual_collapsing_string_loops} shows the resulting evolution of all such strings in our QCD epoch simulation, where we extract string locations every $\Delta\eta=0.08$.

\subsection{Measuring the Domain Wall Area}
\label{sec:DWs}
We extract the domain wall area analogously to the string length. This is done by counting links where the axion field value $\theta=\textrm{atan2}(\psi_2, \psi_1)$ wraps from $\pi$ to $-\pi$ from one cell to another. The physical domain wall area is then $\mathcal{A}_\textrm{dw}=2/3 N_\textrm{link}(\eta \Delta x)^2$, where the factor $2/3$ is the same over-counting correction factor that is also present when computing the string length~\cite{Fleury:2015aca}. Since domain walls are not necessarily refined to the finest level we perform this analysis self-consistently at the coarse level. We then define the domain wall area parameter analogously to the string length parameter $\xi$ as
\begin{equation}
    \xi_\textrm{dw} = \frac{\mathcal{A}_\textrm{dw} t}{\mathcal{V}},
\end{equation}
where $\mathcal{V} = (\eta L)^3$ is our physical simulation volume. An illustration of $\xi_\textrm{dw}$ in our QCD epoch simulation is shown in Fig.~\ref{fig:xi_qcd}.

\section{Systematics: Impact of the Emission Spectrum Fitting Region}
\label{SM:FittingVariations}

Our fiducial analysis, the analyses of \cite{Gorghetto:2018myk, Gorghetto:2020qws}, and the analysis of \cite{Saikawa:2024bta} all differ in terms of the range of wavenumbers considered while fitting the instantaneous emission spectrum $F(k/H)$. In this section, we provide comprehensive results varying the lowest and highest wavenumbers in our fitting regions. We set the lowest wavenumber allowed in the fit by $k_\mathrm{IR} = x_\mathrm{IR} H$, where $H$ is the physical Hubble rate. Similarly, we set the highest wavenumber allowed in the fit by $k_\mathrm{UV} = x_\mathrm{UV} m_r$. In our fiducial analysis, we take $x_\mathrm{IR}= 50$ (as in Ref. \cite{Saikawa:2024bta}) and $x_\mathrm{UV} = 1/16$, corresponding to an identical set of choices made in \cite{Buschmann:2021sdq}. By comparison, \cite{Gorghetto:2018myk, Gorghetto:2020qws} consider $x_\mathrm{IR}$ as small as $30$ and $x_\mathrm{UV}$ as large as $1/4$.

\begin{figure}[t!]
    \includegraphics[width=1.0\textwidth]{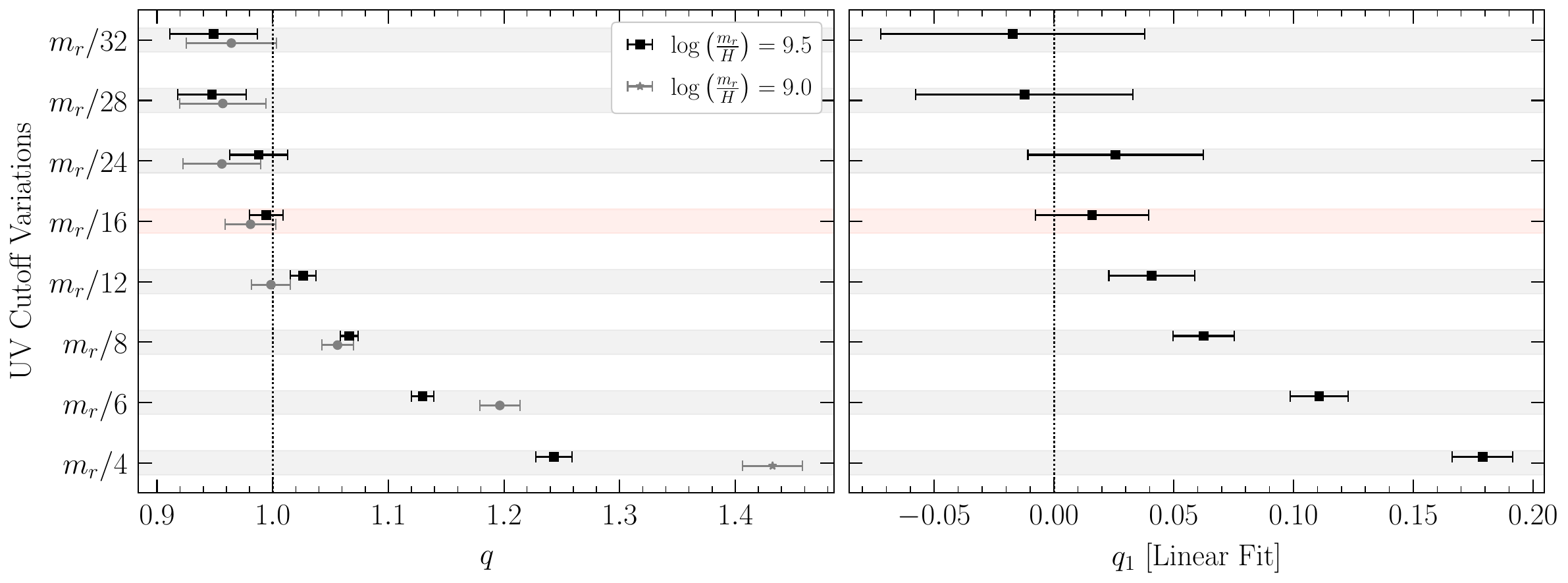}
    \caption{(\textit{Left}) A comparison of the best-fit values for the index $q$ of the instantaneous axion emission at $\log(m_r/H)=9.5$ (black) and $\log(m_r/H) = 9$ (gray), corresponding to two of the latest times in our simulation. Shown are several choices of the UV cutoff, with the choice made in our fiducial analysis highlighted with a light red band. Smaller values of the UV cutoff are more conservative. (\textit{Right}) Slope $q_1$ of the linear fit to the evolution of $q$. We observe no evidence for a linear increase of $q$ with $\log(m_r/H)$ for a UV cutoff at or below $m_r/16$.}
    \label{fig:UVVariations}
\end{figure}

\begin{figure}[t!]
    \includegraphics[width=0.8\textwidth]{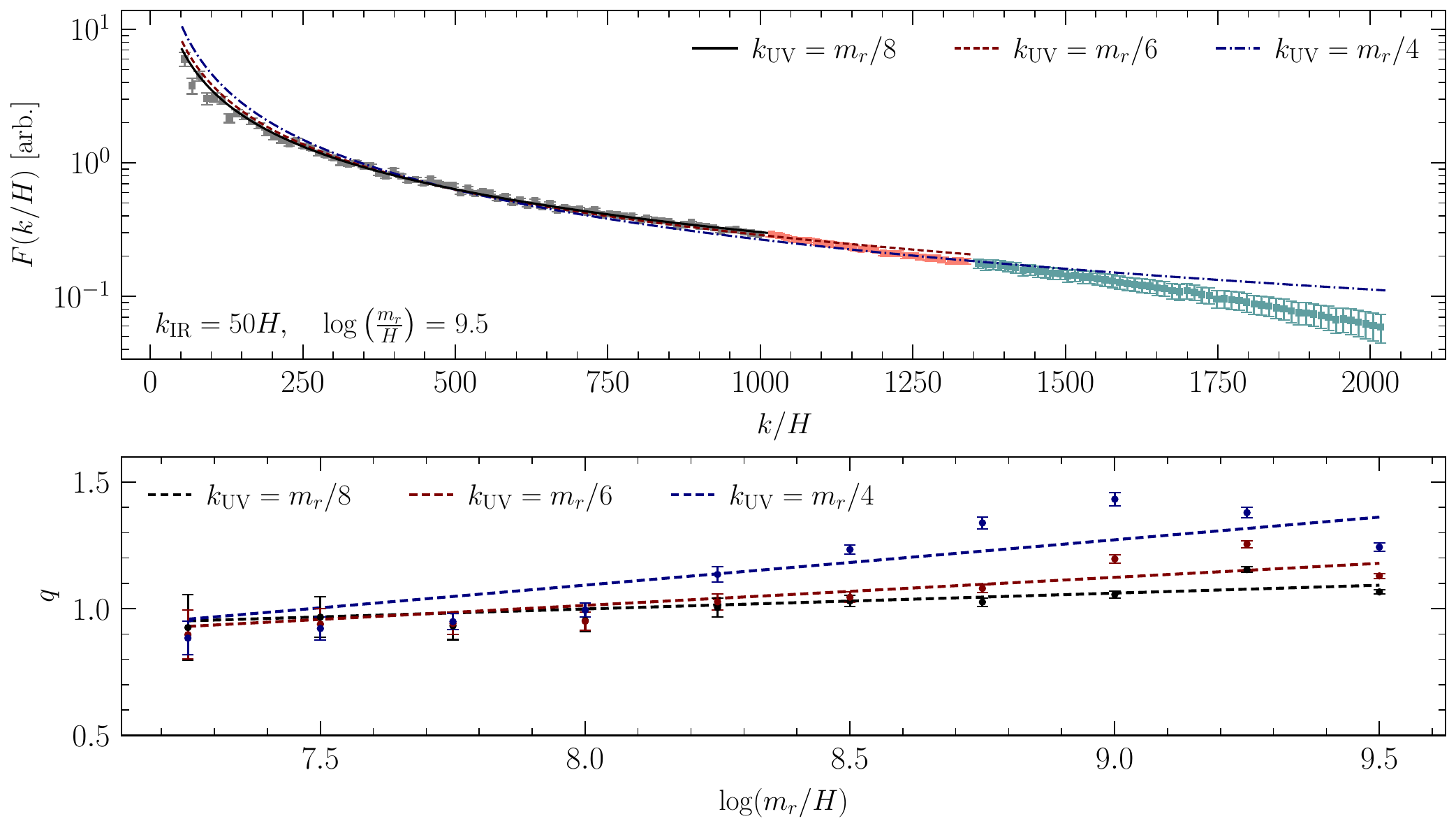}
    \caption{As in Fig.~\ref{fig:FiducialSpectrum}, but examining the instantaneous emission spectrum and time evolution of the emission spectrum index for large values of our UV cutoff. All other analysis choices are held fixed to those of our fiducial analysis. In the top panel, the data are downsampled by a factor of four for the purposes of visualization.}
    \label{fig:LargeUVCutoff}
\end{figure}

In Fig.~\ref{fig:UVVariations}, we inspect the impact of our choice of UV cutoff on our modeling of the emission spectrum index when all other details of the fiducial analysis are held fixed. Larger values of the UV cutoff are more aggressive as they extend the simple power-law modeling across a broader range, and, critically, to momenta near the radial mode mass, receiving nontrivial contributions from radial mode dynamics that could contaminate the spectrum. As we observe in Fig.~\ref{fig:UVVariations}, no appreciable evidence for a $q$ growing linearly with $\log(m_r/H)$ is found until a UV cutoff at least as large as $m_r / 8$. At larger UV cutoffs, the radial mode masking may in principle play a role in determining the spectrum, but its effect was found to be negligible in \cite{Buschmann:2021sdq}.  We inspect the emission spectra and associated best-fit indices for these large values of the UV cutoff in Fig.~\ref{fig:LargeUVCutoff}. We find increasing levels of systematic mismodeling that bias the best-fit index to a given instantaneous spectrum which grows in effect with increasing UV cutoff.

We note that our definition of the UV cutoff in an AMR context is somewhat more subtle than for a static lattice. At the end of our simulation, while the string width $\sqrt{2}m_r$ is resolved at our finest level of refinement by four lattice sites, it is under-resolved by a factor of eight at the base level where we measure the emission spectrum. As a result, the most aggressive choices of UV cutoff, like one at $m_r / 4$, will admit wavenumbers up to nearly the Nyquist mode of our lattice, where discretization effects are certain to be important. To rectify this, we impose an additional constraint such that, independent of our UV cutoff, no momenta within a factor of four of the Nyquist momentum are included within our fitting range. This also ensures a more fair comparison of our $m_r/4$ analysis with that \cite{Gorghetto:2018myk, Gorghetto:2020qws}, which takes a UV cutoff of $m_r/4$ but maintains at least unit resolution of the string core on their static lattice. We note that in both prior static lattice simulations and on the coarsest level of our adaptive mesh simulations, at late times, $m_r$ is roughly at the Nyquist frequency, meaning that modes near the UV cutoff of $m_r/4$ are likely quite strongly affected by discretization choices. We do not expect these choices to strongly affect the data for our more conservative choices of UV cutoff.

\begin{figure}[t!]
    \includegraphics[width=1.0\textwidth]{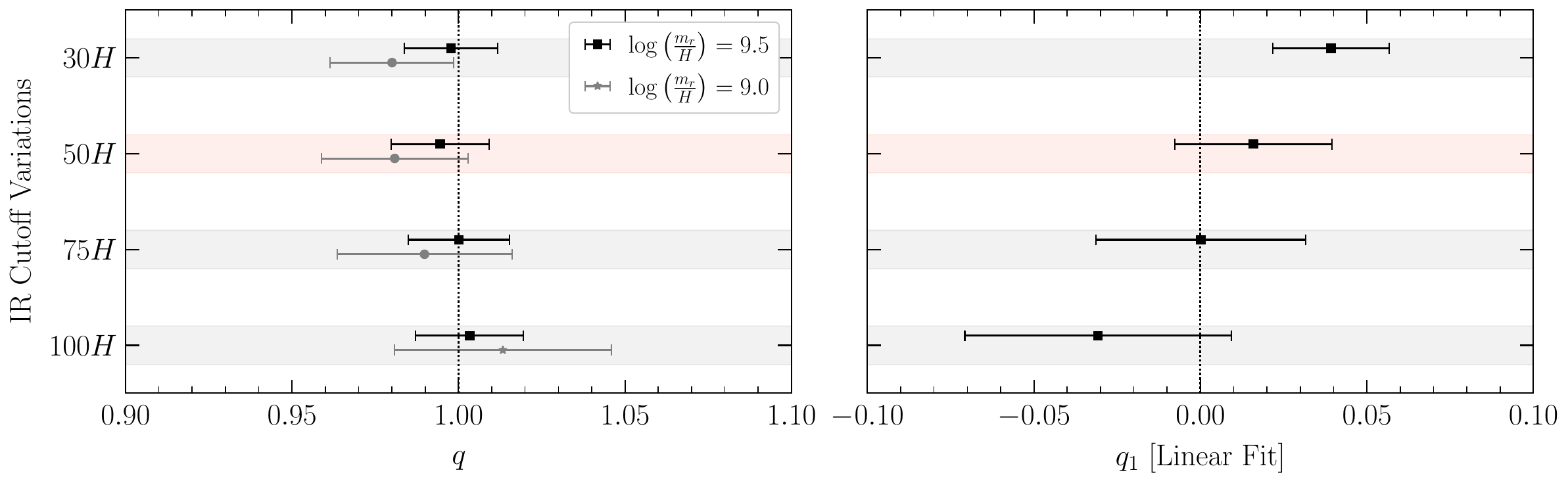}
    \caption{As in Fig.~\ref{fig:UVVariations}, but comparing different choices of the IR cutoff.}
    \label{fig:IRVariations}
\end{figure}

Similarly, in Fig.~\ref{fig:IRVariations}, we study the results when all details of the fiducial analysis are held fixed with the exception of varying $x_\mathrm{IR}$ between $\{30, 50, 75, 100\}$. A lower choice of IR cutoff risks systematic mismodeling of the instantaneous emission spectrum by a simple power law model as low momentum axion emission is cut off by the interstring spacing. Additionally, the smaller density of states of low wavenumbers of our rectilinear lattice means lower wavenumber data is intrinsically noisier. Taken together, a larger value of the IR cutoff is a more conservative one. As we observe in Fig.~\ref{fig:IRVariations}, more aggressive smaller values of the IR cutoff result in estimated linear growth of the emission spectrum index, while more conservative larger values for the IR cutoff result in little to no evidence for a time-evolving index.

In Tab.~\ref{tab:ir_variations} and Tab.~\ref{tab:uv_variations}, we provide the tabulated results from the systematic variations to the UV and IR cutoffs examined in this section. Other than the variations to those cutoffs, all details of the fiducial analysis are held fixed. We note that thanks to increased simulation volume and dynamical range relative to that of the previously state-of-the-art simulation in \cite{Buschmann:2021sdq}, fit uncertainties have been reduced at least threefold and, in the case of large $x_\mathrm{UV}$, by as much as an order of magnitude.

\begin{table}[t!]
\ra{1.3}
\centering
\begin{tabularx}{\textwidth}{p{0.2\textwidth}*{5}{P{0.2\textwidth}}}
\hline
  Coefficient & $x_\mathrm{IR} = 30$ & $\mathbf{x_\mathbf{IR} = 50}$ & $x_\mathrm{IR} = 75$ & $x_\mathrm{IR} = 100$\\ \hline
  $q_1$ & $0.04 \pm 0.02$ & $\mathbf{0.02 \pm 0.02}$ & $0.00 \pm 0.03$ & $-0.03 \pm 0.04$   \\ \hline
  $q_0$ & $0.65 \pm 0.16$ & $\mathbf{0.87 \pm 0.22}$ & $1.03 \pm 0.01$ & $1.31 \pm 0.37$ \\ \hline
  $q_0^{\rm const.}$ & $1.01 \pm 0.01$ & $\mathbf{1.01 \pm 0.01}$ & $1.02 \pm 0.01$ & $1.03 \pm 0.01$
  \\ \hline
\end{tabularx}
\caption{Results of the fits to the spectral evolution holding all our fiducial analysis choices fixed but for various IR cutoffs $x_\mathrm{IR}$. We provide the fits and uncertainties for the $q_1$ and $q_0$ in the linearly growing index model and the best fit for $q_0^{\rm const.}$ in the constant index model. Our fiducial choice of $x_\mathrm{IR} = 50$ is shown in bold.}
\label{tab:ir_variations}
\end{table}

\begin{table}[t!]
\ra{1.3}
\centering
\begin{tabularx}{\textwidth}{p{0.1\textwidth}*{9}{P{0.105\textwidth}}}
\hline
  Coefficient & $x_\mathrm{UV} = 1/4$ & $x_\mathrm{UV} = 1/6$ & $x_\mathrm{UV} = 1/8$ & $x_\mathrm{UV} = 1/12$
  & $\mathbf{x_\mathrm{UV} = 1/16}$ & $x_\mathrm{UV} = 1/24$ & $x_\mathrm{UV} = 1/28$  & $x_\mathrm{UV} = 1/32$ \\ \hline 
  $q_1$ & $0.18 \pm 0.01$ & $0.11 \pm 0.01$ & $0.06 \pm 0.01$ & $0.04 \pm 0.02$ & $\mathbf{0.02 \pm 0.02}$ & $0.03 \pm 0.04$ & $-0.01 \pm 0.05$ & $-0.02 \pm 0.06$ \\ \hline
  $q_0$ & $-0.34 \pm 0.11$ & $0.13 \pm 0.11$ & $0.49 \pm 0.12$ & $0.66 \pm 0.16$ & $\mathbf{0.87 \pm 0.21}$ & $0.77 \pm 0.33$ & $1.10 \pm 0.41$ & $1.14 \pm 0.50$ \\ \hline
  $q_0^{\rm const.}$ &  $1.23 \pm 0.01$ & $1.13 \pm 0.01$ & $1.07 \pm 0.01$ & $1.03 \pm 0.01$ & $\mathbf{1.01 \pm 0.01}$ & $1.00 \pm 0.01\quad$ & $0.98 \pm 0.02$ & $0.98 \pm 0.02$  \\ \hline
\end{tabularx}
\caption{As in Tab.~\ref{tab:ir_variations}, but for varying UV cutoff $x_\mathrm{UV}$ with all other parameters fixed to their fiducial values. Our fiducial choice of $x_\mathrm{UV} = 16$ is shown in bold.}
\label{tab:uv_variations}
\end{table}

\section{Systematics: Pre-evolution to Remove Transients}
\label{SM:Preevolution}
The results of this Letter differ appreciably from those of \cite{Gorghetto:2018myk, Gorghetto:2020qws, Saikawa:2024bta}. While none of these works implement AMR techniques to maintain the high resolution of strings, they also differ in their procedure of generating a string network for studying axion string emission. We consider the possibility here that these differences in the initialization procedure are the origin of the discrepant measurements of $q$ across the works.

Our simulations begin with a thermal state before the breaking of the PQ symmetry, with strings realized dynamically as the shrinking thermal mass shifts the minimum of the scalar field potential. By contrast, \cite{Gorghetto:2018myk, Gorghetto:2020qws, Saikawa:2024bta} do not include a thermal mass for the PQ scalar, corresponding to starting their simulations deep in the broken phase of the theory well after strings should have first formed. It is unclear how to generate principled initial conditions for the complex scalar at this time, and so these works take an alternate approach of generating initial conditions appropriate for the PQ scalar in the unbroken phase and evolve the unphysical field dynamics under strong damping until strings form in the simulation and reach a target string length (typically corresponding to matching the claimed attractor solution of \cite{Gorghetto:2018myk, Gorghetto:2020qws} realized absent the external damping). This strategy is often referred to as \textsl{pre-evolution}. 

We perform an AMR simulation initialized with pre-evolution and apply our analysis procedure to extract $q$, allowing us to directly assess the impact of the initialization procedure on the results presented in the main text. We perform our simulation in a box with size $\bar L= 96$ resolved by 2,048$^3$ lattice sites at the base level; this enables a simulation out to $\log(m_r / H) = 8.5$. All other simulation parameters are the same as in our primary simulation. We generate an initial state of independent Gaussian noise at each lattice site in the $\psi_1$ and $\psi_2$ fields, and we evolve them with the equations of motion
\begin{equation}
\psi_i''+\frac{3}{\eta}\psi_i'-\frac{\bar\nabla\psi_i}{\eta^2\eta_i}+\frac{\eta_i^2}{\eta^2}\psi_i(\left|\psi\right|^2-1)=0.
\label{eq:DampedEoM}
\end{equation}
These equations of motion were developed in \cite{Benabou:2023ghl} to realize the desired additional damping by replacing $R(t)\propto t^{1/2}$ with $R(t)\propto t$ while also providing a tuneable parameter $\eta_i$ such that the width of the string on the lattice as evolved under ~\eqref{eq:DampedEoM} would match the width of the string on the lattice as evolved under the fiducial equations of motion at conformal time $\eta_i$. This then allows us to use the pre-evolved state at an arbitrary time $\eta_\textrm{pre}$ as the initial condition for our standard simulations at time $\eta_i$. During this transition, we account for the change of physics by rescaling the field derivatives by $\eta_\textrm{pre}/\eta_i$.

We perform a simulation following this procedure, which we compare to our primary simulation performed without any pre-evolution. Note that this pre-evolved simulation uses a smaller box and has a more limited dynamical range, meaning that its individual statistical power is lesser. We choose $\eta_i \approx 3.77$, corresponding to $\log(m_r /H) = 3$, and we evolve under damping until the string length reaches its predicted value under the logarithmically growing attractor solution of $\xi=0.3$ at this time \cite{Saikawa:2024bta}. After reaching the targeted string length, we transition to our standard equations of motion, taking the conformal time to be $\eta_i$.  The pre-evolution phase ran for about 8 hours on 128 nodes of the Perlmutter GPU cluster, which provides 512 NVIDIA A100 (40 GB) GPUs, 8,192 CPU cores, and 66 TB of aggregate memory. The subsequent evolution ran on 256 Perlmutter GPU nodes for about 5 hours. 

\begin{figure}[t!]
    \includegraphics[width=0.6\textwidth]{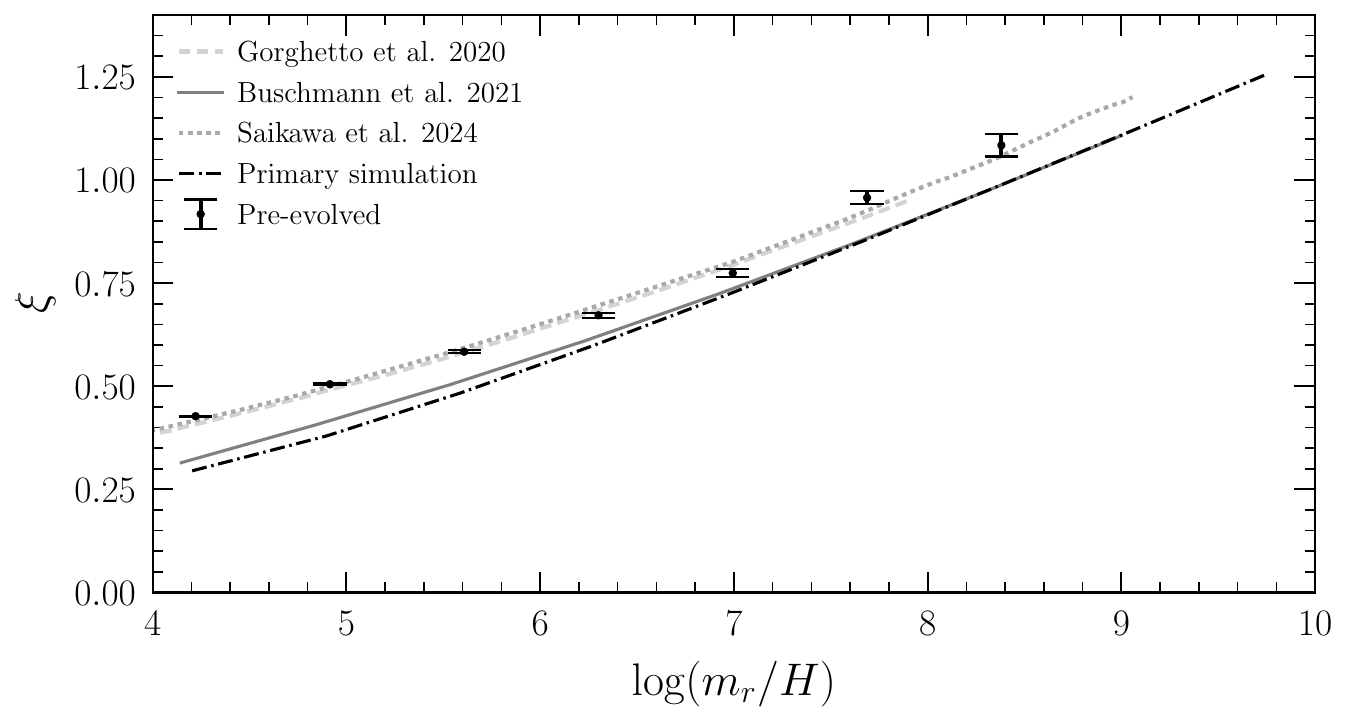}
    \caption{Comparison of the string length per Hubble volume $\xi$ between our main result (black, dash-dotted) and a simulation using a pre-evolved initial state tuned to be close to the scaling solution (data points). This plot illustrates consistency between our result and that of works from Gorghetto et al.~\cite{Gorghetto:2020qws} and Saikawa et al.~\cite{Saikawa:2024bta}, which also used a pre-evolved initial state.}
    \label{fig:xi_alt}
\end{figure}

We first study the time-evolution of the string length in our pre-evolved simulation as compared with that found in our primary simulation and prior results in the literature \cite{Gorghetto:2018myk, Gorghetto:2020qws, Buschmann:2021sdq, Saikawa:2024bta}. The results are shown in Fig.~\ref{fig:xi_alt}, where we find good consistency between the growth of the string network in our pre-evolved simulation and other works that also implement a pre-evolution procedure, namely \cite{Gorghetto:2020qws, Saikawa:2024bta}. 

\begin{figure}[t!]
    \includegraphics[width=.6\textwidth]{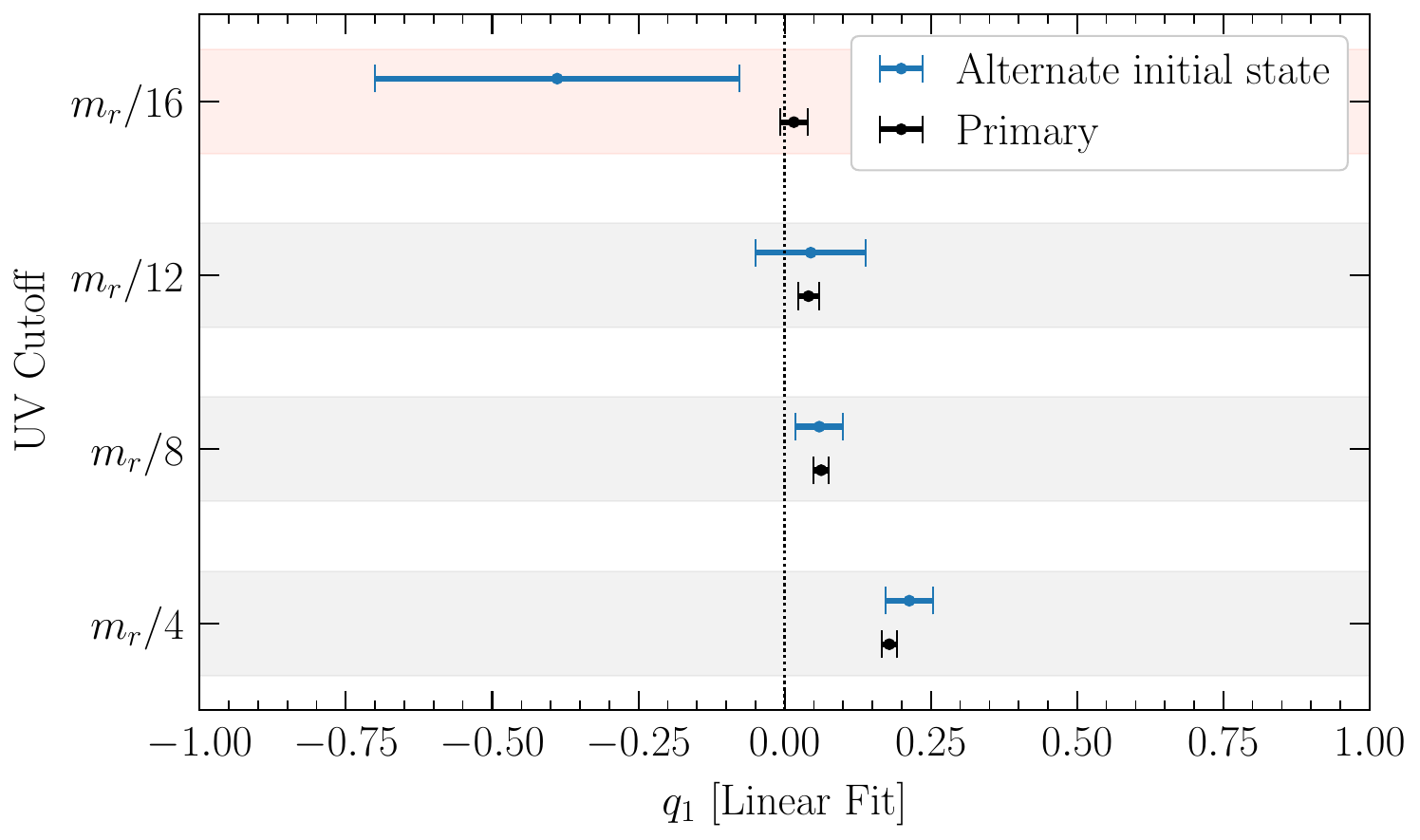 }
    \caption{Comparison of spectrum linear fit slope parameter $q_1$ from our simulation using the alternate initial state procedure with pre-evolution (blue), to our primary large AMR simulation (black). Note that the simulation with an alternate state is the same as shown in Fig. \ref{fig:xi_alt}. Here we assume the same IR cutoff as in our fiducial spectrum ($k_\mathrm{IR}/H=50$). The UV cutoff corresponding to our fiducial spectrum is shaded in light red. The numerical values of the fit indices are tabulated in Table.~\ref{tab:ir_variations_preevolution_test}.}
    \label{fig:random_ic_qfit}
\end{figure}

We go on to measure $q$ and evaluate its possible time-evolving behavior in the pre-evolved simulation following the analysis procedure described in the main text. We compare the results of the linearly growing emission index shown in Fig.~\ref{fig:random_ic_qfit}, finding results that are compatible at the level of the statistical errors. In general, we observe the same trend as in SM Sec.~\ref{SM:FittingVariations}, with larger UV cutoffs for the fitting region resulting in more rapid growth of the emission spectrum index. Based on the good compatibility of the growing emission spectrum index between our primary simulation and our pre-evolved simulation, we conclude that our assessment of the systematic bias of the fit at large UV cutoffs is a robust one.

\begin{table}[t!]
\ra{1.3}
\centering
\setlength{\tabcolsep}{10pt}      
\renewcommand{\arraystretch}{1.3} 
\begin{tabularx}{\textwidth}{p{0.1\textwidth}*{9}{P{0.1805\textwidth}}}
\hline
  Coefficient & $\mathbf{x_\mathbf{UV} = 1/16}$ & $x_\mathrm{UV} = 1/12$ & $x_\mathrm{UV} = 1/8$ & $x_\mathrm{UV} = 1/4$\\ \hline
  $q_1$ & $\mathbf{0.05  \pm  0.45}$ $\mathbf{-0.39  \pm  0.31}$   & $\,\,\,\,\,\,0.21  \pm  0.09\quad$ $0.04  \pm  0.09$  & $\,\,\,\,\,\,0.21  \pm  0.05\quad$ $0.06  \pm  0.04$ & $\,\,\,\,\,\,0.28  \pm  0.03\quad$ $0.21  \pm  0.04$   \\ \hline
  $q_0$ & $\,\,\,\,\,\mathbf{0.73  \pm 3.58}\quad$ $\mathbf{4.09  \pm 2.51}$ & $\,\,\,\,\,-0.56  \pm 0.69$ $0.63  \pm 0.74$ & $\,\,\,\,\,\,\,-0.54  \pm 0.35\quad$ $0.53  \pm 0.31$& $ \,\,\,-0.95  \pm 0.21$ $ -0.52  \pm 0.3$ \\ \hline
  $q_0^{\rm const.}$ & $\mathbf{\,\,\,\,\,1.15  \pm  0.07}\quad$ $\mathbf{0.96  \pm  0.04}$ & $\,\,\,\,\,\,1.05  \pm  0.04\quad $ $0.98  \pm  0.03$ & $\,\,\,\,\,\,\,1.07  \pm  0.04\quad$ $1.00  \pm  0.01$& $\,\,\,\,\,\,1.13  \pm  0.06\quad$ $1.05  \pm  0.05$
  \\ \hline
\end{tabularx}
\caption{Fits and uncertainties for the $q_1$ and $q_0$ in the linearly growing index model and the best fit for $q_0^{\rm const.}$ in the constant index model, for our simulations prepared via the alternate initial state procedure as described in SM Sec.~\ref{SM:Preevolution} and SM Sec.~\ref{sec:res_test}.  We perform these alternate simulations at two different grid resolutions, labeled `Low-res' and `High-res' in Fig. \ref{fig:qfit_res_comparison}, corresponding to the top and bottom quantities in each cell, respectively. Note that these spectral fits are compared to our primary simulation in Figs. \ref{fig:random_ic_qfit} and \ref{fig:qfit_res_comparison}. Here we assume our fiducial choice of IR cutoff $k_\mathrm{IR}/H=50$ and vary the UV cutoff (with the fiducial choice bolded). 
}
\label{tab:ir_variations_preevolution_test}
\end{table}

For completeness, in Tab.~\ref{tab:ir_variations_preevolution_test}, we provide the best-fits and $1\sigma$ errors for the fit parameters to the constant and linearly growing emission index models for each of the variations of the UV cutoff considered here.

\section{Systematics: Impact of Resolution}
\label{sec:res_test}
By implementing AMR, our simulations realize a nearly constant (to within a factor of 2) resolution of strings, while simulations that use a static lattice are doomed to run out of resolution as the physical width of the string shrinks on the comoving lattice. For instance, in \cite{Gorghetto:2018myk, Gorghetto:2020qws, Saikawa:2024bta},  the majority of simulations are run until the string width is resolved by a single lattice site. This corresponds to the $\eta$ at which $1/(m_r \Delta x)$ is unity. 

Not only does our AMR approach allow us to maintain resolution in a computationally efficient manner, but it also enables us to directly test the impact of resolution by loosening our string tagging criteria. We perform identical pre-evolved simulations to those studied in SM Sec.~\ref{SM:Preevolution}, but in our tagging criteria, we tag strings only at levels where they are expected to be resolved by one or fewer lattice sites. This means both simulations will have the same base resolution with identical initial states and only differ in the number of refinement levels. We refer to this simulation as our \textsl{low-resolution} simulation, while the previous simulation of SM Sec.~\ref{SM:Preevolution} is referred to as our \textsl{high-resolution} simulation. Note that we chose to perform this comparative test between pre-evolved simulations to sharply inspect the resolution criteria in the existing literature. The low-resolution simulation ran around 25 times faster than the high-resolution simulation, requiring a total of 32 Perlmutter GPU nodes with 128 NVIDIA A100 (40 GB) GPUs for about 90 minutes.

\begin{figure}[t!]
    \includegraphics[width=1.0\textwidth]{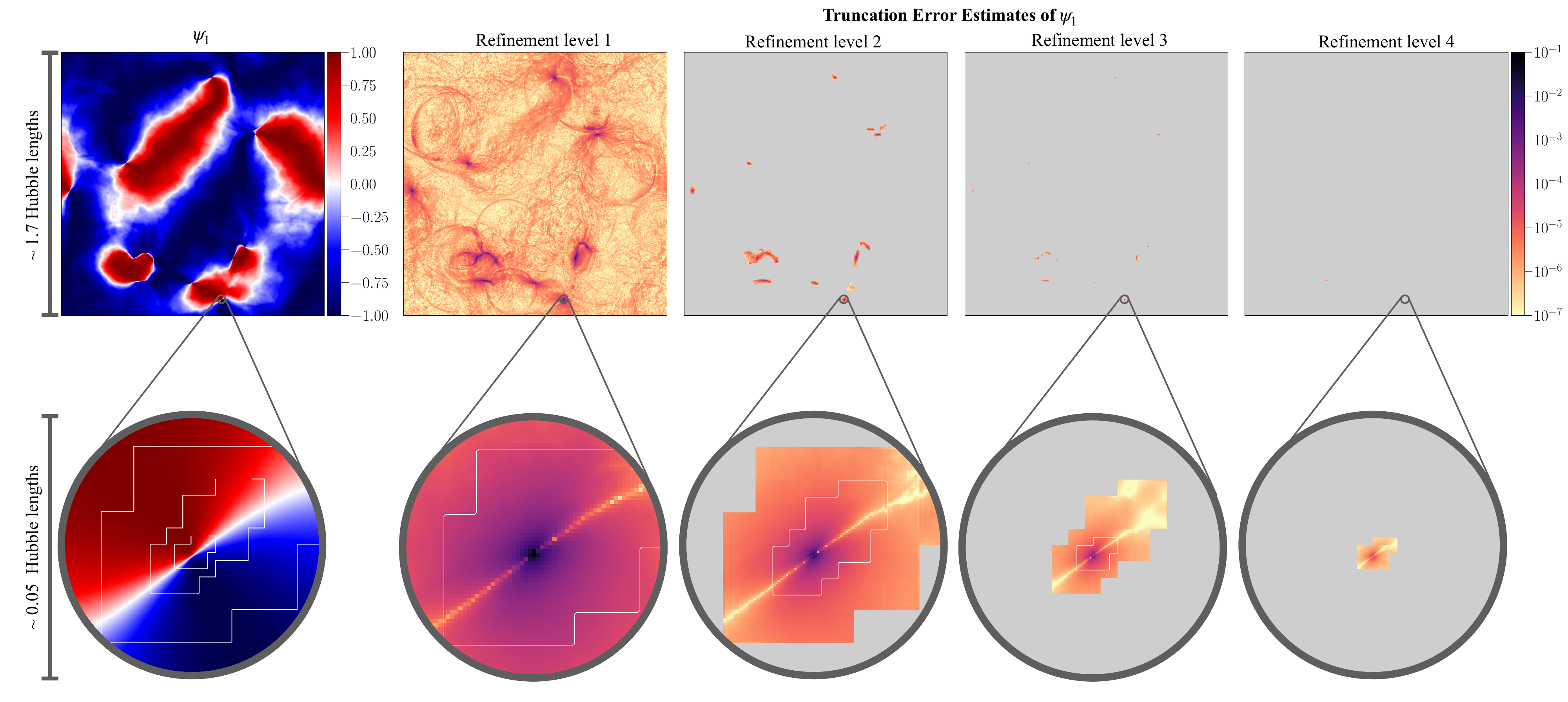}
    \caption{Illustration of truncation error estimates in the high-resolution simulation based on a pre-evolved initial state. The far-left panel is a slice through the $\psi_1$ field component with the other four panels containing the corresponding truncation error estimates at different refinement levels. The bottom row is a zoom-in centered around a string piercing the slice. Overlayed are coarse-fine boundaries (white lines). Areas in gray do not contain data as they are not covered by the respective refinement level.
    }
    \label{fig:FieldRes}
\end{figure}

We present truncation error estimates of the high-resolution simulation on a slice through the simulation volume in Fig.~\ref{fig:FieldRes}. Truncation error estimates correspond to the absolute difference between refinement level $\ell$ and $\ell-1$ after evolving both levels independently for $\Delta \eta_{\ell-1} = 2\Delta \eta_\ell$. See SM Sec.~\ref{sec:TEE} for a description of how truncation error estimates are computed and utilized in our simulation. From Fig.~\ref{fig:FieldRes} it becomes clear that the majority of the simulation volume is captured much better at a given resolution than the string core, with errors several orders of magnitude smaller. At the first refinement level, the typical error does not exceed the $10^{-5}$ level away from strings, whereas the errors around the string core almost approach unity. For every extra refinement level, the error estimate drops by roughly an order of magnitude or more, eventually bringing it down to an acceptable level at the string core. This highlights how important resolution around the string core is and how AMR is the correct tool to simulate the network. Since we resolve the string core width by four grid sites in our primary simulation whereas other groups (\textsl{e.g.}~\cite{Gorghetto:2018myk, Gorghetto:2020qws, Saikawa:2024bta}) often reduce this criterion to just one grid site per string core width, it implies that we effectively use two extra refinement levels. We can thus infer that our numerical error is roughly two orders of magnitude smaller.

\begin{figure}[t!]
    \includegraphics[width=0.6\textwidth]{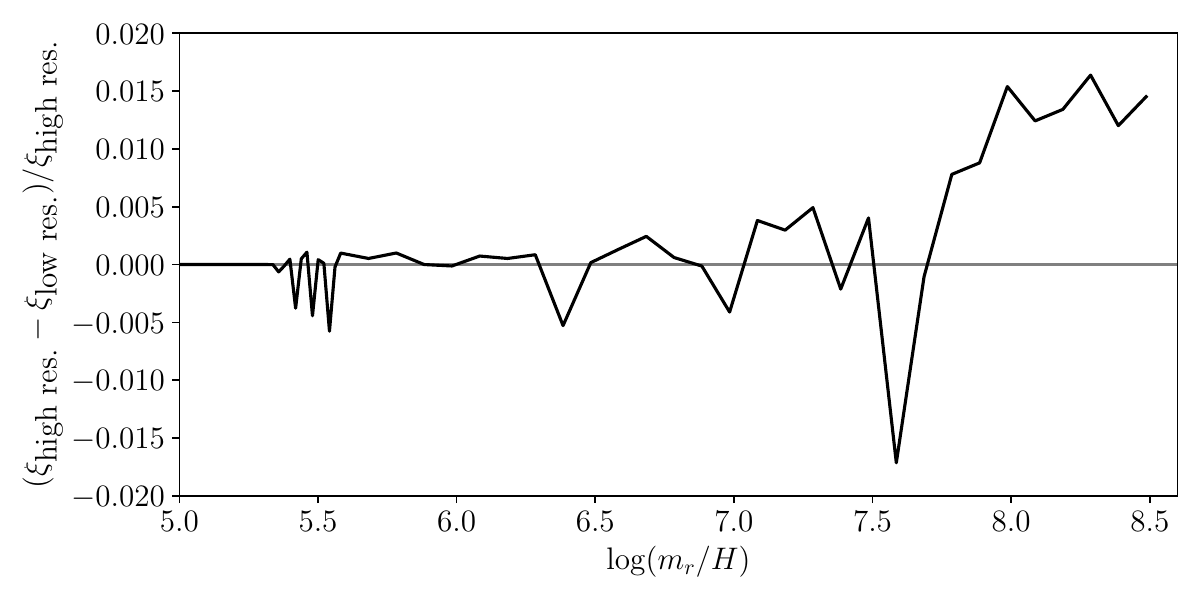}
    \caption{The relative difference between the high-resolution and low-resolution pre-evolved simulations. Only very small differences at the percent level are observed. See text for more details.}
    \label{fig:xi_res}
\end{figure}

We compare the time-evolving string length in the low-resolution and high-resolution simulation, with results shown in Fig.~\ref{fig:xi_res}. We find modest discrepancies at the 2\% level. The absolute difference is slowly increasing with time but with no clear trend in either direction.
 
\begin{figure}[t!]
    \includegraphics[width=1.0\textwidth]{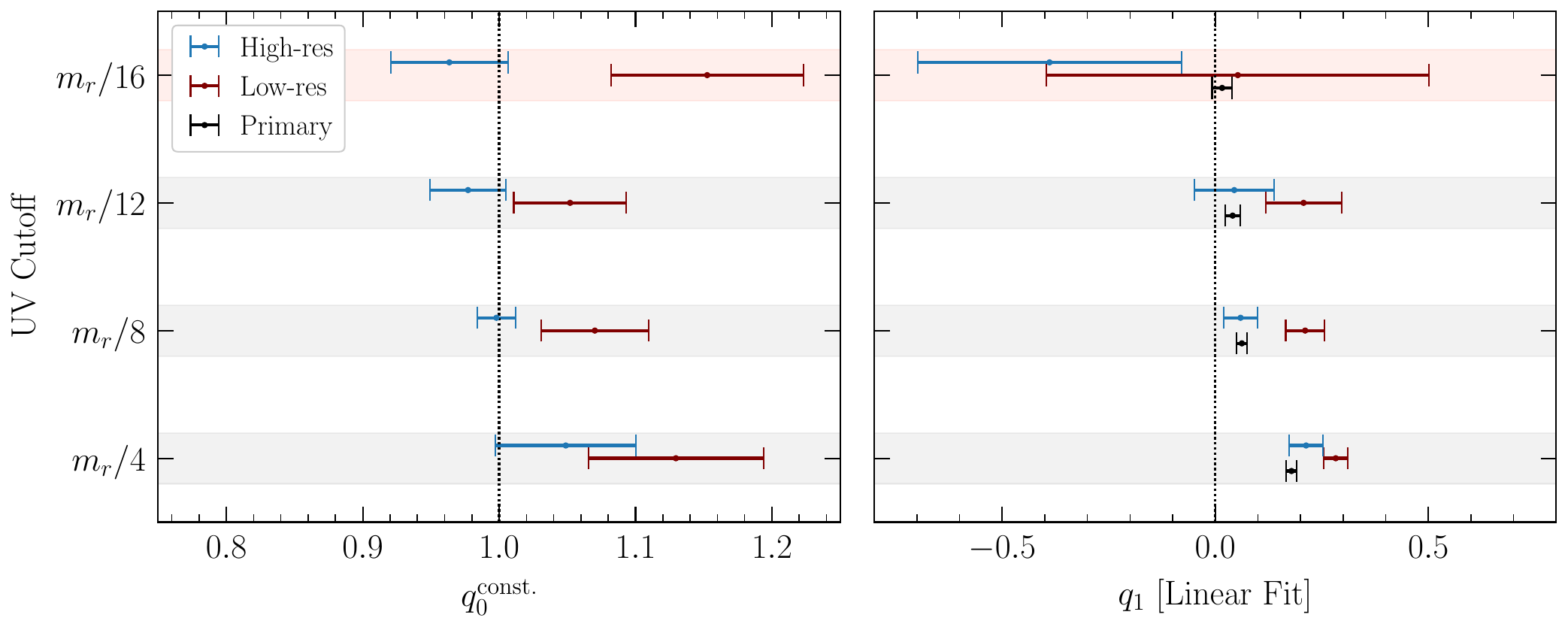}
    \caption{ 
    As in Fig. \ref{fig:random_ic_qfit}, but comparing two simulations at differing grid resolutions which are generated with the alternate initial state procedure (for details see text). The simulation labeled `High-res' is the same as shown in Fig. \ref{fig:random_ic_qfit}. Note that we do not directly compare the constant fit parameter $q_0^{\mathrm{const.}}$ between the primary simulation and those generated using the alternate initial state as the primary simulation differs in simulation volume from the latter.}
    \label{fig:qfit_res_comparison}
\end{figure}

As before, we measure $q$ in the low-resolution simulation and compare our findings with those of the high-resolution simulation and our primary simulation. The results are depicted in Fig.~\ref{fig:qfit_res_comparison} and provided in detail in Tab.~\ref{tab:ir_variations_preevolution_test}. Strikingly, we find that low-resolution simulations systematically prefer a larger emission spectrum index in the case of the constant model and a more rapidly growing emission spectrum index in the logarithmically growing model. 

One possible reason for the difference in the spectrum is illustrated in Fig.~\ref{fig:SingleStringDx}. Here, we performed two simulations of a single circular string in the QCD epoch, which is described in detail in Sec.~\ref{sec:SingleString}. One simulation uses our high-resolution setup, the other is based on our low-resolution settings. Both simulations are otherwise identical. From Fig.~\ref{fig:SingleStringDx} it becomes apparent that the low-resolution string is not sufficiently resolved to properly track its motion. The string collapses more slowly and loses its circular shape. Facets diagonally to the underlying grid are developing. This is due to inaccuracies in the finite-difference Laplacian caused by the limited resolution, which systematically reduces the speed of the string along the grid lines. The effect is large enough for visible kinks to develop that radiate extra energy. This extra artificial source of radiation is likely polluting the measurement of $q$. Even though this comparison was done in the QCD epoch, the same numerical arguments apply to PQ epoch simulations.

\section{Axions through the QCD Phase Transition}
\label{sec:string_emission_during_QCD_PT_analytic}
Understanding the fate of axions radiated by strings through the transient epoch during which the axion acquires its zero temperature mass is critical for accurately estimating the axion relic abundance. For example, it is possible that axions produced from the collapse of strings and domain walls during the QCD phase transition contribute appreciably to the DM relic abundance.

\begin{figure*}[t!]
    \includegraphics[width=0.245\textwidth]{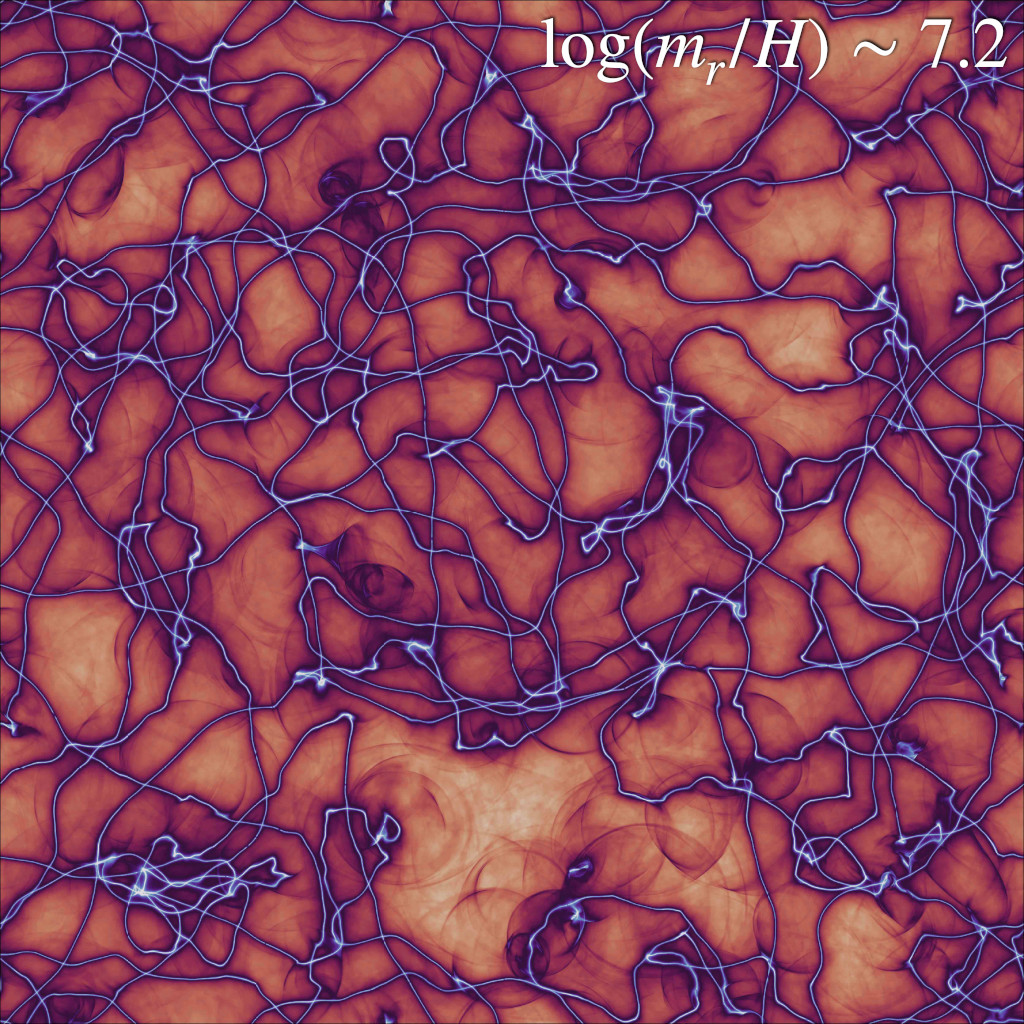}
    \includegraphics[width=0.245\textwidth]{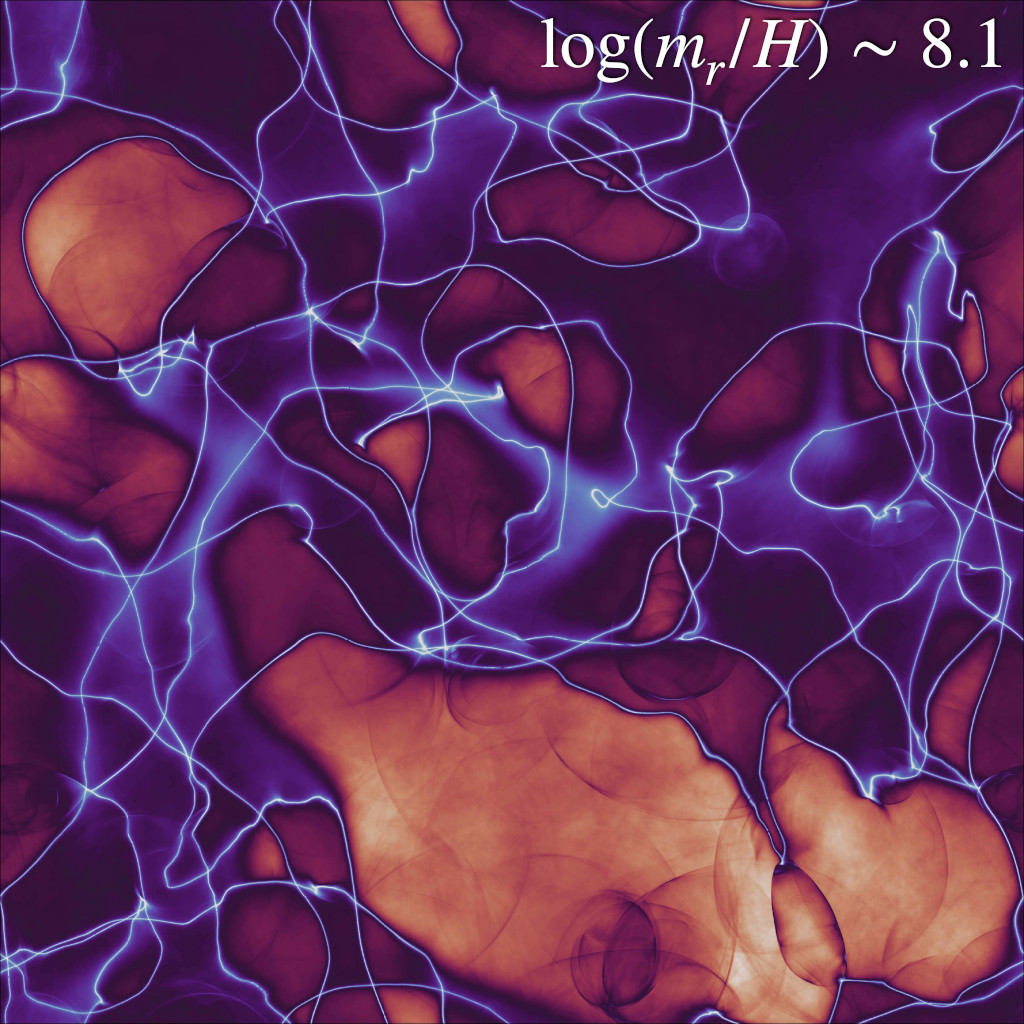}
    \includegraphics[width=0.245\textwidth]{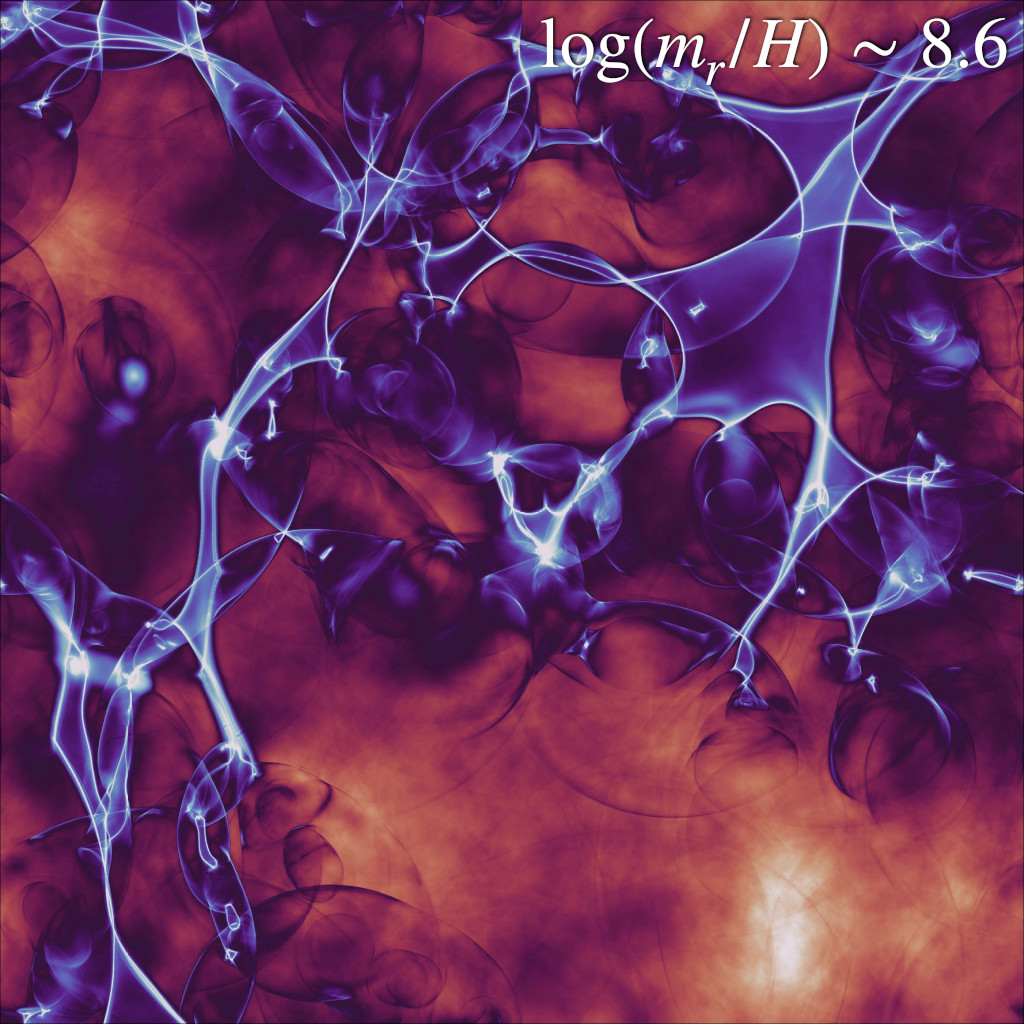}
    \includegraphics[width=0.245\textwidth]{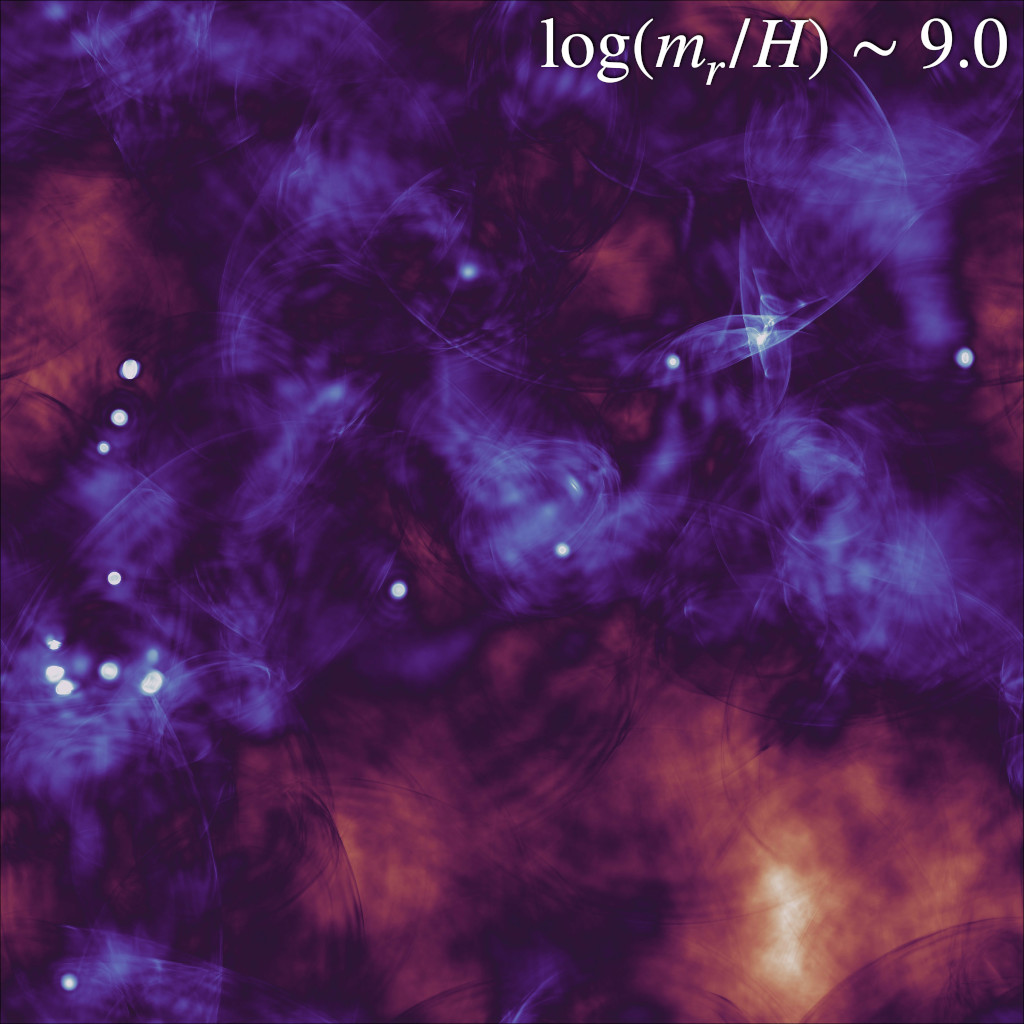}
    \caption{3D$\rightarrow$2D projection of the total axion energy density $\rho_{a} = \frac{1}{2}\dot{a}^2+\frac{1}{2}(\nabla a)^2 + V(a)$ from a simulation of the QCD phase transition. From left to right: (1) Axion string network before the QCD phase transition. (2) The beginning of the phase transition. The axion mass starts growing and domain walls form. (3) The extra tension from the domain walls causes the network to collapse. (4) The QCD phase transition is complete. All strings have vanished and just a few oscillons are visible as bright point-like spots, though we do not investigate these features in this work. An animated version of this simulation can be found \href{https://tinyurl.com/AxionStringsQCD}{here}.
    }
    \label{fig:QCDSim}
\end{figure*}

In this section, we evaluate the efficiency of strings and domain walls in producing axions at late times using analytic arguments calibrated against a simulation of a string-domain-wall network.  The QCD-epoch simulation that we run for this analysis is described in App.~\ref{app:QCD}. The simulation takes an initial state containing axion strings with no axion mass and turns on a rapidly growing axion potential (with $N_{\rm dw} = 1$), such that domain walls form and collapse the network.    In Fig.~\ref{fig:QCDSim} we illustrate a few snapshots of the simulation, showing the initial string network, the formation of domain walls bounded by these strings, and the subsequent collapse of the defect network under its tension. Oscillons, also referred to as axitons, appear in the final stages of the network collapse. We defer a more detailed study of them to future work.

It is useful to define a few important time periods during the epoch of the QCD phase transition to understand the dynamics of string-wall network collapse. First, we define $t_*$ as the time at which the axion begins to oscillate when $m_a(t_*) = 3 H(t_*)$. In our simulations, this time occurs at $\eta \approx 32$ or equivalently $\log(m_r/H) \approx 7.3$. Prior estimates of the axion abundance in {\it e.g.}~\cite{Buschmann:2021sdq} neglected axion emission by strings after this somewhat arbitrary time. More physically, strings will continue to emit axions, though possibly with reduced efficiency, until the total collapse of the defect network. 

The next relevant time is $t_{m_a}$, which defines when the axion mass begins to play a role in determining the string emission spectrum. In particular, once $m_a\gg H$, the axion provides an IR cutoff to the string emission spectrum, suppressing emission at momenta $k < m_a$. To estimate this effect, let us assume a simple, analytic functional form for the instantaneous emission spectrum $F[k/H]$. In particular, note that for $t < t_*$ we may reproduce our fiducial result by approximating $F[k/H] \propto 1/(k/H)$ for $x_{\rm L} < k / H < m_r / H$, where we take $x_{\rm L} = \delta_1 \sqrt{\xi}$, and zero otherwise. (Throughout this section we assume the strings radiate a conformal spectrum of axions; {\it i.e.}, we assume $q = 1$ for definiteness.) That is, we assume simply that $F[k/H]$ is a power-law for all $k/H$ above some critical value $x_{\rm L}$, which slowly moves to the UV like $\sqrt{\xi}$ as $\xi$ grows with time. Now accounting for the time-dependent axion mass we simply use the above functional form but with $x_L = {\rm max}( \delta_1 \sqrt{\xi}, m_a(t) / H )$. Then we have $t_{m_a}$ defined as the time at which $m_a/H = \delta_1 \sqrt{\xi}$. Note that from our primary simulation we measure $\delta_1 \approx 8.6$ and $\xi_* \in (11,15)$. For concreteness in this section, we fix $\delta_1 = 8.6$ and $\xi_* = 13$, while $\xi_* \approx 0.8$ for the simulations which we study.

\begin{figure}[t!]    \includegraphics[width=1\textwidth]{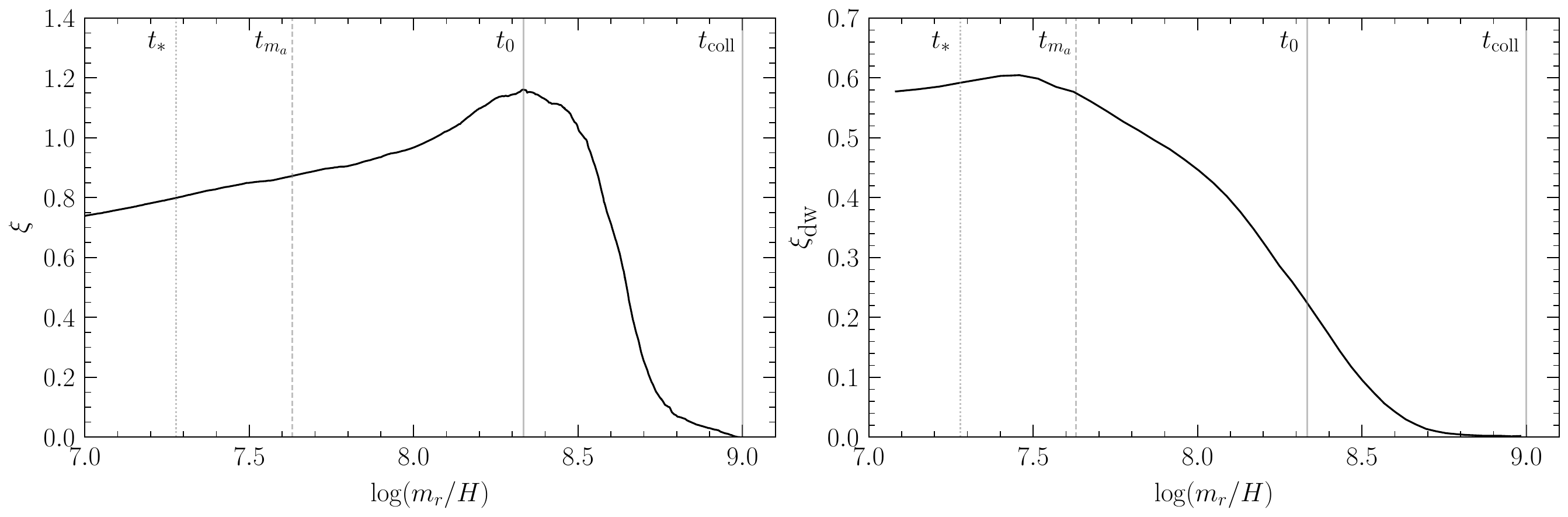}
    \caption{The string length parameter $\xi$ and the domain wall area parameter $\xi_\textrm{dw}$ during the QCD phase transition simulation. We indicate the time at which the axion mass becomes dynamical $t_\mathrm{*}$,  and the estimated time $t_{m_a}$ when the IR cutoff of the emission spectrum becomes set by the axion mass rather than the inter-string separation distance. We further indicate the time $t_0$ when $\xi$ begins to decrease due to the shrinking of string loops under domain wall tension, and when the network has collapsed completely, $t_\mathrm{coll}$.
    }
    \label{fig:xi_qcd}
\end{figure}

At later times, after $t_{m_a}$, the string network begins to collapse, and we define $t_0$ to be the time at which the string length parameter $\xi$ begins to shrink. Finally, the collapse completes at $t_\mathrm{coll}$. After this time, the axion field is (with the exception of some oscillons), to good approximation, linear, and no additional processes contribute to the axion abundance. In Fig.~\ref{fig:xi_qcd}, we illustrate $\xi$ as a function of time as measured in our QCD-epoch simulation, with the four critical times indicated. We also show the corresponding domain wall area parameter $\xi_\textrm{dw}$ as defined in Sec.~\ref{sec:DWs}.

Given an axion mass parametrization $m_a(t)$ and string length parameter $\xi$, the times $t_*$ and $t_{m_a}$ are analytically calculable, whereas computing $t_0$ and $t_\mathrm{coll}$ is less straightforward. We begin by recognizing that, at $t_*$, in addition to abundant subhorizon-scale defects, there will also exist defects that are superhorizon-scale at this time. These superhorizon defects are comparatively less numerous but appear with nonzero probability due to the horizon-scale statistics associated with the axion field configuration via the Kibble mechanism, see, \textit{e.g.} \cite{Dunsky:2024zdo} for more details. Further evidence for the presence of these superhorizon defects in our simulations is presented in Sec.~\ref{sec:SingleString}, particularly in Fig.~\ref{fig:StringEvolutionEmission} and the surrounding discussion.

The evolution of defects at times after $t_*$ when the axion field becomes dynamic depends sharply on their size. A subhorizon string loop that bounds a domain wall will rapidly collapse (within roughly a Hubble time) under the joint string and domain wall tension.\footnote{Note that this differs from the time-evolution of relatively large but still subhorizon string loops at early times, which do not collapse rapidly absent the domain wall tension.} By comparison, a string loop bounding a domain wall that is superhorizon at $t_*$ will not begin to efficiently collapse until it enters the horizon, and in the intervening time, its physical size will grow due to the growth of the scale factor. As a result, in order to estimate axion production from the defect network around the time of the collapse, we treat subhorizon and superhorizon defects independently. Throughout this section, we make the simplifying assumption that string loops of physical length $\ell$ are circular and bound by a domain wall with radius $\ell/2\pi$. A defect is then taken to be superhorizon if its physical diameter $d$ is greater than $H^{-1}$. We also assume, for simplicity, that domain walls within the horizon collapse at the speed of light, while domain walls outside the horizon are frozen in comoving coordinates. 

\subsection{Collapse of Subhorizon Defects}
We begin with defects which are initially subhorizon at $t_*$. To put an upper bound on their axion production, we take $\xi_*$,  the string length at $t_*$, to be fully accounted for by subhorizon defects. Our collapse ansatz specifies then that the physical radius of the defect evolves as
\begin{equation}
r(t)= r(t_*) - (t-t_*) \,,
\label{eq:subhorizoncollapse}
\end{equation}
starting at $t_*$ and ending at $t_\mathrm{end} = r(t_*) + t_*$. We now consider the production efficiency of both these strings and their associated domain walls. 

\subsubsection{Emission from Subhorizon Strings}
We assume that strings radiate axions with an emission spectrum $F(k,t) = 1/[k \log(k_\mathrm{UV} / k_\mathrm{IR}(t))]$ at $k$ between an IR cutoff at $k_\mathrm{IR}(t)$ and a UV cutoff of $m_r$. In our simplified collapse model, the rate at which a loop emits axions is given by
\begin{equation}
\frac{dE_a}{dt} = 2 \pi^2 f_a^2 \log(m_r/H).
\end{equation}
The instantaneous axion emission spectrum associated with a single loop is then given by 
\begin{equation}
\frac{dN_a}{dt dk} =  \frac{2 \pi^2 f_a^2 \log(m_r / H)}{k \sqrt{k^2 + m_a(t)^2} \log(m_r/k_\mathrm{IR}(t))} \theta(r(t)),
\end{equation}
where $\theta$ is the Heaviside theta function. By integrating over $k$ and $t$, we find the total number of axions emitted is
\begin{equation}
N_a(\ell) =  \int_{t_*}^{t_* + \ell/2\pi} dt \int_{k_\mathrm{IR}(t)}^{m_r} dk\frac{dN_a}{dt dk} \,,
\end{equation}
where $k_\mathrm{IR}$ is set by either the inter-string spacing or the axion mass. That is,
\begin{equation}
    k_\mathrm{IR}(t) = \mathrm{max}\left[m_a(t), \delta_1 \sqrt{\xi_*} H(t) \right] \,,
\end{equation}
with turnover between the two cutoffs occurring at $t_{m_a}$.

Following {\it e.g.}~\cite{Buschmann:2021sdq}, the number density of closed string loops was found to be $dn_\ell / d\ell \propto 1/\ell$; we expect this to still hold true at $t_*$ when the axion mass has just entered the horizon but has not yet had time to affect the network dynamics.  Then, subject to our assumption that all string loops at $t_*$ are subhorizon (having a length less than $2 \pi t_*$), the physical number density of strings at time $t_*$ with length $\ell$ is given by
\begin{equation}
\frac{dn_\ell}{d\ell} = \frac{\xi_*}{2\pi t_*^3 \ell}.
\end{equation}
We can evaluate the number density of subhorizon-string-emitted axions at $2t_*$, at which time all subhorizon defects have collapsed, by 
\begin{equation}
\delta n_a^\mathrm{string, sub}(2 t_*) =  \left[ \frac{R(t_*)}{R(2 t_*)}\right]^3 \int_0^{2\pi t*} d\ell \frac{dn_\ell}{d\ell}N_a(\ell) \,,
\end{equation}
where the ratio of scale factors arises from redshifting the number density from $t_*$ to $2 t_*$. We compare this to the number density at $2 t_*$ of axions emitted by strings at times before $t_*$, which is given by
\begin{equation}
n_a(2t_*) = \frac{8 \pi f_a^2 \xi_*^{1/2} H(t_*)}{\delta_1}\left[ \frac{R(t_*)}{R(2 t_*)} \right]^3.
\end{equation}
These number densities are useful, as from them we can compute the ratio of the energy density of axions produced by the collapse of subhorizon string loops to the number density of axions produced by string emission prior to $t_*$ by
\begin{equation}
    \frac{\delta \rho_a^\mathrm{string, sub}}{ \rho_a } \equiv \frac{\delta n_a^\mathrm{string, sub}(2 t_*)}{  n_a(2t_*)  } = \frac{\delta_1 \xi_*^{1/2}}{4t_*^2}\int_0^{2\pi t_*} \frac{d\ell}{\ell} \int_{t_*}^{t_* +\ell/2\pi} dt \frac{\log (m_r / H)}{  \log (m_r/ k_\mathrm{IR})} \int_{k_\mathrm{IR}}^{m_r} \frac{dk}{k  \sqrt{k^2+m_a^2}} \,.
  \label{eq:sub}
\end{equation}
In the physical hierarchy, this integral is easy to evaluate as all subhorizon loops will collapse by $2 t_*$, while $t_{m_a} > t_*$. Hence, the axion mass never plays an important role in cutting off the string emission. In this case, we obtain
\begin{equation}
    \frac{\delta \rho_a^\mathrm{string, sub}}{ \rho_a } = 0.65\,.
\end{equation}
On the other hand, a similar case for the simulated hierarchy is somewhat more complicated as $t_{m_a} \approx 1.4 t_*$, meaning that we must account for the axion mass in the IR cutoff of the string emission spectrum. Treating the cases of loop emission before and after $t_{m_a}$ appropriately, we obtain 
\begin{equation}
    \frac{\delta \rho_a^\mathrm{string, sub}}{ \rho_a } = 0.75\,.
\end{equation}
Hence, we see in both the simulated and physical hierarchies, we expect string emission after the time of axion oscillation to make an $\mathcal{O}(1)$ contribution to the late-time axion abundance.

\subsubsection{Emission from Subhorizon Domain Walls}
Next, we consider axion production from the collapse of domain walls that are subhorizon at time $t_*$. First, however, we describe the domain walls themselves in more detail. An infinitely wide and flat domain wall has the axion profile in the $z$-direction (perpendicular to the domain wall)
\es{}{
a(z) = 4 f_a \tan^{-1} \exp(m_a z) \,,
}
where the domain wall is at $z = 0$. The energy density in the domain wall is proportional to $|\partial_z a(z)|^2$. Let us now Fourier transform $\partial_z a(z)$ and take the magnitude squared of the result to gain insight into the $k$-modes available for the domain wall to produce upon its radiation. Denoting $\rho_{\rm dw}(k)$ as the Fourier transform of the energy density in the domain wall in the $z$-direction, we estimate 
\begin{equation}
    \rho_{\rm dw}(k) \propto {\rm sech^2}\left( \frac{k \pi }{2 m_a} \right).
\end{equation} 
By inspection, $\rho_\mathrm{dw}(k)$ peaks at $k \lesssim m_a$, suggesting that most axions emitted from domain wall decay are nonrelativistic or mildly semi-relativistic, in which case their energies are approximately $m_a$. In the interest of placing an upper bound on axion production from domain wall collapse, we assume below that domain walls emit zero-momentum axions.  In Sec.~\ref{sec:single_string} we provide additional evidence that domain walls radiate semi-relativistically by simulating the collapse of a single circular string and domain wall. However, we emphasize that the assumption that the domain walls radiate nonrelativistically is a crucial one that should be checked more rigorously in future work, as it plays a key role in determining the DM abundance from domain wall decay. 

Now, consider at time $t_*$ a single string loop of physical length $\ell$ which bounds a domain wall of physical radius $r=\ell/2\pi$. The energy of the domain wall is given by
\begin{equation}
E(t) = 8 \pi r(t)^2  m_a f_a^2 \,.
\end{equation}
Then, the rate at which some number $N_a$ axions are emitted by the shrinking domain wall is given by
\begin{equation}
\frac{dN_a}{dt} = 16 \pi r(t) f_a^2 \,, 
\end{equation}
where we crucially assume all of the radiated axions are nonrelativistic with energy equal to $m_a$. Integrating this expression to the time when the loop radius has shrunk to zero, we find a total of $N_a(\ell) = 2\ell^2 f_a^2 /\pi$ axions emitted by the shrinking domain wall. As in the case of strings, we integrate this emission over the loop distribution to evaluate the number density of these axions at time $2 t_*$ as
\begin{equation}
\delta n_a^\mathrm{dw, sub} = \left[ \frac{R(t_*)}{R(2 t_*)}\right]^3 \int_0^{2\pi t*} d \ell \frac{dn_\ell}{d\ell}N_a(\ell) \,.
\end{equation}
This yields the result
\begin{equation}
\frac{\delta \rho_a^{\mathrm{dw}, \mathrm{sub}}}{\rho_a} = \frac{\delta_1}{2 \pi} \xi_*^{1/2} \,.
\label{eq:subhorizon_zeromomentum}
\end{equation} 
Hence, we find that for our simulations, where $\xi_* \approx 0.8$ and $\delta_1\approx 8.6$, we expect the collapse of subhorizon domain walls to contribute roughly equally to the string emission. In the physical case, the collapse of subhorizon domain walls may be as much as 5 times more efficient in producing axions as compared to string emission prior to $t_*$. (See Tab.~\ref{table:summary_analytic_estimates_CLEAN}.)

\subsection{Collapse of Superhorizon Defects}
We now proceed to consider the case of superhorizon defects. For simplicity, we treat all superhorizon defects as having a single characteristic size. While a spectrum of superhorizon defect sizes might be physically expected, larger and larger defects grow exponentially unlikely, and so we expect quantities associated with the spectrum of superhorizon defects to be dominated by a single typical scale.

Since these superhorizon domain walls are initially frozen out, their physical radius evolves as $r(t) \propto t^{1/2}$ until a time $t_0$ defined by when $r(t_0) = t_0$; after this time they are within the horizon and they collapse relativistically. We parametrize this physical radius as 
\begin{equation}
r(t) \approx
\begin{cases}
    r_\mathrm{dw}\left(\frac{t}{t_*}\right)^{1/2} & \text{if } t_* < t < t_0 \\
     r_\mathrm{dw}\left(\frac{t_0}{t_*}\right)^{1/2} - (t-t_0) = 2 t_0 - t& \text{if } t_0 < t < t_\mathrm{coll} \\
    0 & \text{if } t_\mathrm{coll} < t \,, 
\end{cases}
\label{eq:superhorizoncases}
\end{equation}
where $r_\mathrm{dw}$ is the physical radius of the domain wall at time $t_*$. The time $t_0$ is then defined by
\begin{equation}
    r_\mathrm{dw}\left(\frac{t_0}{t_*}\right)^{1/2} = t_0.
\end{equation}

We determine $r_\mathrm{dw}$ via the observed ratio of $t_0/t_*$ in our QCD-epoch simulation, finding $r_\mathrm{dw} \approx 1.7 t_*$. Moreover, we can determine the number density of these superhorizon defects from the value $\xi_0 \approx 1.2$, the string length parameter at time $t_0$. The physical number density at $t_0$, of these superhorizon defects is then
\begin{equation}
\frac{dn}{d\ell} = \frac{\xi_0}{2\pi t_0^3} \delta(\ell - 2 \pi t_0) \,.
\end{equation}
Further, while we have heavily calibrated against our QCD-epoch simulation, we note that this model for superhorizon defects makes two nontrivial predictions. First, at times shortly before $t_0$, the string length parameter $\xi$ should grow. This is because the frozen-out superhorizon defects, which are coming to dominate the total string length, drive the growth of $\xi$ linear in $t$. Additionally, our relativistic collapse model also predicts that these superhorizon defects collapse by $t_\mathrm{coll} = 2 t_0$. Both these predictions are seen in Fig.~\ref{fig:xi_qcd}.

\subsubsection{Emission from Superhorizon Strings}
Now we consider the emission of axions from the collapse of initially superhorizon strings. Since $r_\mathrm{dw} \approx 1.7 t_*$, we have $t_0 \approx 2.9 t_*$. At this time, whether in the reduced hierarchy of the simulation or in the physical scenario, the axion mass always plays the dominant role in setting the IR cutoff on the emission spectrum. Analogous to the case of subhorizon string emission, we can compute the number density of axions produced by emission from strings that are superhorizon at time $t_*$. These strings emit at times beginning at $t_0$ when they enter the horizon until $2t_0$ when the collapse has completed. We thus compute the axion number density at $2 t_0$ as
\begin{equation}
\delta n_a^\mathrm{string, sup}(2 t_0) =  \frac{ \pi f_a^2 \xi_0}{t_0^3} \left[ \frac{R(t_0)}{R(2 t_0)}\right]^3 \int_{t_0}^{2 t_0}\frac{ \log(m_r / H)}{ \log(m_r/k_\mathrm{IR}(t))} \int_{m_a}^{m_r} \frac{dk}{k \sqrt{k^2 + m_a^2}}\,,
\end{equation}
convolving the strings' emission spectrum with the delta function number density of strings. As before, we compare the abundance of these axions to the abundance of axions from emission prior to $t_*$, given by
\begin{equation}
n_a(2t_0) = \frac{8 \pi f_a^2 \xi_*^{1/2} H(t_*)}{\delta_1}\left[ \frac{R(t_*)}{R(2 t_0)} \right]^3 \,,
\end{equation}
after redshifting to time $2 t_0$. Taking $n = 6.68$, $\xi_* = 0.8$, and $\xi_0=1.2$ to calculate this quantity for our simulation, we find
\begin{equation}
    \frac{\delta \rho_a^\mathrm{string,sup}}{\rho_a} = 0.06 \,,
\end{equation}
while in the physical extrapolation, with $\xi_*\approx 13$ and $n = 8.16$, we have
\begin{equation}
    \frac{\delta \rho_a^\mathrm{string,sup}}{\rho_a} = 0.007 \,.
\end{equation}
These results encode that at late times, the rapidly growing axion mass renders string emission highly inefficient.

\subsubsection{Emission from Superhorizon Domain Walls}
We now determine the number of axions emitted by the collapse of the superhorizon domain walls. The calculation is identical to that of the subhorizon domain walls except that the collapse process starts at $t_0$ rather than $t_*$. We then have
\begin{equation}
N_a = 16 \pi f_a^2 \int_{t_0}^{2 t_0}  r(t) = 8 \pi f_a^2 t_0^2 \,,
\end{equation}
yielding
\begin{equation}
\frac{\delta \rho_a^{\mathrm{dw}, \mathrm{sup}}}{\rho_a} \approx 0.65 \delta_1 \xi_*^{-1/2} 
\end{equation}
when comparing both evaluated at $t = 2 t_0$. In our simulations, where $\xi_* =0.8$, this suggests the collapse of initially superhorizon domain walls outproduces string emission by a factor $\sim 6.3$. For the physical scenario of $\xi_* \approx 13$, this falls to approximately $1.6$. (See Tab.~\ref{table:summary_analytic_estimates_CLEAN}.)

\subsection{Summary of axion production in the QCD epoch}
\label{SMSec:QCDCollapseSim}
We summarize our results for the efficiency of axion production at times after $t_*$ in Tab.~\ref{table:summary_analytic_estimates_CLEAN}. In general, we find that domain walls may make an appreciable or even dominant contribution to the late-time axion abundance as compared to string emission. This is primarily because domain walls may radiate axions into low-momentum modes, though this assumption should be checked more carefully in future work (see also Sec.~\ref{sec:single_string}).

\begin{table}[t!]
\centering
\setlength{\tabcolsep}{7pt} 
\renewcommand{\arraystretch}{1.3}
\begin{tabular}{@{}lcccccccc@{}}
\toprule
Scenario & $t_{m_a}/t_*$ & $t_0/t_*$ & $t_\mathrm{coll}/t_*$ & $\delta \rho^{\mathrm{string, sub}}/\rho_a$ & $\delta \rho^{\mathrm{string, sup}}/\rho_a$  & $\delta \rho^{\mathrm{dw},\,\mathrm{sub}}/\rho_a$  & $\delta \rho^{\mathrm{dw},\,\mathrm{sup}}/\rho_a$ & $\delta \rho^\mathrm{tot}/\rho_a$ \\ \midrule \hline
Simulation & 1.4  & 2.9 & 5.8  & 0.75 &  0.06  & 1.2 & 6.3 & 8.3 \\  \hline
Physical & 2.16  &2.9  & 5.8  & 0.65 &  0.007 & 4.9 & 1.6 & 7.2 \\ 
\bottomrule
\end{tabular}
\caption{
Summary of our analytic estimates for the energy densities in axions produced at times $t>t_*$ from strings and domain walls, relative to the energy density $\rho_a$ in axions emitted at times $t<t_*$. We split these into contributions from string loops/walls that are subhorizon at the oscillation time $t_*$ and those which are superhorizon (denoted by superscripts `sup', `sub', respectively). The parameters in the first (second) row correspond to our QCD epoch simulation (physical scenario). We measure $\xi_*=0.8$ and $\log(m_r/H_*)=7.34$ in our QCD epoch simulation and fix $\xi_*=13$ and $\log(m_r/H_*)=70$ in the physical scenario.}
\label{table:summary_analytic_estimates_CLEAN}
\end{table}

However, we qualify this in several ways. First, we have assumed domain walls radiate zero-momentum axions, while radiating semi-relativistic axions may reduce their production efficiency by an $\mathcal{O}(1)$ factor. Moreover, we have assumed string loops that bound domain walls are circular, which has the effect of maximizing the area at a fixed string length. Non-circular configurations would reduce the domain wall energy available for axion emission, leading to a further reduced production efficiency. As a result, the domain wall production efficiency we have estimated in this section should be considered an upper bound and represents an important direction for future study.

We go on to study the efficiency of axion-production after $t_*$ by comparing the differential number density spectrum $\partial n_a/ \partial k$ in our QCD-epoch simulation with the equivalent spectrum obtained from the string-only simulation of \cite{Buschmann:2021sdq}. Since these two simulations are identical up to the inclusion of the QCD potential, this provides the opportunity to sharply and self-consistently examine the estimates made in this section. To extract the number density spectrum from our QCD simulations, we use
\begin{equation}
\frac{\partial \rho_a}{\partial k} = \frac{1}{2}\left[\frac{\partial}{\partial k}|\dot{\tilde a}(k)|^2 + \frac{1}{2}\left(k^2 + m_a(\eta)^2 \right) \frac{\partial}{\partial k}|\tilde a(k)|^2 \right] \approx \frac{\partial}{\partial k}|\dot{\tilde a}(k)|^2\,,
\label{eq:AxionEnergyDensitySpectrum}
\end{equation}
which we relate to a physical number density by
\begin{equation}
    \frac{\partial n_a}{\partial k} = \frac{1}{\sqrt{k^2 + m_a(\eta)^2}} \frac{\partial \rho_a}{\partial k}.
\label{eq:PhysicalNumberDensity}
\end{equation}
In practice, it is convenient to work with the differential spectrum of comoving density by comoving momentum, which is constant as a function of $\eta$ when there is no active axion production.

\begin{figure}[t!]    
\includegraphics[width=0.8\textwidth]{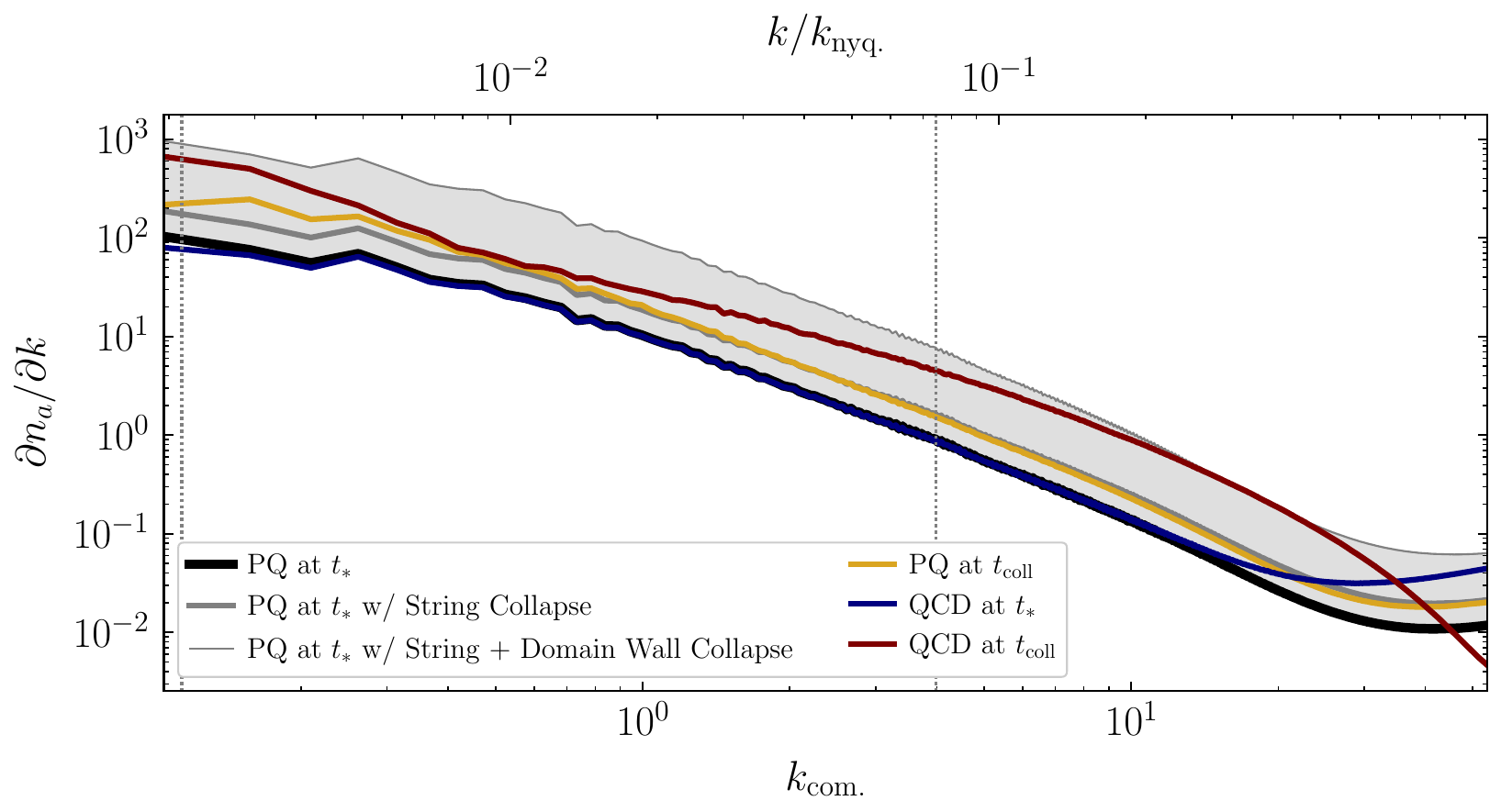}
\caption{The number density spectra realized in our QCD epoch simulation and the otherwise identical string-only simulation of \cite{Buschmann:2021sdq}. The left (right) vertical dotted gray lines indicate the axion mass at time $t_*$ ($t_\mathrm{coll.}$) in these comoving units, \textit{i.e.}, they are the values $a(\eta) m_a(\eta)$. The heavy black line, corresponding to the string-only simulation of \cite{Buschmann:2021sdq} at time $t_*$, and the navy line, corresponding to the QCD epoch simulation of this work at time $t_*$, are highly similar, with some mild discrepancy at large $k$. This is to be expected, as the axion mass has only recently begun to play an important role in the dynamics at time $t_*$. At $t_\mathrm{coll}$, when all defects have collapsed and axion production has ceased, the number density spectrum realized by our simulation is shown in maroon. Based on our parametric estimates for axion production in this section, we present two additional spectra. The heavy gray line indicates the spectrum obtained by enhancing the spectrum from the PQ simulation at $t_*$ by a factor of $1.81$. This factor represents the estimated production of axions by the populations of subhorizon and superhorizon strings at times between $t_*$ and $t_\mathrm{coll}$. The lighter gray line indicates the spectrum obtained by enhancing the spectrum from the PQ simulation at $t_*$ by a factor of $9.3$. This factor represents the estimated production of axions by both strings and domain walls at times between $t_*$ and $t_\mathrm{coll}$. By comparison, the number density spectra at $t_\mathrm{coll}$ in our PQ simulation realized by undisrupted string emission is shown in yellow. Note that our calculation of the number of axions produced by domain walls assumes a primarily nonrelativistic spectrum, so applying this enhancement factor to the PQ spectrum is not strictly correct but is useful in determining where axion production is most efficient during the times between $t_*$ and $t_\mathrm{coll}$. The heavy black and light gray lines derived from the PQ-simulation spectrum at time $t_*$ are used to define the lightly-shaded gray band.}
\label{fig:NumberDensityComparison}
\end{figure}

In Fig.~\ref{fig:NumberDensityComparison}, we compare these differential comoving number densities between the two simulations. By integrating these comoving number densities, we find the comoving axion abundance at $t_\mathrm{coll}$ in our QCD simulation to be roughly 4.5 times larger than the axion abundance at $t_*$ in our equivalent PQ simulation. Integrating only modes with $k_{com} > 10 a(t_*) H(t_*)$ to avoid standard misalignment, which dominantly contributes to modes at or below the comoving horizon at $t_*$, given by $a(t_*) H(t_*)$, this ratio falls to 3.5. 

By comparison, based on a scale-invariant emission spectrum, we expect the strings to radiate axions with an efficiency that enhances the axion production by only a total factor of $\sim$1.8, suggesting that domain walls are contributing a roughly equal amount to the late time abundance of axions as strings. This is less than our simple estimate of domain walls making a contribution as much as $7.5$ times larger than string emissions prior to $t_*$ under the assumption that they emit nonrelativistic axions. This suggests that domain walls are emitting somewhat relativistic axions with $\mathcal{O}(1)$ boost factors. As we will see in later sections, we see further evidence that axions emitted by domain walls are at least somewhat relativistic. However, even if domain walls are emitting axions with $\mathcal{O}(1)$ boost factors, our estimates when extrapolated to the physical hierarchy still indicate that axion emission by domain walls will contribute nearly half of their late time abundance.

\section{Collapse of a String-Domain Wall Loop}
\label{sec:single_string}
\begin{figure}[t!]    
\includegraphics[width=1\textwidth]{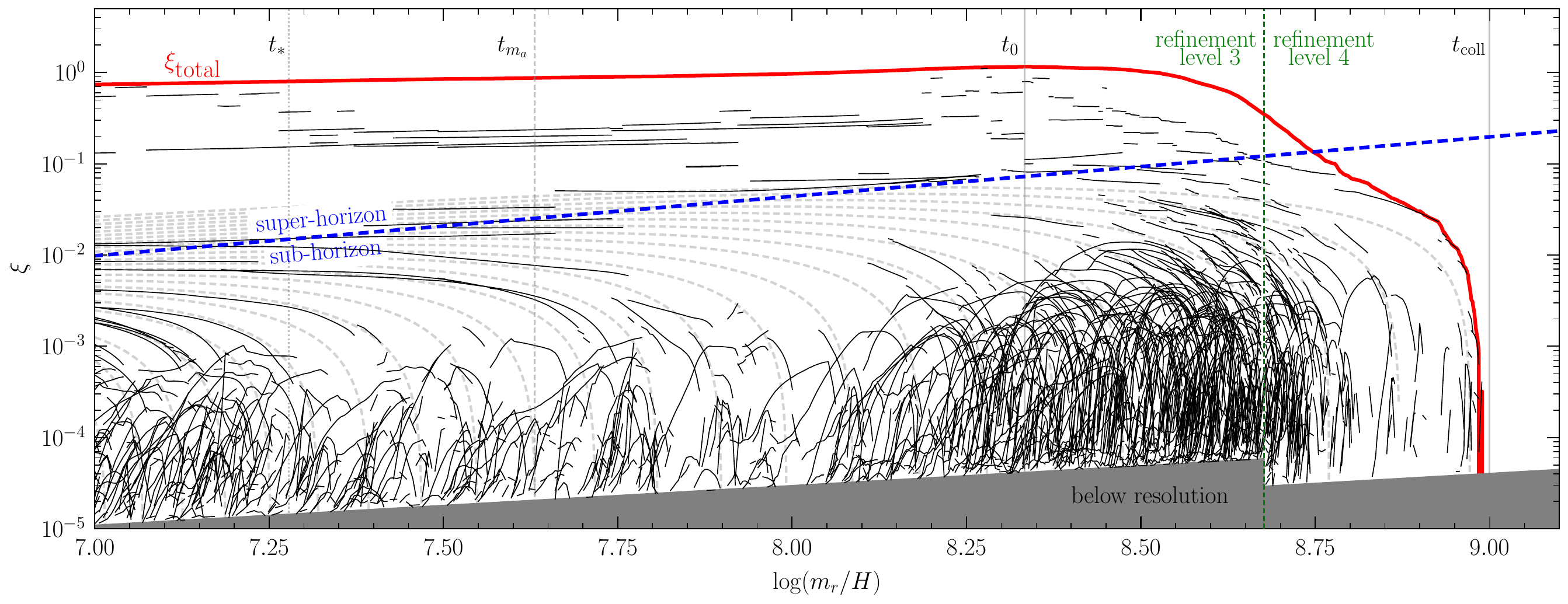}
    \caption{String length evolution in our QCD epoch simulation. Black lines are individual strings with the total string length indicated in red. The black lines often stop or appear suddenly due to string reconnections and break-ups. The results are overlayed with gray dashed lines of a few examples of the string collapse model described in the text. The threshold between superhorizon and subhorizon loops is indicated by the dashed blue line and the simulation transitions from three to four refinement levels at the vertical green dashed line. Only strings with at least twenty string-plaquette piercings are taken into account as indicated by the grayed-out area at the bottom of the plot. The reference times $t_*$, $t_{m_a}$, $t_0$, and $t_\textrm{coll}$ are highlighted as well.}
    \label{fig:individual_collapsing_string_loops}
\end{figure}

We study in further detail the collapse of a loop of string bounding a domain wall in an expanding background. Let us consider at time $\eta$, a string loop bounding a domain wall that has comoving radius $r$. We parametrize the spacetime location of a domain wall element with  $(\eta, R, \phi)$ via the general map to the four-vector
\begin{equation}
    \begin{pmatrix}\eta \\ R \\ \phi
    \end{pmatrix} \rightarrow 
    \begin{pmatrix}
    \frac{\eta^2}{2 f_a} \\ R r(\eta) \cos \phi \\ R r(\eta)\sin\phi
    \\ 0
    \end{pmatrix} \,,
\end{equation}
where $\phi$ is the azimuthal coordinate taking values in $[0, 2\pi)$, $r(\eta)$ is the maximum comoving radius of the domain wall at time $\eta$, and $R$ is the relative location on the domain wall taking values in $[0, 1]$. (Note that $R$ should not be confused with the scale factor.) This parametrization specifies an induced metric $g_\mathrm{dw}$ on the domain wall, from which we derive the domain wall action
\begin{equation}
S_\mathrm{dw} = -\int d\tau \sigma(\tau)  \int_0^1 dR  \int_0^{2\pi} d\phi \sqrt{|g|} = - \int d\tau \left[ 8 \pi f_a^2 \eta ^3 m_a (\eta ) r(\eta )^2\right].
\end{equation}
The associated string parametrization is 
\begin{equation}
    \begin{pmatrix}\eta \\ \phi
    \end{pmatrix} \rightarrow 
    \begin{pmatrix}
    \frac{\eta^2}{2 f_a} \\  r(\eta) \cos \phi \\  r(\eta)\sin\phi
    \\ 0
    \end{pmatrix} \,,
\end{equation}
from which we obtain the string action
\begin{equation}
    S_\mathrm{string} = - \int d\tau \mu(\tau) \int_0^{2\pi} d\phi \sqrt{|g|} = -\int d\tau \left[ 2 \pi ^2 f_a^2 \eta ^2 \log\left(\frac{m_r}{H}\right) r(\eta ) \sqrt{1 -  r'(\eta )^2 } \right] \,.
\end{equation}
For simplicity, we will approximate $\log(m_r/H)$ to be constant. From this action, we may derive the equation of motion for $r(\eta)$ with the Euler-Lagrange equations, obtaining
\begin{equation}
r'' = \frac{\left(\eta +2 r r'\right) \left(r'^2-1\right)}{\eta  r} -\frac{8 \eta  m_a(\eta ) \left(1-r'^2\right)^{3/2}}{\pi  \log(m_r/H)}.
\label{eq:loopCollapse}
\end{equation}
For a circular loop of string, the comoving radius $r(\eta)$ is related to the string length parameter by $\xi = \pi \eta^2 r(\eta)/ 2 H_1^2 L^3$ where $L$ is the comoving box length. We compare the $\xi$ associated with the numerical solution of ~\eqref{eq:loopCollapse} to the string evolution measured in our QCD epoch simulation in Fig.~\ref{fig:individual_collapsing_string_loops}. Specifically, we identify strings and track their individual contributions to $\xi$ with the $i^\mathrm{th}$ string contributing $\xi_i$ and the total string length parameter being $\xi = \sum_i \xi_i$.

We observe subhorizon strings collapsing on relatively short time scales whereas superhorizon strings decay very little until $t_0$ besides frequent reconnections. After $t_0$, the superhorizon loops undergo a rapid decay driven by the domain wall tension. This quick decay is often accompanied by the shedding of smaller loops near kinks, which can be visually seen by the simultaneous emergence of many subhorizon loops. While these effects are not captured by ~\eqref{eq:loopCollapse}, the collapse of individual strings is fairly well described by our simple isolated loop model. See also the top left panel of Fig.~\ref{fig:StringEvolutionEmission}.

We also note that a key feature of our analytic treatment of collapsing strings and domain walls at times was the assumption of defects that were superhorizon at $t_*$. These superhorizon defects are clearly visible in Fig.~\ref{fig:StringEvolutionEmission} as those string trajectories that lie above the dashed blue line defining the horizon diameter for circular string loops. These large loops demonstrate the expected $t^{1/2}$ growth until times as late as roughly $t_0$, at which time they begin to collapse efficiently.

\section{Simulation of a single axion string and domain wall}
\label{sec:SingleString}
To improve our understanding of axion emission in the QCD epoch, in particular the emission from domain walls, we perform a simulation of a single axion string that comes to bound a collapsing domain wall. To do so we engineer an initial state that will result in a perfectly circular string by setting
\begin{equation}
\begin{split}
\psi_1(\mathbf{x})&=\frac{2}{1 + e^{-(\left|\mathbf{x}-N/2\right| - N/4)/20}} - 1, \qquad 
\psi_2(\mathbf{x})=\sin\left(2\pi\frac{k}{N}\right) \,,
\end{split}
\end{equation}
with $N$ the number of coarse-level cells in each spatial direction, and $\mathbf{x}=(i,j,k)^T$ the index space with $i,j,k=\{0,\dots, N-1\}$. The diameter of the string is exactly half of the simulation volume at the beginning of the simulation, and the domain wall forms on the inside of the string. The simulation starts as a normal PQ phase simulation with an identical setup to our primary simulation with $\bar L=200$. Since we are dealing with an isolated string, however, we can reduce the coarse level to 1,024$^3$ grid cells compensated by a larger number of refinement levels. To form a domain wall we adiabatically turn on the axion mass using the same parameterization as in our QCD simulation of ~\eqref{eq:axion_mass_parameterization} with $N=19$, $\eta_*=50$, and $n=6.68$. The adiabatic period is described by a logistic function 
\begin{equation}
m_a(\eta) \rightarrow \frac{1}{1 + e^{-f (\eta - \eta_0)}} m_a(\eta=35)\,,
\end{equation}
with $f=3$ and $\eta_0=1$. This implies that beyond this brief adiabatic regime, the axion mass remains constant. We check that neither the exact initial state setup nor the exact parameterization of the adiabatic regime significantly affects the result. Due to its smaller size, we were able to run this simulation on GPUs. We used 128 nodes of the Perlmutter GPU cluster, which provides 512 NVIDIA A100 (40 GB) GPUs, 8,192 CPU cores, and 66 TB of aggregate memory. The total run time was around 3.5 hours.

\begin{figure}[t!]
    \includegraphics[width=1\textwidth]{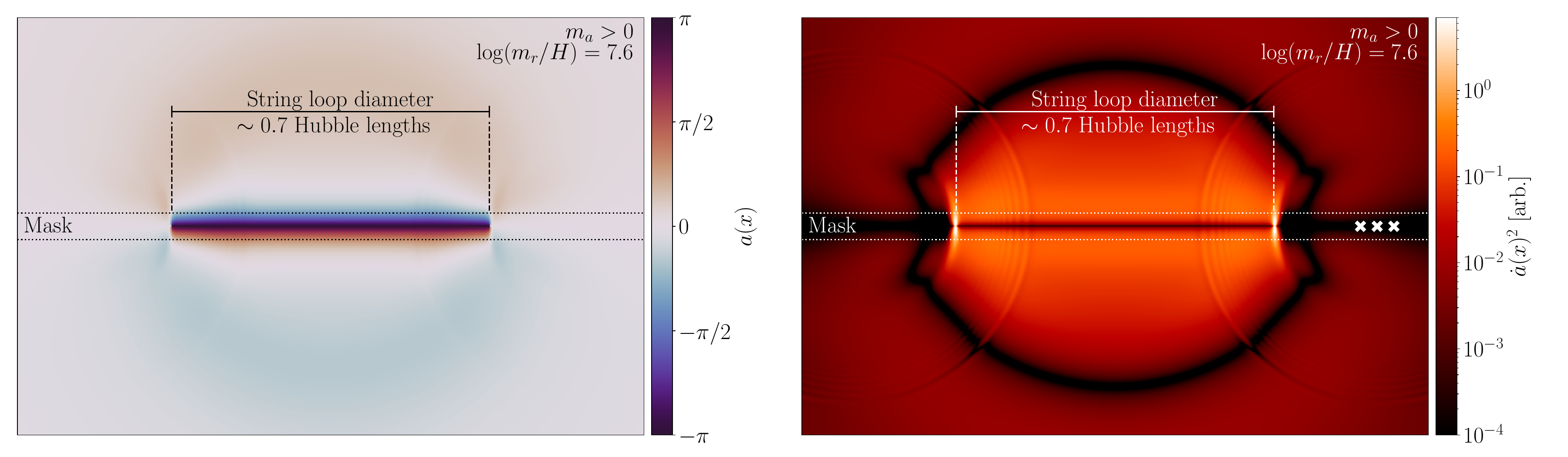}\\
    \includegraphics[width=1\textwidth]{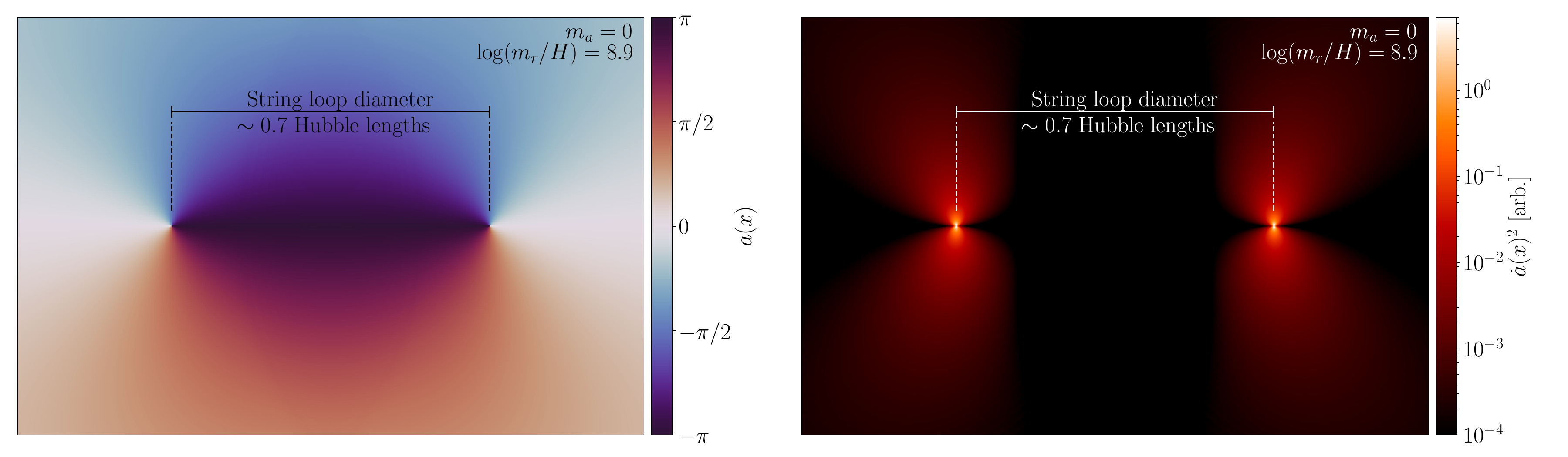}
    \caption{Cut through the axion field $a(x)$ \textsl{(left column)} and the axion kinetic energy $\dot{a}(x)^2$ \textsl{(right column)} of a single circular string in a simulation with constant non-zero $m_a$ \textsl{(top row)} and $m_a=0$ \textsl{(bottom row)}. The string is piercing the slice twice at a distance of roughly 0.7 Hubble lengths. A cut through the domain wall is visible in the top row. The dotted horizontal lines indicate the mask encompassing the plane of the domain wall. Both simulations are based on the same initial state. Note that the string in the $m_a>0$ case reaches a radius of 0.7 Hubble lengths more quickly due to the extra domain wall tension. As a result, the top row is a snapshot at an earlier time with $\log(m_r/H)\sim7.6$ and the bottom row at $\log(m_r/H)\sim 8.9$.}
    \label{fig:SingleStringCut}
\end{figure}

\begin{figure}[t!]
    \includegraphics[width=0.75\textwidth]{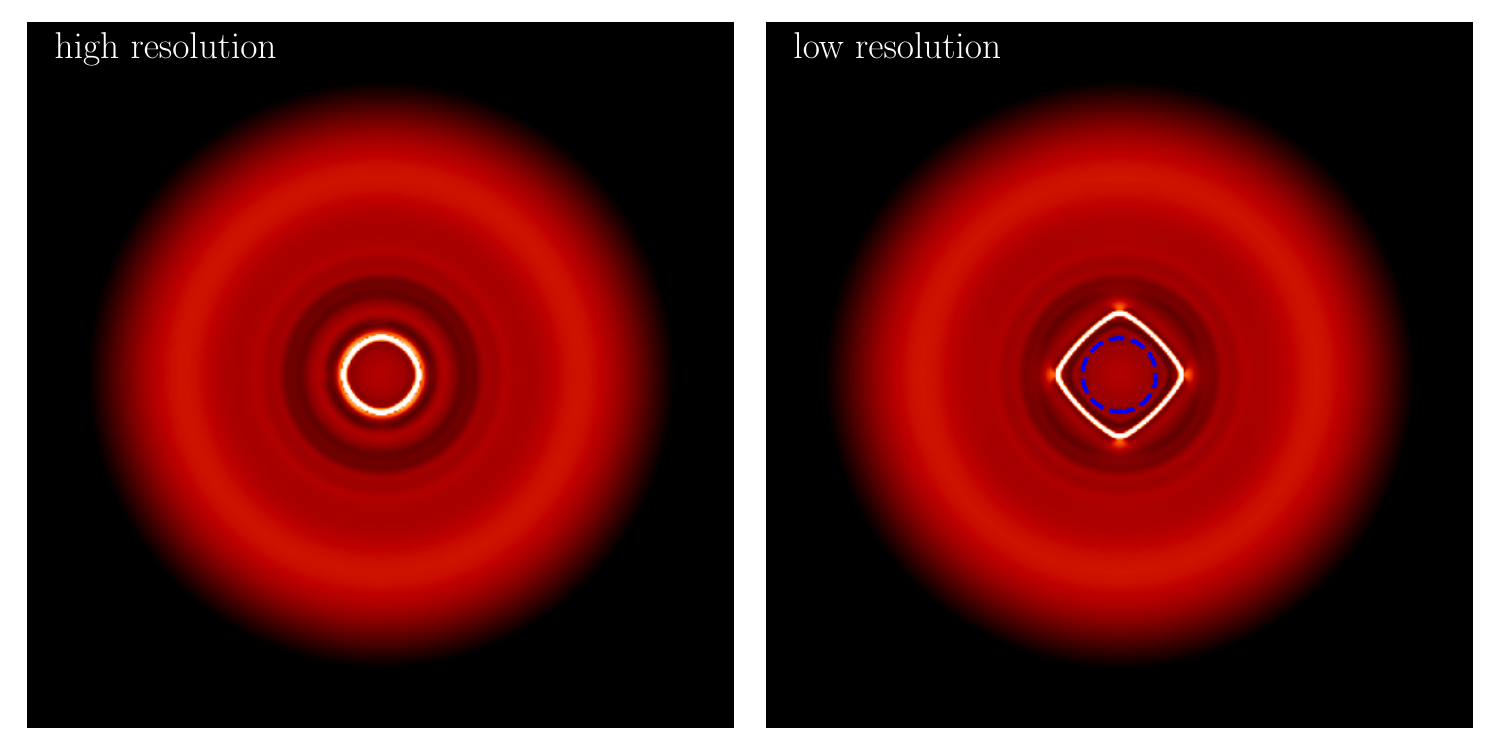}
    \caption{Comparison of a single circular string in the QCD epoch between a high-resolution setup \textsl{(left)} and a low-resolution setup \textsl{(right)} shortly before the strings final collapse at $\log(m_r/H)\sim8.7$. The 3D$\rightarrow$2D projection of $\dot{a}(x)^2$ on a logarithmic color scale is shown. Each panel spans roughly 0.9 Hubble lengths. The contour of the high-resolution string is overlayed in blue in the low-resolution panel. In the low-resolution simulation, the string collapses more slowly and loses its circular shape with facets developing diagonally to the underlying grid. This introduces artificial kinks that produce extra radiation.
    }
    \label{fig:SingleStringDx}
\end{figure}

We extract the axion emission spectra in the usual way but additionally mask out the domain wall by setting $\dot{a}^2 \rightarrow 0$ within the planar slab with width $6/m_a$ centered on the domain wall, which has characteristic width $1/m_a$. This masking region is illustrated in Fig.~\ref{fig:SingleStringCut}. A top-down projection is shown in Fig.~\ref{fig:SingleStringDx} for two simulations with different resolution criteria. We use the high-resolution simulation in our analysis. 

\begin{figure}[t!]
    \includegraphics[width=0.8\textwidth]{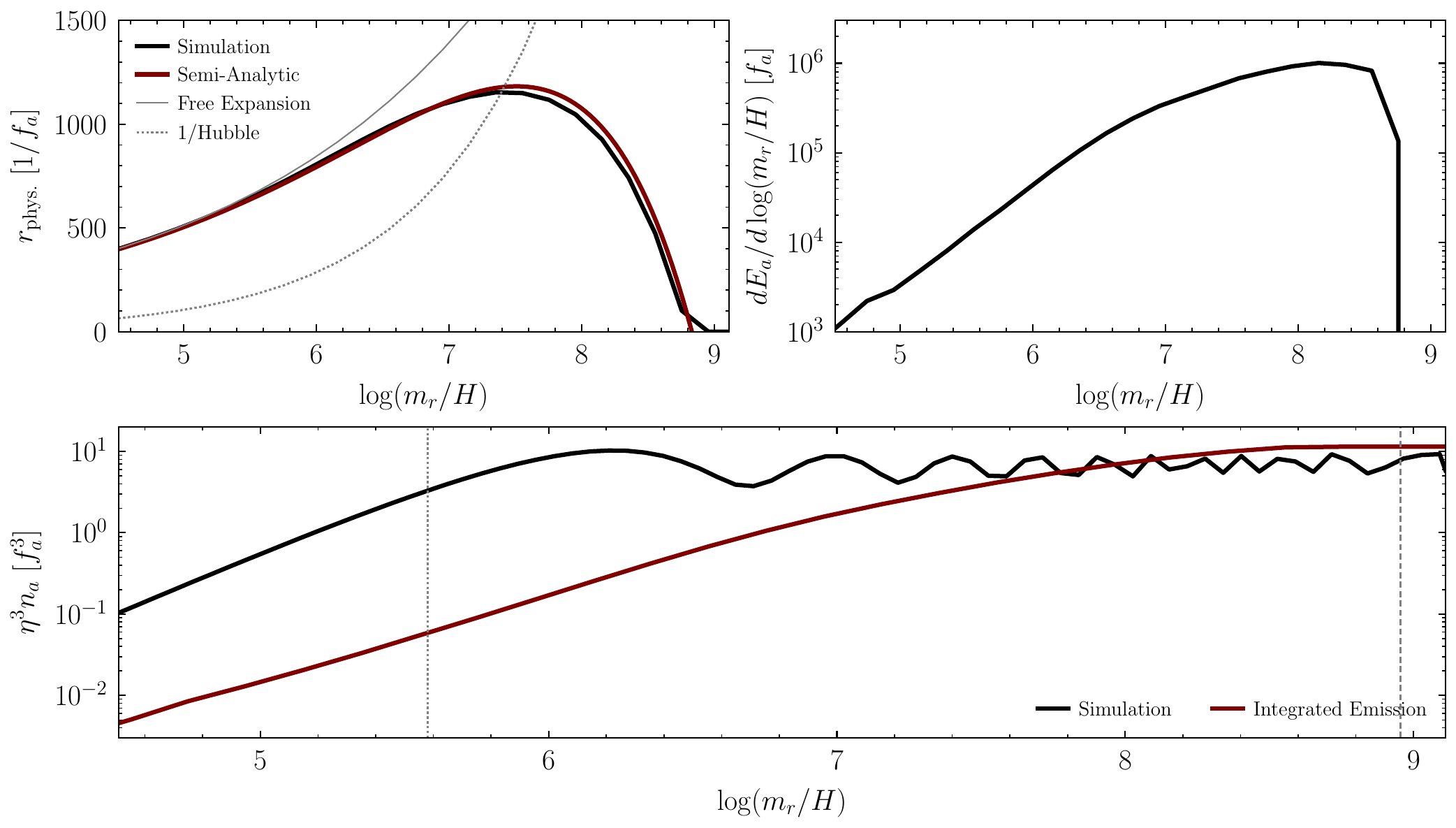}
    \caption{(\textit{Top Left}) The time evolution of the physical radius in our single string loop simulations. In solid black, we depict the physical radius of the string loop measured directly in the simulation, which we compare to the evolution of the string assuming free expansion with Hubble flow (gray) and inverse Hubble which defines the horizon size (dotted gray). When the radius of the string loop is much larger than the horizon, the size of the loop grows with $\eta$ and only begins to contract efficiently once $r_\mathrm{phys} H < 1$. We also compare the observed loop evolution with that expected by solving ~\eqref{eq:loopCollapse} as shown in red, which demonstrates qualitatively good agreement. (\textit{Top Right}) The instantaneous rate of axion emission per logarithmic interval was obtained via numerical evaluation of ~\eqref{eq:WallEmissionRate}. We see that the rate of axion emission is expected to peak around $\log(m_r/H) \approx 8.2$. (\textit{Bottom}) A comparison of the total number density of axions in our string loop simulation (black) as compared to the prediction from integrating axion emission associated with a contracting string loop under the assumption of a nonrelativistic emission spectrum. By multiplying by a factor of $\eta^3$, we compare comoving number density, which is constant at times when axions are not produced efficiently. The integrated emission number density exceeds that observed in our simulations, suggesting that the emission of axions from a shrinking domain wall is at least quasi-relativistic. See text for details.}
    \label{fig:StringEvolutionEmission}
\end{figure}

One means of examining the axion emission spectrum from domain walls is to compare the number density of axions realized in our simulation with the number density one would predict associated with the shrinking of the domain wall given an assumed emission spectrum. Consider at some time $t_0$ a domain wall of physical radius $r(t_0)$. Then the energy associated with that domain wall is
\begin{equation}
    E_\mathrm{wall} = 8 \pi m_a f_a^2 r(t_0)^2 \,.
\end{equation}
If the wall were not contracting then its energy would grow like $a(t)^2 \propto (t/t_0)$. We define the free wall energy
\begin{equation}
E_\mathrm{wall}^\mathrm{free}(t | r(t_0), t_0) = 8 \pi m_a f_a^2 r(t_0)^2 \left(\frac{t}{t_0} \right)\,,
\end{equation}
leading to an instantaneous rate of change in the domain wall energy of 
\begin{equation}
\frac{\partial E_\mathrm{wall}^\mathrm{free}}{\partial t} = \frac{8 \pi m_a f_a^2  r(t_0)^2}{t_0} \,.
\end{equation}
On the other hand, from our simulation, we numerically extract $r(t)$ and $\dot r(t)$, so that the change in the wall energy is
\begin{equation}
\frac{\partial E_\mathrm{wall}^\mathrm{sim}}{\partial t} =  16 \pi m_a f_a^2  r(t_0) \dot r(t_0) \,.
\end{equation}
The energy loss computed in our simulation is given by the difference of the emission into axions less the free expansion energy gain. Hence, from our simulation, we can compute the emission into axions as 
\begin{equation}
\frac{\partial E_a(t)}{\partial t} = -\frac{\partial E_\mathrm{wall}^\mathrm{free}}{\partial t} - \frac{\partial E_\mathrm{wall}^\mathrm{sim}}{\partial t} = 16 \pi m_a f_a^2 r(t) \left[\frac{r(t)}{2 t}- \dot{r}(t) \right] \,.
\label{eq:WallEmissionRate}
\end{equation}
Assuming that axions emitted by the wall collapse are nonrelativistic, the instantaneous rate of axion emission is 
\begin{equation}
\frac{\partial N_a}{\partial t} = 8 \pi f_a^2 r(t) \left[\frac{r(t)}{ t}- 2 \dot{r}(t) \right] \,.
\end{equation}
We evaluate this numerically and divide it by the physical volume of our simulation to determine the physical number density $n_a$. We then compare this integrated axion emission to the number density realized in our simulations, with the results shown in Fig.~\ref{fig:StringEvolutionEmission}.

It is clear from the results in Fig.~\ref{fig:StringEvolutionEmission} that the majority of the axions in our simulation volume were not produced by domain wall emission. Namely, the time-dependence of the number produced by domain wall emission does not match the time dependence of axions in the simulation, which saturates much earlier. The majority of axions produced in this simulation are likely generated by spurious modes realized in the initial conditions or the transient as the axion enters the horizon. On the other hand, the number of axions produced by domain wall collapse under the assumption of emission into nonrelativistic modes would nonetheless overproduce the observed axion number density. This suggests that the axions which are emitted from the domain wall must be at least somewhat relativistic as a harder emission spectrum would produce a smaller axion number density.

\begin{figure}[t!]
    \includegraphics[width=0.6\textwidth]{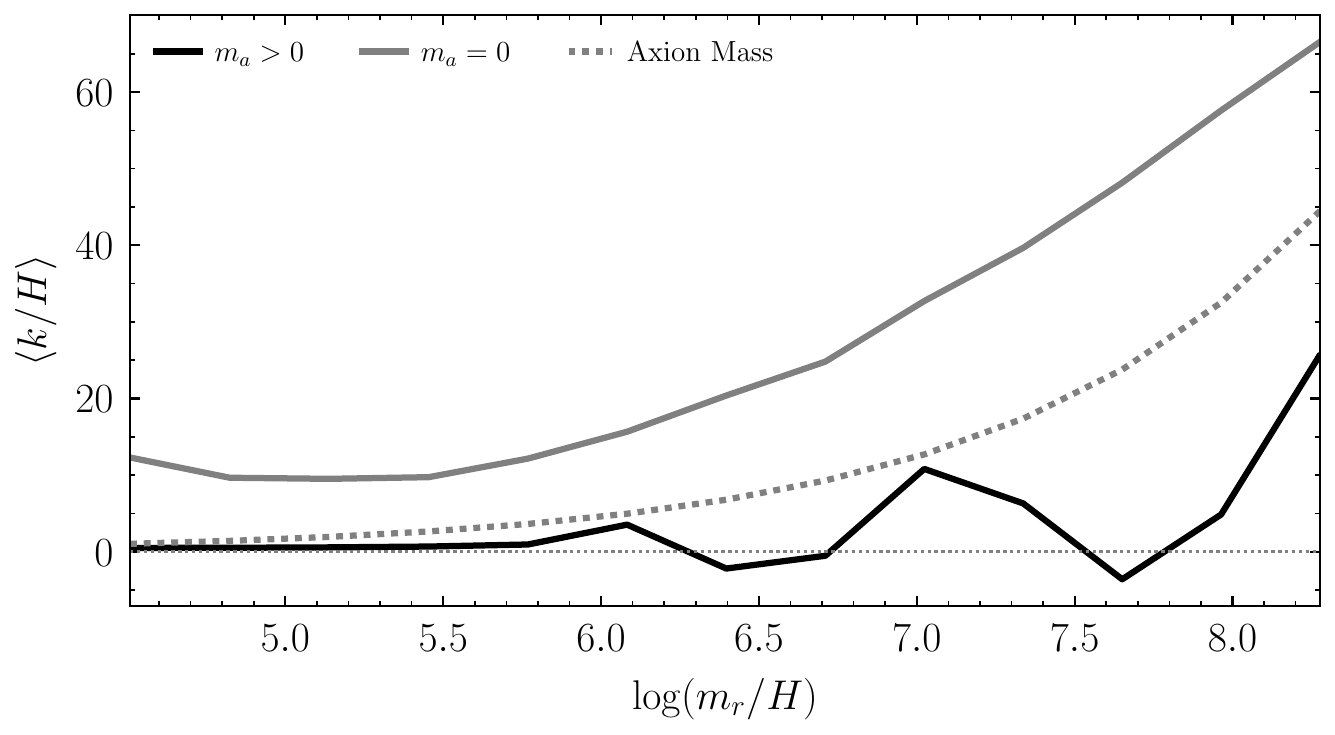}
    \caption{A comparison of the typical momentum of axion emission computed for two simulations using an identical initial state, but with one simulating $m_a =0$ (gray) while the other simulates  $m_a > 0$ (black), such that a domain wall forms and collapses by the end of the run. 
    In dotted gray, we show the ratio of the axion mass used in our $m_a \neq 0$ simulation to the Hubble expansion rate.}
    \label{fig:ExpectationValue}
\end{figure}

Another means of examining the axion production in the domain wall collapse is to study the typical momentum with which axions are emitted. To compute this, as in the main text, we evaluate $F(k/H)$ at each time in our wall collapse simulation and compute $\langle k/H \rangle = (\int dk (k/H) F(k/H)) / (\int dk F(k/H))$. For comparison we also perform a string-only simulation, \textit{i.e.}, with $m_a = 0$, using an identical initial condition, and similarly calculate $\langle k/H \rangle$. The results are shown in Fig.~\ref{fig:ExpectationValue}.
Broadly, $\langle k/H \rangle$ is reduced in the domain wall simulation with $m_a > 0 $ as compared to the string-only simulation with $m_a = 0$ and is typically less than the axion mass. This suggests that axions produced in our domain wall simulations are typically non- or semi-relativistic, though we caution that these results could be somewhat contaminated by unphysical modes produced from transient effects of the initial state. 

\section{Simulating axion number non-conservation during QCD phase transition}
\label{SMSec:NumberDensitySim}
The cosine potential experienced by the axion is an intrinsically nonlinear one that supports number density-changing processes. We consider the role of these processes on axions emitted by strings by performing an axion-only simulation with the equations of motion
\begin{equation}
    \theta'' + \frac{2}{\eta}\theta' - \bar\nabla^2 \theta + \frac{\eta^2 m_a(\eta)^2}{f_a^2} \sin\theta = 0 \,,
\end{equation}
where $\theta$ is a compact field varying between $0$ and $2\pi$ related to the axion field $a$ by $\theta \equiv a/f_a$. Note we have otherwise maintained a consistent set of simulation units and axion mass parametrization from other simulations considered in this work. 

We generate initial conditions at $\eta \approx 33$ when the axion begins to oscillate in a box of comoving sidelength $\bar L \approx 66$ such that the simulation contains 8 Hubble volumes at the time of oscillation. We take the axion to oscillate at $\log(m_r/H_\mathrm{osc}) \approx 65$, and assume an emission history between $\log(m_r/H) = 1$ and $\log(m_r/H_\mathrm{osc})$ described by a normalized instantaneous emission spectrum of the form
\begin{equation}
    F(k/H, t) \propto \frac{1}{k/H(t)} \,,
\end{equation}
with support at $k/H$ between $6 \xi^{1/2}$ and $m_r/H$. We adopt our logarithmically growing fit to describe the time evolution of $\xi$. We then obtain the total energy in axions by integrating this emission history as 
\begin{equation}
    \frac{\partial \rho_a}{\partial k}(t) = \int dt' \frac{\Gamma(t')}{H(t')} \left(\frac{R(t')}{R(t)} \right)^3 F\left[\frac{k R(t)}{R(t') H(t')}, t'\right] \,,
\end{equation}
where $\Gamma(t) = 8 \pi H(t)^3 f_a^2 \xi(t) \log(m_r / H(t))$ is the string decay rate. 
We may relate the energy density spectrum to the axion and axion derivative spectrum via \eqref{eq:AxionEnergyDensitySpectrum}
assuming energy is shared equally between the kinetic term set by $|\dot{\tilde{a}}|^2$ and the gradient and mass energy term set by $(k^2 + m_a^2) |\tilde{a}|$. However, we allow the phases of the $a(k)$ and $\dot a(k)$ to be random and uncorrelated. 

We generate our initial conditions on a $512^3$ lattice without adaptive meshing, zeroing out support in the momentum spectrum of $a(k)$ and $\dot a(k)$ for momentum within a factor of three of the Nyquist momentum. This ensures that our simulations are initially well-resolved. We then simulate from $\eta_i \approx 33$ to $\eta = 76$, corresponding to the end time of our full QCD simulation after defects have fully collapsed. Note that because we have generated initial conditions directly from the axion field, domain walls but not strings may be realized in our field configuration.

\begin{figure}[t!]
    \includegraphics[width=0.95\textwidth]{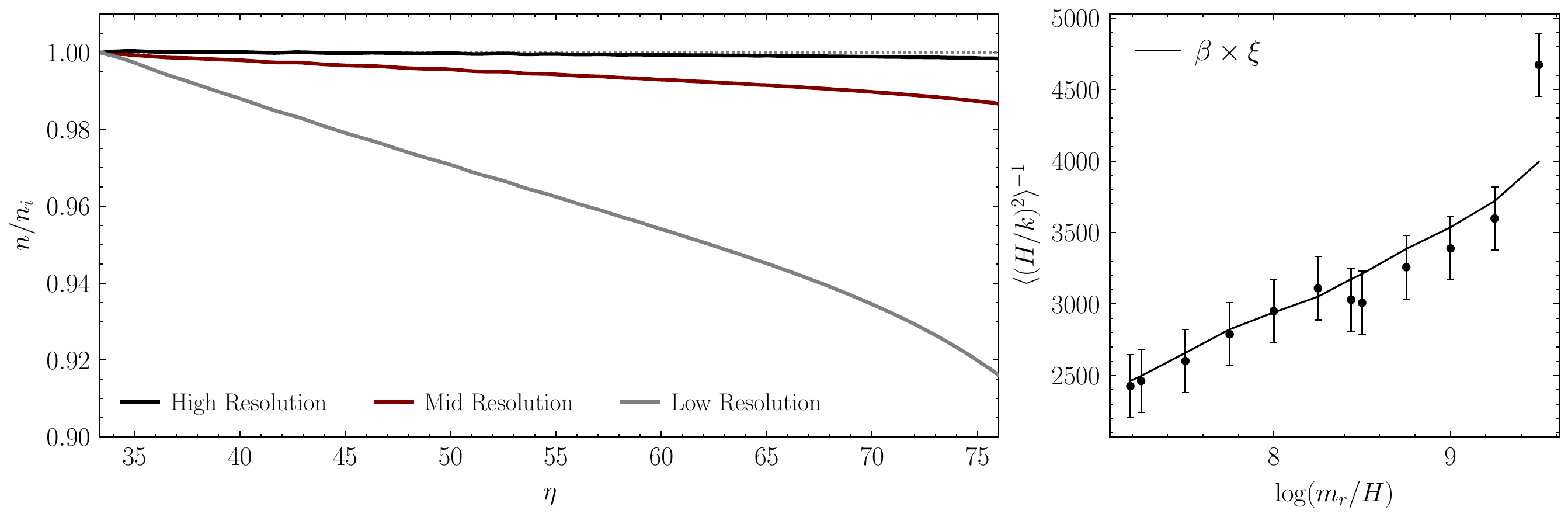}
    \caption{(\textit{Left}) The time-evolution of the comoving axion number density relative to the initial axion number density in three simulations of an identical initial state at three choices of lattice resolution. Perfect number conservation would be expected to result in a constant line at $1$, while decreases in the comoving number density are indicative of number density loss, associated with either discretization error or number-changing physical processes. For more details, see text. (\textit{Right}) The inverse expectation value $\langle (H/k)^2\rangle^{-1}$ (data points) of the axion emission spectrum $F$ for our fiducial spectrum with $k_\mathrm{IR}/H = 50$ (plotted in Fig. \ref{fig:FiducialSpectrum}). Here we assume the spectral index is at its upper limit at $1\sigma$, $q=1.02$. We fit $\langle (H/k)^2\rangle^{-1}=\beta \xi$ for a constant $\beta$ (best fit model in solid), finding $\beta= 3226.2 \pm 65.2$. We fix  $\log(m_r/H_\mathrm{QCD})=70$.}
    \label{fig:StringInjection}
\end{figure}

From our simulation, we calculate the comoving axion number density via \eqref{eq:PhysicalNumberDensity}, including both field and field derivative terms in $\partial \rho_a/\partial k$. If number density is indeed conserved over these simulations, then we expect this comoving number density to be constant. In the left panel of Fig.~\ref{fig:StringInjection}, we depict the ratio of the comoving number density evaluated in the simulation to the initial comoving number density. The results using the $512^3$ lattice are shown in gray, where we observe $\mathcal{O}(10\%)$ loss in the axion number by the end of our simulation. To test the origin of this loss, we upsample the initial state to a $1024^3$ lattice, finding the red line, which exhibits only $1\%$ loss in the number density. We then perform one additional simulation with a $2048^3$ lattice, which achieves a resolution comparable to the coarse-level resolution of our network collapse simulation. The number density evolution is shown in black and reveals entirely negligible loss.

We conclude that the nonlinear potential is inefficient in changing the number density of axions emitted by strings. However, these simulations establish that appreciable spatial/temporal resolution is an important criterion for ensuring convergence, even in the axion-only scenario.

Following the method of Ref. \cite{Buschmann:2021sdq} we also directly verify that nonlinearities in the axion field are small in our primary simulation. We fit $\langle (H/k)^2\rangle^{-1}=\beta \xi$ for a constant $\beta$ (see the right panel of Fig. \ref{fig:StringInjection}) and estimate the nonlinearity of the axion field $\langle (a/f_a)^2 \rangle \approx (4\pi/\beta)\log(m_r/H) \le 0.27 $ for $\log(m_r/H) \le 70$, consistent with our measurement of number density loss at the sub-percent level.

\end{document}